\documentclass[twocolumn]{aastex63}

\usepackage{graphicx}
\usepackage{wrapfig}
\usepackage[figuresleft]{rotating}
\usepackage{epstopdf}
\usepackage{float}
\usepackage{amsmath}
\usepackage{multirow}
\usepackage[mathscr]{eucal}
\usepackage{hyperref}
\newcommand{\Msun}{M_{\odot}}
\newcommand{\Lsun}{L_{\odot}}

\newcommand{\kms}{\mbox{${\rm km~s^{-1}}$}}
\newcommand{\kmsMpc}{\mbox{km s$^{-1}$ Mpc$^{-1}$}}

\newcommand{\p}{^\prime}

\begin{document}

\title{The 10~Mpc Volume}

\
\author[0000-0002-9291-1981]{R. Brent Tully}
\email{tully@ifa.hawaii.edu}
\affil{Institute for Astronomy, University of Hawaii, 2680 Woodlawn Drive, Honolulu, HI 96822, USA}

\author[0000-0002-5259-2314]{Gagandeep S.\ Anand}
\affiliation{Space Telescope Science Institute, 3700 San Martin Drive, Baltimore, MD 21218, USA}

\author[0000-0003-0509-1776]{H\'el\`ene M. Courtois}
\affil{University of Lyon, UCB Lyon 1, CNRS/IN2P3, IUF, IP2I Lyon, France}

\author[0000-0003-0307-4344]{Igor D. Karachentsev}
\affiliation{Special Astrophysical Observatory, Russian Academy of Sciences, Nizhniy Arkhyz, Karachai-Cherkessia 369167, Russia}

\author[0009-0007-3326-4191]{Valentina E. Karachentseva}
\affiliation{Main Astronomical Observatory, National Academy of Sciences of Ukraine, Kiev, 03143, Ukraine}

\author[0000-0002-5514-3354]{Ehsan Kourkchi}
\email{ehsan@ifa.hawaii.edu}
\affil{Institute for Astronomy, University of Hawaii, 2680 Woodlawn Drive, Honolulu, HI 96822, USA}

\author[0000-0001-9110-3221]{Dmitry I. Makarov}
\affiliation{Special Astrophysical Observatory, Russian Academy of Sciences, Nizhniy Arkhyz, Karachai-Cherkessia 369167, Russia}

\author[0000-0003-0736-7609]{Lidia N. Makarova}
\affiliation{Special Astrophysical Observatory, Russian Academy of Sciences, Nizhniy Arkhyz, Karachai-Cherkessia 369167, Russia}

\author[0000-0003-2038-0488]{Daniel Pomar\`ede}
\affil{Institut de Recherche sur les Lois Fondamentales de l'Univers, CEA Universit\'e Paris-Saclay, 91191 Gif-sur-Yvette, France}

\author[0000-0002-3234-8699]{Edward J. Shaya}
\affiliation{Astronomy Department, University of Maryland, College Park, MD 20743, USA}

\begin{abstract}
Accurate distances are now known for most unobscured galaxies brighter than $M_B = -13$~mag within 10~Mpc from Hubble Space Telescope observations of their luminous stars.
Complementary mid-infrared observations with the Wide-field Infrared Explorer provide global stellar mass estimates for these galaxies.
Consequences are detailed knowledge of the properties of the Universe in a volume-limited space. Galaxies tend to lie within groups, within filaments. The filaments experience bulk motions in response to well defined gravitational attractors such as the Virgo Cluster and repulsors such as the Local Void. The groups are inevitably dominated by a single galaxy, with 10 galaxies between one and nine Mpc containing most of the stellar mass within the sample. There is a curious lack of galaxies in an interval fainter than these dominant galaxies.
The dominant galaxies in halos tend to have gas reservoirs and ongoing star formation, a property extending to dwarf galaxies in separate halos.
Secondary galaxies in halos are frequently gas depleted and lack evidence of star formation.
With one notable exception, all galaxies in the complete sample have well formed red giant branches indicative of early-onset star formation.

\end{abstract}

\keywords{
\href{http://astrothesaurus.org/uat/590}{Galaxy distances (590)}; \href{http://astrothesaurus.org/uat/1560}{Spiral galaxies (1560)}; \href{http://astrothesaurus.org/uat/758}{Hubble constant (758)}; \href{http://astrothesaurus.org/uat/690}{H I line emission (690)}; \href{http://astrothesaurus.org/uat/205}{Catalogs (205)}; \href{http://astrothesaurus.org/uat/395}{Distance measure (395)};
}

%%%%%%%%%%%%%%%%%%%%%%%%%%%%%%%%%%%%%%%%%%%%%%%%%%%%%%%%%%%%%%%%%%%%%%%%%
%%%%%%%%%%%%%%%%%%%%%%%%%%%%%%%%%%%%%%%%%%%%%%%%%%%%%%%%%%%%%%%%%%%%%%%%%
%%%%%%%%%%%%%%%%%%%%%%%%%%%%%%%%%%%%%%%%%%%%%%%%%%%%%%%%%%%%%%%%%%%%%%%%%
\section{Introduction}
\label{sec:intro}

It will be a remarkable legacy of Hubble Space Telescope (HST) that, within 10~Mpc, essentially every unobscured galaxy more luminous than extreme dwarfs has been observed in a manner sufficient to resolve the brightest Population I and II stars and to measure a distance with an accuracy of 5\% \citep{2021AJ....162...80A}. With little ambiguity, all massive halos in the volume can be identified and there are reliable linkages of all significant galaxies within those halos \citep{2017ApJ...843...16K, 2021AN....342..999K}. The HST observations resolve stars to at least as faint as $M_I \sim -3$ for all these galaxies, providing a record of the ages and metal content of stellar populations.

These nearby galaxies lie in structures that are now well specified and avoid regions that are equally well specified. Departures from Hubble expansion are clearly documented, both at the local scales associated with halos and on the large scales of streaming in response to $10 -100$~Mpc structures \citep{2017ApJ...850..207S}.

In addition to information from resolved stellar populations with HST imaging, mid-infrared photometry for the same galaxies provides total stellar mass estimates \citep{2026arXiv260605780T}.
It is to be expected that our small part of the universe is not unusual. Properties that can be gleaned in detail nearby inform studies that are obscured by distance.

%Fig 1
\begin{figure*}
\centering
\includegraphics[width=0.99\linewidth]{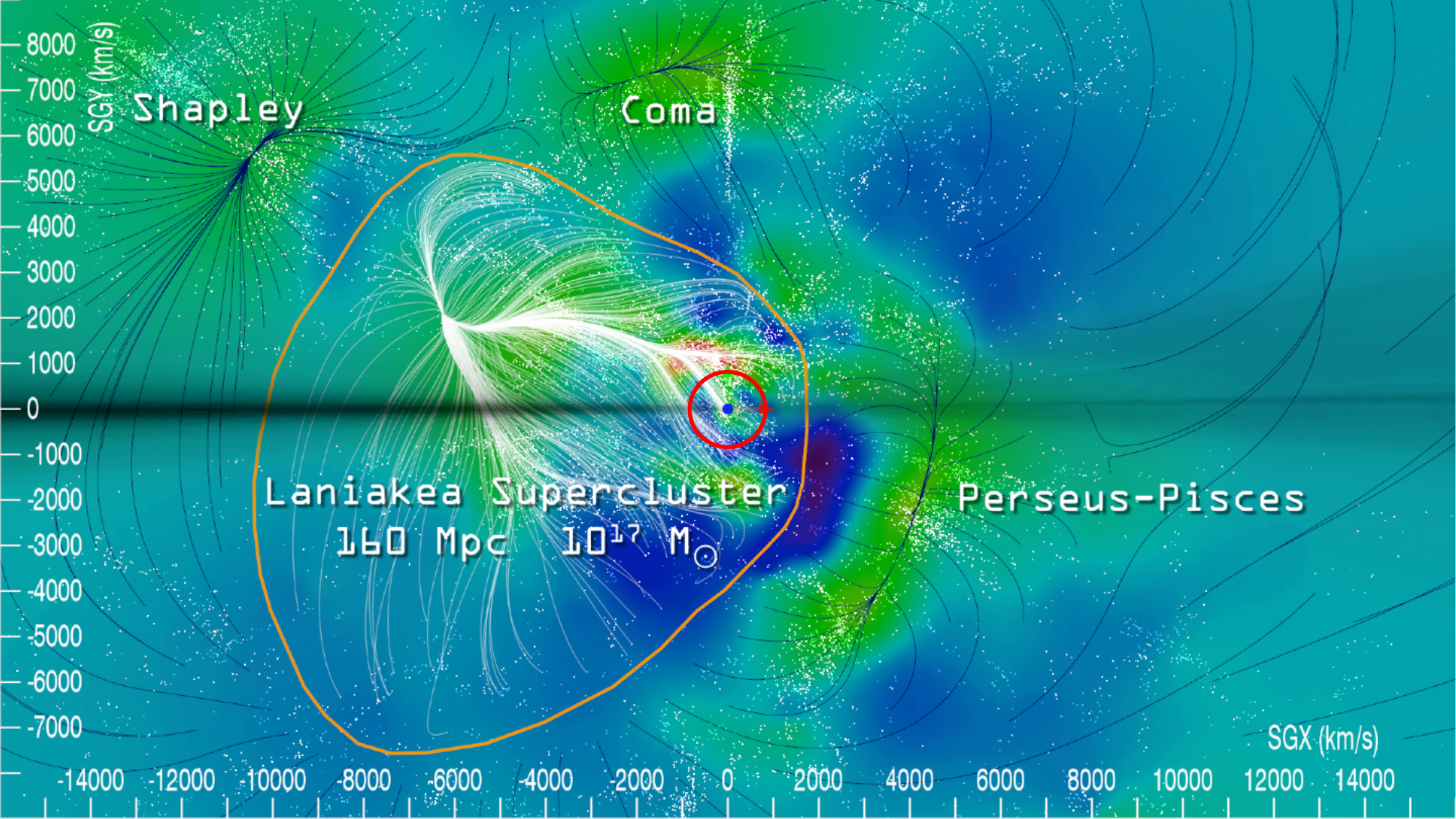}
\caption{
A slice on the supergalactic equator through a section of the universe extending across $\sim 0.1c$ on the long axis. Densities are encoded in shades from blue in voids to greens and red at higher densities. The Milky Way Galaxy is at the origin and the dark horizontal band through our position identifies the zone of obscuration of the Milky Way. Black and white curved lines outline peculiar velocity flow patterns. The Laniakea Supercluster lies within the orange band. The adjacent large scale structures, Perseus-Pisces, Coma, and Shapley, are identified. The 10~Mpc region of interest for this discussion lies within the red circle. Image from \citet{2014Natur.513...71T}.
}
\label{fig:laniakea}
\end{figure*}

This article will touch on a broad range of issues.
To start, in Section~\ref{sec:completeness} there is an evaluation of the completion of the HST imaging coverage of galaxies within 10~Mpc.  Section~\ref{sec:10mpc} delves into the filamentary structural elements of the volume within 10~Mpc, including in Subsection~\ref{sec:distances} the use of accurate distances to extract line-of-sight peculiar velocities and, in Subsection~\ref{sec:vpec}, the bulk motions of separate structural components.
There is a strong tendancy for galaxies to gather in groups and these entities are described in Section~\ref{sec:groups}.
Then, in Section~\ref{sec:photometry} there is a description of mid-infrared observations that provide measures of the stellar content of galaxies within 10~Mpc.
The halo luminosity and mass functions of the groups are discussed in Section~\ref{sec:halofn}. The disposition of the mass in stars is discussed in Section~\ref{sec:pops} and global galaxy metallicity properties are discussed in Section~\ref{sec:metals}. Morphological variations of galaxies with environment are touched on in Section~\ref{sec:environment}. Section~\ref{sec:young} introduces a galaxy within 10~Mpc that may only contain young stars.  Section~\ref{sec:summary} is a summary.
Let us begin with an overview of where we live, then drill down to details.

%%%%%%%%%%%%%%%%%%%%%%%%%%%%%%%%%%%%%%%
%%%%%%%%%%%%%%%%%%%%%%%%%%%%%%%%%%%%%%%

\subsection{The Big Picture}
\label{sec:bigpic}

G\'erard \citet{1953AJ.....58...30D} was first to identity what became called the Local Supercluster, dominated by the Virgo Cluster. Observations of a dipole anisotropy in the cosmic microwave background were taken as evidence that our Local Group participates in a large scale peculiar velocity flow of $\sim 600$~\kms\ \citep{1979ApJ...234L..83S}. Such a flow was found in the pattern of velocities associated with a sample of elliptical galaxies, with a vaguely defined source called the ``Great Attractor" \citep{1987ApJ...313L..37D}. More recent mapping of velocity fields have clarified that we live in an extensive supercluster, given the name Laniakea, with the Great Attractor region at its core \citep{2014Natur.513...71T}.

A representation of the Laniakea Supercluster is seen in Figure~\ref{fig:laniakea}.  Flow patterns, represented by the white and black curves, were extracted from a model of peculiar velocities \citep{2012ApJ...744...43C, 2013AJ....146...69C}. The white curves within the envelope of Laniakea converge in the region of the Great Attractor. The details of structure in this core region have been discussed by \citet{2022ApJ...927..168S}. %Shaya et al. (2021).

It is seen that we live near an edge of Laniakea Supercluster. The boundary that separates us from the adjacent Perseus-Pisces structure lies in a substantial void \citep{1986ApJ...306L..55H}. It is now appreciated that this void is part of the Local Void that encroaches to the edge of the Local Group \citep{2019ApJ...880...24T}.

The density surface in red just above our location in Figure~\ref{fig:laniakea} centers on the Virgo Cluster (density contours reach higher values nearby where the data constraints are best). The Virgo Cluster lies at 16~Mpc, so 60\% farther than the confines of the present study. The smaller Fornax Cluster lies at a distance of 19~Mpc at roughly the position of the second letter ``e'' in the label supercluster. Velocity field information reveals three dominant influences on our local 10~Mpc patch from larger scales: (i) a flow toward 10 o'clock in Figure~\ref{fig:laniakea} arising from large scale influences such as the Great Attractor and Shapley concentration, (ii) a pull from the Virgo Cluster, and (iii) a push from the Local Void~\citep{2008ApJ...676..184T}. Flow patterns have been modeled in considerable detail by \citet{2017ApJ...850..207S, 2022ApJ...927..168S}.

%%%%%%%%%%%%%%%%%%%%%%%%%%%%%%%%%%%%%%%
%%%%%%%%%%%%%%%%%%%%%%%%%%%%%%%%%%%%%%%

\section{Sample Completeness}
\label{sec:completeness}

\begin{figure}
\centering
\includegraphics[width=0.99\linewidth]{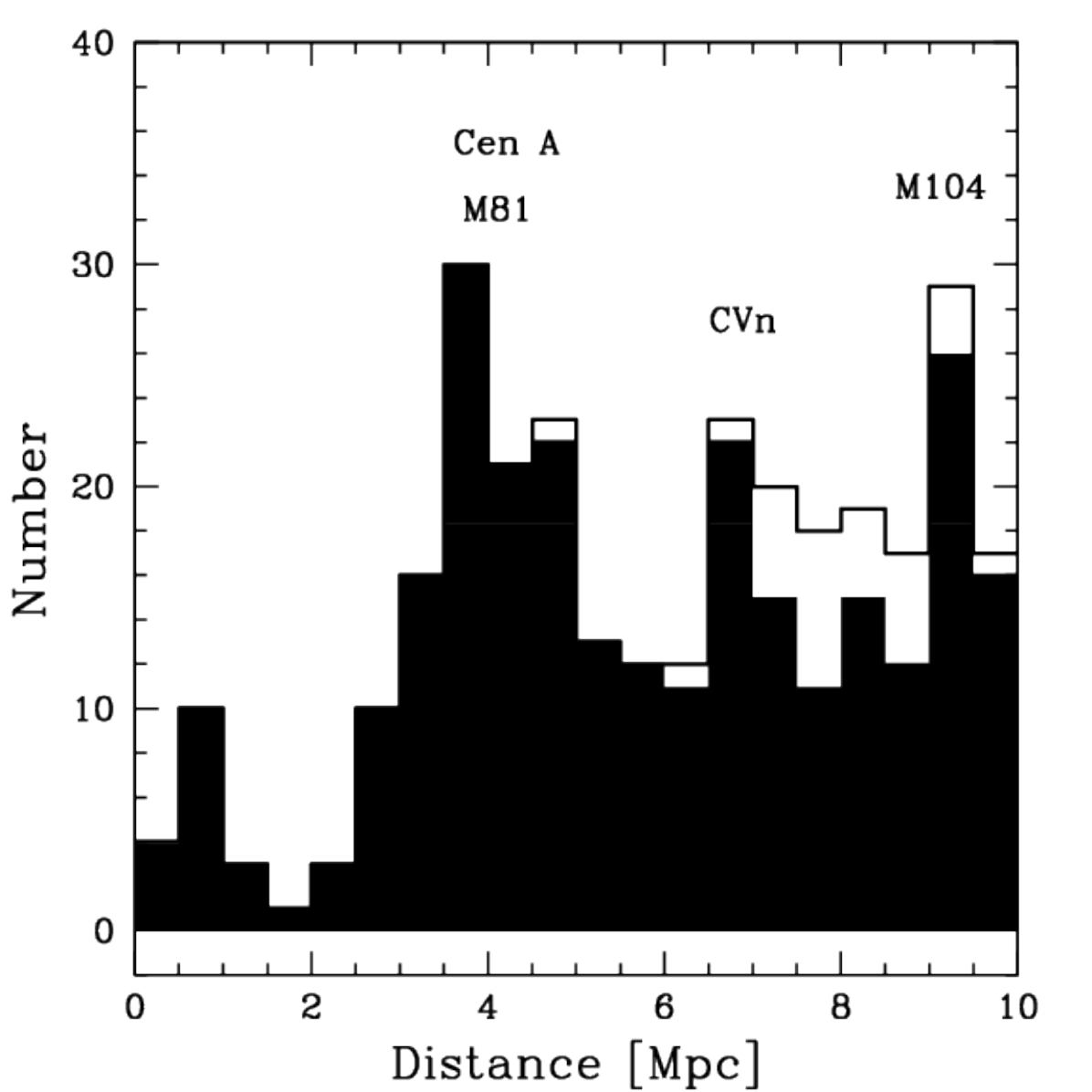}
\caption{
Histogram of numbers of galaxies as a function of group distances.  Filled histogram: TRGB or other precision distance; open histogram: uncertain distances.  Peaks attributable to groups identified by labels.
}
\label{fig:histdn}
\end{figure}

In the discussions to follow concerning luminosity and mass functions, both of halos and of the constituents of halos, we need an understanding of our sample completeness.
A histogram of the run of the galaxies with $M_B < -13$ with distance is seen in Figure~\ref{fig:histdn} (all galaxies within a common group (Section~\ref{sec:groups}) are assigned the same distance). The open spaces in the histogram are attributable to galaxies without precision distances.  Our knowledge of accurate distances to galaxies above our threshold and within 10 Mpc is only 90\% complete (missing $\sim 30$ of $\sim 300$ possible).  The onset of losses from a complete sample with good distances commences at 7~Mpc.  The good distance coverage is 98\% complete within 7~Mpc, with an expectation that we lack distance confirmation for only $\sim 3$ systems brighter than the magnitude cut.  In the interval $7-10$~Mpc, about $20\%$ of our coverage is uncertain.

%Fig 2

Can we have confidence in these claims of completeness?  The initial discovery sources are diverse.  Fundamental contributions to our knowledge of galaxy populations were reported in the series of papers describing searches for images of dwarf galaxies on the Palomar and ESO/SERC sky survey prints and films \citep{1959PDDO....2..147V, 1968CoBAO..39...61K, 1999A&AS..135..221K, 2000A&AS..146..359K, 2000A&AS..145..415K}.  The nearby nature of many dwarfs has been confirmed with the measurement of accurate velocities with HI observations, either pointed observations of known candidates~\citep{1981ApJS...47..139F, 2000A&AS..147..187H, 2000A&AS..141..469H, 2025AJ....170...23N} or blind HI surveys at Parkes and Arecibo observatories \citep{2001MNRAS.322..486B, 2018ApJ...861...49H} and most recently with the FAST All Sky HI Survey \citep{2024SCPMA..6719511Z, 2024A&A...684L..24K, 2026arXiv260409082N}.
Programs were begun to confirm proximity and acquire accurate distances with HST observations at the beginning of this century \citep{2000A&A...363..117K} and are ongoing \citep{2021AJ....162...80A, 2025hst..prop18070T}.

These days, additional dwarf galaxies are regularly being discovered.  However our faint limit of $M_B = -13$ was carefully chosen.  Although many dwarfs are being found, almost all are fainter than this limit.  An unobscured galaxy brighter than $M_B=-13$ within 10~Mpc is inevitably visible on the sky atlas images from the previous century.  Recent surveys provide confirmation.  For example, wide-field surveys carried out with the Canada-France-Hawaii Telescope (CFHT) and the MegaCam imager of regions around 10 of the major nearby galaxies \citep{2020ApJ...891..144C} resulted in 93 new discoveries but only two that are (slightly) brighter than our faint limit and within 10~Mpc (at 7--9~Mpc).
Of 53 galaxies observed in the course of HST SNAP programs 17158 and 17797 (PI Bell) only three had resolved stellar populations that revealed distances within 10~Mpc~\citep{2025arXiv250821494K}.
As will be discussed, a galaxy with $M_B=-13$ has a mass of $\sim 10^{10}~\Msun$.  New-found galaxies suspected to lie within 11~Mpc are regularly added to the on-line \textit{Updated Nearby Galaxies Catalog}\footnote{\url{https://www.sao.ru/lv/lvgdb/}}~\citep[UNGC, ][]{2012AstBu..67..115K, 2013AJ....145..101K}.

It can be anticipated that galaxies satisfying our sample criteria are lost in the zone of Milky Way obscuration.  Indeed, lying almost on the Galactic plane, the Maffei Group and the Circinus galaxy and any companions it might have are excluded.  A limit for inclusion at $\vert b \vert = 10^{\circ}$ is an approximation, but HST observations have been more governed by awareness of candidates than by a latitude limit.  Our sample includes three galaxies within 7~Mpc with TRGB distances acquired solely with infrared observations \citep{2019ApJ...880...52A}.  Given the very clumpy distribution of galaxies and the restricted volume, statistics per unit volume have little sense; i.e., only relative quantities in galaxy properties hold interest.  It is likely that in regions of significant obscuration a larger galaxy might be identified where a smaller galaxy would be lost.  Two substantial galaxies in our full sample flirt with the zone of obscuration, each at $b \sim 11$. One is the nearby IC\,342, while the other, NGC\,6946, is beyond 7~Mpc.  In each case, the known distribution of companions is asymmetric about the larger galaxy, favoring higher galactic latitudes, as seen in Figures~\ref{fig:xy_loc} and \ref{fig:xy_peg} respectively.  We conclude that, in relative counts between large and small galaxies, we will miss a small number of dwarfs due to obscuration.  The effect on relative statistics for the volume wihin 7~Mpc is expected to be minor.

The Local Volume Database~\citep[LVDB, ][]{2025OJAp....8E.142P} provides a complementary inventory of nearby galaxies to those studied here.  Whereas we strive for completion within 10~Mpc for galaxies brighter than $M_B = -13$ and carry an incomplete sampling of fainter dwarfs, the LVDB is maintaining a record of all known systems that potentially can be called galaxies within 3~Mpc, with an incomplete inclusion of galaxies at larger distances.

%%%%%%%%%%%%%%%%%%%%%%%%%%%%%%%%%%%%%%%
%%%%%%%%%%%%%%%%%%%%%%%%%%%%%%%%%%%%%%%

\section{Structures and Peculiar Velocities Within 10 Mpc}
\label{sec:10mpc}

There are two principal structures within the 10~Mpc volume.  The dominant one, in which we live, was called the Coma-Sculptor Cloud in the Nearby Galaxies Atlas \citep{1987nga..book.....T} and Catalog \citep{1988ngc..book.....T} but renamed the Local Sheet \citep{2008ApJ...676..184T}, the name we will use here.  As that name implies, this structure has the morphology of two long and one short axes.  As a consequence of living within this structure that we view edge-on, most nearby galaxies lie in a narrow band on the sky.  This phenomenon, plus the continuance of it in the distribution of galaxies in the environment of the Virgo Cluster, led \citet{1956VA......2.1584D} to define the supergalactic coordinate system.  This coordinate system is the natural one for displaying nearby structure and, remarkably, even for structure on 100~Mpc scales \citep{1986ApJ...303...25T}. In the supergalactic coordinate system, the Milky Way plane lies roughly at $SGY=0$ (slightly tilted at $6.32^{\circ}$).  Galaxies with $SGY>0$ (modulo the small tilt) lie in the Galactic north while those with $SGY<0$ lie in the Galactic south.

%Fig 3
\begin{figure*}[]
\centering
\includegraphics[width=0.99\linewidth]{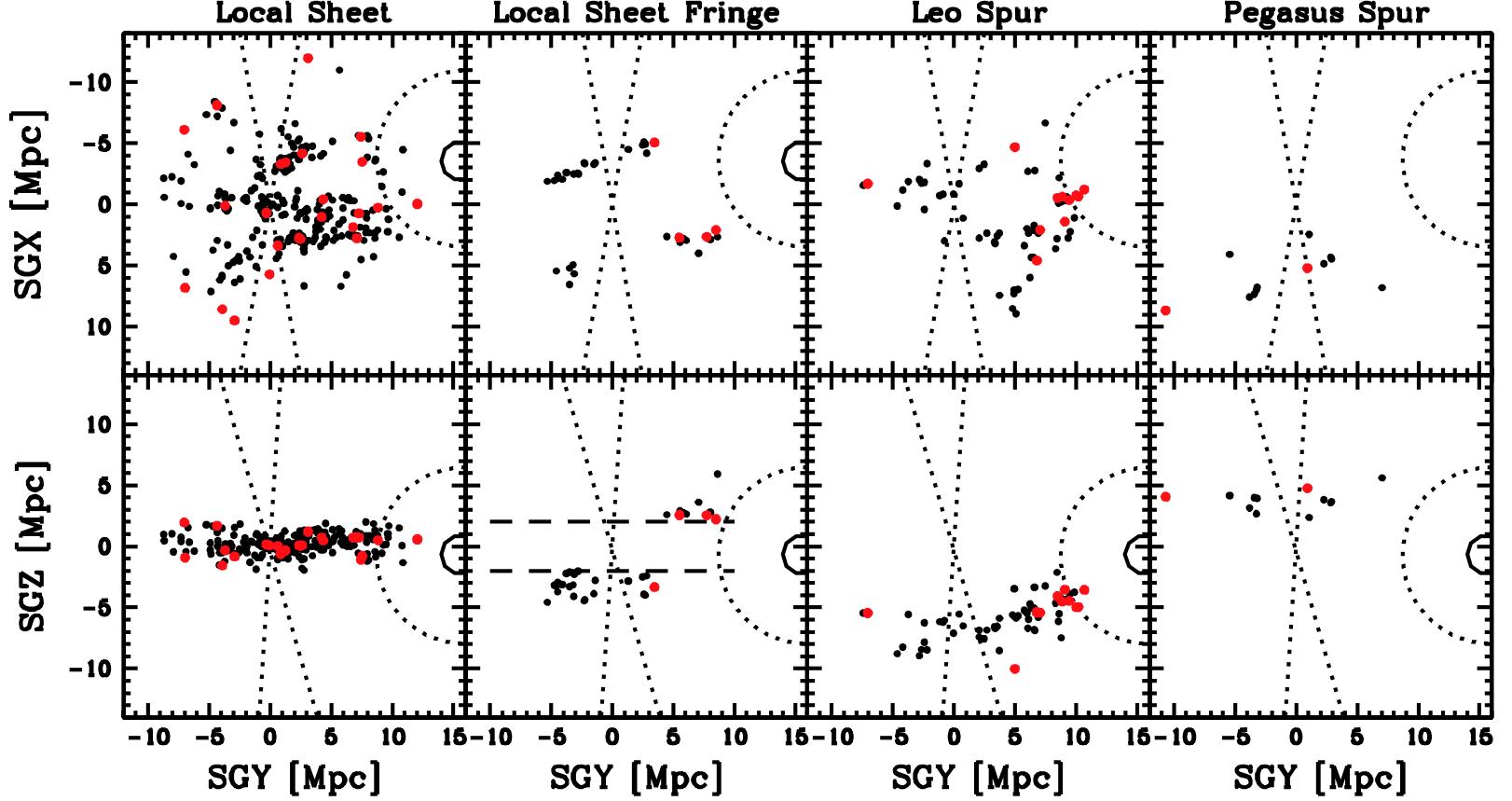}
\caption{
Three structures within 10~Mpc: the Local Sheet in the left panels, the Leo Spur in the third panels, and the Pegasus Spur in the right panels.  Galaxies that slightly stray from the Local Sheet are shown in the second panels.  In the top row the views are from the supergalactic north pole. The views on the bottom row are from $+SGX$.  Individual galaxies are represented by dots at positions determined by TRGB distances. Galaxies more luminous than $L_K=10^{10}~\Msun$ are shown in red.  The heavy hemi-circles locate the Virgo Cluster while the dotted larger hemi-circles outline the outer edge of the Virgo Cluster infall regime.  Dotted wedges outline Galactic latitudes $b\pm10^{\circ}$.  The $6.32^{\circ}$ tilt of the Galactic plane in the supergalactic coordinate system is manifested in the lower panels.
}
\label{fig:filam}
\end{figure*}

The sub-dominant structure in the 10~Mpc patch was partitioned into three pieces in the Nearby Galaxies Atlas and Catalog: as the Leo Spur and Antlia Cloud in the Galactic north and the Dorado Cloud in the south.  It was clearly shown by \citet{2019ApJ...880...52A} that these three components merge within the zone of obscuration.  This combined structure, that we will simply call the Leo Spur, lies in a plane roughly parallel to the Local Sheet at $SGZ\sim -6$~Mpc.\footnote{Our characterization of nearby structure by two components, the Local Sheet and the Leo Spur, is adequate for a discussion of the 10~Mpc region but a larger view reveals a more complex picture.  The main backbone of the Leo Spur runs to positive $SGX$ out of our study volume at $SGY \sim 8$~Mpc and $SGZ \sim -4$~Mpc.  The Dorado Cloud, emanating from the region of the Fornax Cluster in the Galactic south, connects at the edge of our study region through the zone of obscuration with the Antlia Cloud that then runs in the Galactic north to meet the Crater Cloud.  There is a wispy connection between the Leo Spur and Dorado-Antlia filament that is only evident because it is very close to us.  Meanwhile, a part of the Local Sheet at negative $SGX$ including the NGC\,6744 Group can be taken to belong to the Pavo-Ara Cloud. There is considerable arbitrariness in drawing distinctions between adjacent structures in the absence of kinematic discontinuities.}

%Fig 4
\begin{figure*}[]
\centering
\includegraphics[width=0.84\linewidth]{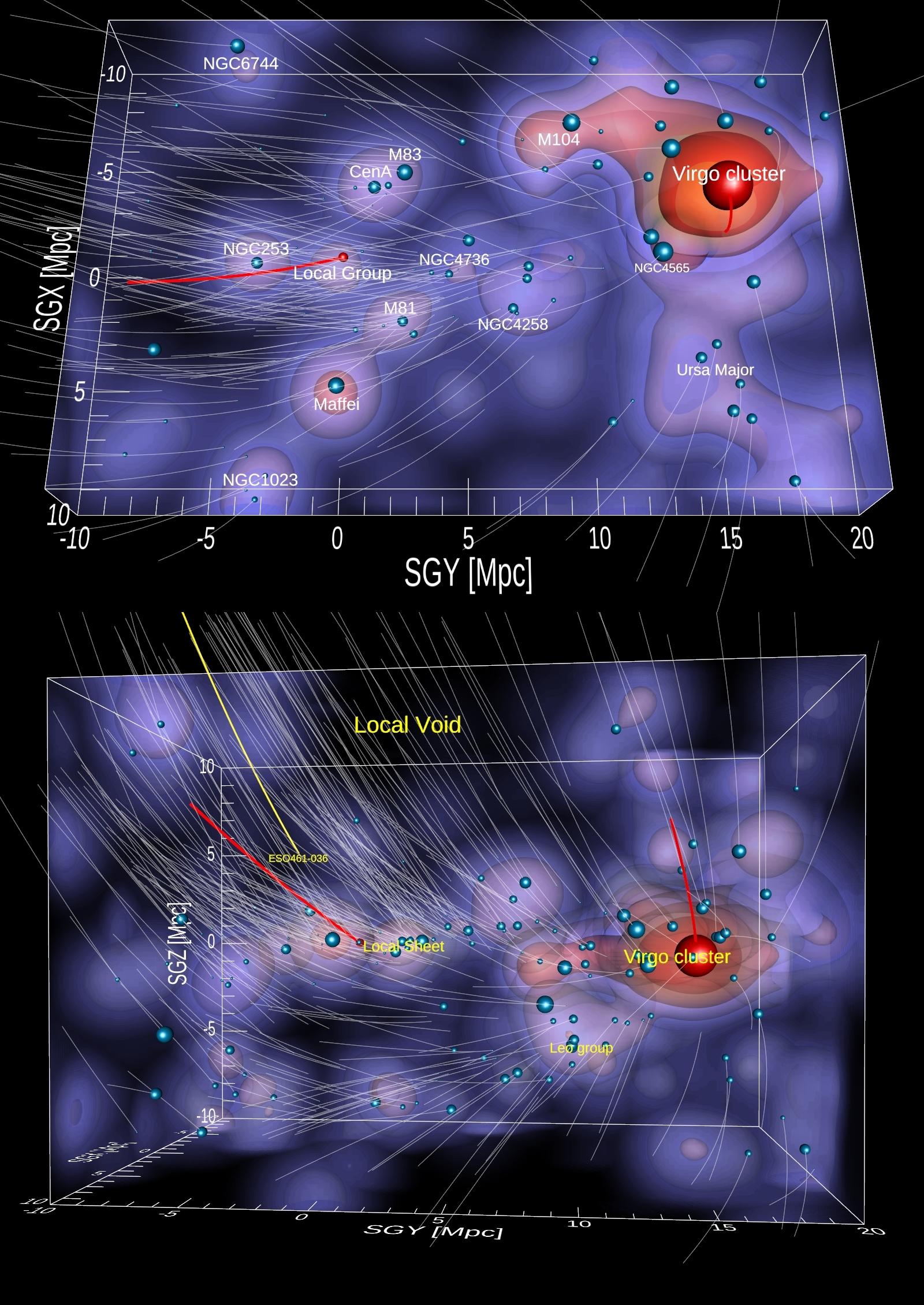}
\caption{
Major components of the local volume extending to include the Virgo Cluster.  Numerical action orbits are superimposed, with the orbits for the Local Group and Virgo Cluster in red and the orbit for the void galaxy ESO\,461-036 in yellow..  Top panel: a view from roughly $+SGZ$. Bottom panel: a view from roughly $+SGX$. The three-dimensional structure can be further explored with an \underline{\href{https://sketchfab.com/3d-models/59f926cc0346468bbadf47392a43948a}{online interactive model}}.
}
\label{fig:LV2}
\end{figure*}

A third structure slightly impinges on the 10~Mpc volume, one called the Pegasus Spur in the Nearby Galaxies Atlas and Catalog.  This filament lies above the Local Sheet at $SGZ\sim+4$~Mpc and extends in $SGY$ at $SGX\sim+4$~Mpc. NGC\,6946 is a member of this filament at its nearest point.

The other major feature of the 10~Mpc region is the {\it absence} of galaxies above the Local Sheet ($SGZ > 0$), a feature identified in the Nearby Galaxies Atlas as the Local Void.  This void commences
immediately at the edge of the Local Group.  The Local Sheet must be viewed as part of a wall bounding the Local Void.

These three structures are illustrated in Figure~\ref{fig:filam}.  Give attention particularly to the views edge-on to the Local Sheet in the bottom panels.  The three local structures are each constrained to narrow $SGZ$ bands.  The sample displayed includes all but two of 480 galaxies, placed on the plots with tip of the red giant branch distances, as will be discussed.  The only three galaxies with known distances outside these three structures, ESO\,461--036, KKR\,18, and ALFAZOA\,J1952+1428, lie in the Local Void.

An influence of importance just outside our study region is the Virgo Cluster (seen at the edge of relevant panels in Figure~\ref{fig:filam}).  The numerical action orbit reconstruction study by \citet{2017ApJ...850..207S} determined the triaxial boundary of the infall region around the Virgo Cluster.  The foreground surface is 7.05~Mpc from the cluster central M\,87, reaching to within $\sim 9$~Mpc of us and barely encroaching on the 10~Mpc shell enclosing the volume of this investigation.

Figure~\ref{fig:LV2} provides two views of our volume of interest, extending in the $+SGY$ coordinate to include the Virgo Cluster.
By happenstance, there are five substantial groups right at the edge of the 10~Mpc sphere centered on us.  Three lie in the equatorial plane.  In the top panel view from the supergalactic north pole  ($+SGZ$ axis): at 2 o'clock there is the NGC\,4594 (Sombrero/M\,104) Group with the dominant galaxy at 9.6~Mpc; at 7 o'clock there is the NGC\,1023 Group at 10.4 Mpc; and at 11 o'clock there is the NGC\,6744 Group at 9.3~Mpc.  Below the plane of the Local Sheet, in the Leo Spur, there is the pair of Leo groups at 10 Mpc associated with NGC\,3379 and NGC\,3627 respectively.  The constituents of these groups and others that lie within $\sim 1$~Mpc of the 10~Mpc boundary are uncertain.

The position of the Local Group surrounded by groups in the Local Sheet at 3--5~Mpc has given rise to the name ``council of giants'' to describe this observed geometry and to speculation whether such a confluence of substantial galaxies in a filament is unusual \citep{2014MNRAS.440..405M, 2020MNRAS.494.2600N, 2024PASA...41...18O}.  In the absence of convincing evidence, we accept a Copernican view that our location is not special.

The white tails on individual objects in Figure~\ref{fig:LV2} trace trajectories from $z=4$ until today found with the numerical action orbit reconstructions of \citet{2017ApJ...850..207S, 2022ApJ...927..168S}.  See \S \ref{sec:vpec} for further discussion.

%It will be noticed with the density profiles in Figure~\ref{fig:LV2} that densities are low to the foreground of the Virgo Cluster and the onsets of either the Local Sheet or the Leo Spur.  This is not an artifact.  There are few galaxies in this space.  For whatever is to be made of it, this relatively evacuated region extends from the Virgo Cluster essentially to the turnaround radius marking cluster infall.

Fortuitously, our view of the Virgo Cluster is not badly confused by foreground projections.
Currently there are many galaxies in the Virgo infall region but mostly they lie close to tangent on the sky in the Virgo Southern Extension \citep{1961ApJS....6..213D} or, to a lesser degree, on the opposite side of the cluster in projection, for example the Coma\,I Group, the largest entity fated to merge with the cluster in the next Hubble time \citep{2022ApJ...927..168S}.

The branching networks of filaments are complex and the association of constituents to specific features can be debated \citep{2016ApJ...833..207K, 2022A&A...657A...9C}.  There is evidence that the Local Sheet continues beyond the NGC\,1023 Group edge of our study, penetrating the Local Void as a wispy filament to reach the vicinity of the Perseus Cluster \citep{2019ApJ...880...24T}.  Viewed on a larger scale, entities in the Local Sheet at the negative $SGX$ and $SGY$ edge of our study region such as NGCs~5398, 5643, 6744, and 7090  can be taken to be part of what was called the Pavo-Ara Cloud in the Nearby Galaxies Atlas and Catalog. Naming convention details are of minor importance.  Of relevance are the evident connections between filamentary structures and their kinematic properties.  It is the measurement of accurate distances, available in profusion only locally, that gives clarity to these linkages.

%%%%%%%%%%%%%%%%%%%%%%%%%%%%%%%%%%%%%%%
%%%%%%%%%%%%%%%%%%%%%%%%%%%%%%%%%%%%%%%

\subsection{Distance and Peculiar Velocity Measurements}
\label{sec:distances}

%Fig 5
\begin{figure}[]
\centering
\includegraphics[width=0.99\linewidth]{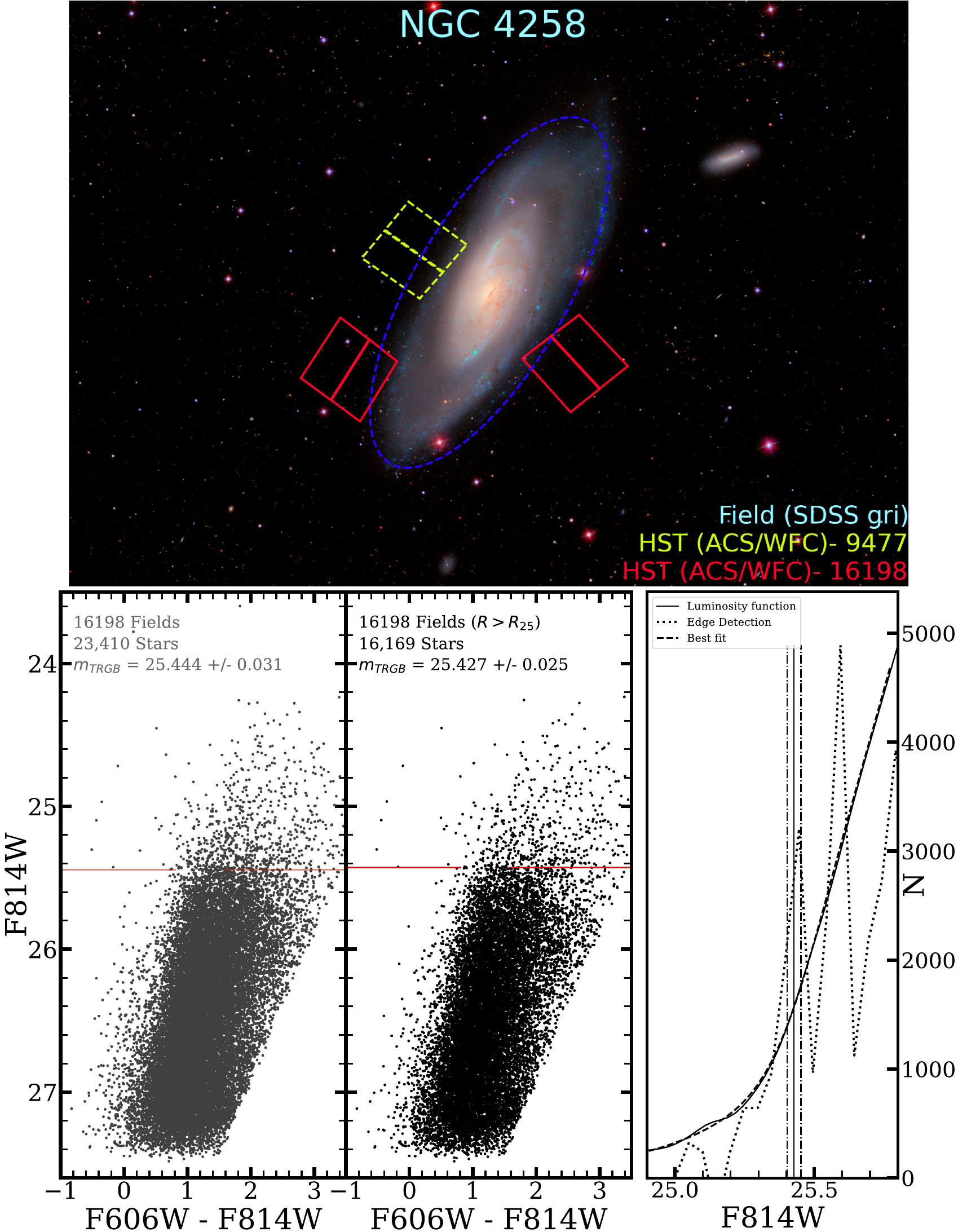}
\caption{
Top. Field of NGC\,4258 with the placement of three ACS fields superimpose. Bottom left and middle. Color magnitude diagrams drawing, at left, from the entire two ACS fields outlined in red in the top panel and, in the middle, restricted to the halo beyond the blue ellipse.  The stellar luminosity function is shown at the right.  Figure extracted from \citet{2022ApJ...932...15A}.
}
\label{fig:n4258}
\end{figure}

The determination of galaxy peculiar velocities, $V_\mathrm{pec}$, depends acutely on the accuracy of galaxy distances since to a good approximation\footnote{The more precise formulation of \citet{2014MNRAS.442.1117D} reduces to this simple form at small distances.}
\begin{equation}
    V_\mathrm{pec} = V_\mathrm{LS} - H_0 d
    \label{eq:vpec}
\end{equation}
where $V_\mathrm{LS}$ is the observed velocity in the Local Group rest frame\footnote{We use the version of Local Group frame introduced by \citet{2008ApJ...676..184T} called the Local Sheet frame.}, $d$ is distance, and we take a value of the Hubble Constant $H_0=75$~\kmsMpc\ (peculiar velocities are rather insensitive to reasonable choices of $H_0$ at these small distances).
It has been stated that distances have $\sim5\%$ accuracy but this condition is not always satisfied.  Among individual cases there are the good, bad and ugly.  Let us first give attention to the good.  Figure~\ref{fig:n4258} presents the case of NGC\,4258.  This galaxy has received particular attention by several research groups because it has an independently determined distance of $7.568\pm0.082(\mathrm{stat})\pm0.076(\mathrm{sys})$~Mpc from observations of a nuclear maser system \citep{2019ApJ...886L..27R}. There are the following tip of the red giant branch (TRGB) measurements: $7.66\pm0.21$~Mpc~\citep{2007ApJ...661..815R}, $7.18\pm0.13(\mathrm{stat})\pm0.40(\mathrm{sys})$~Mpc~\citep{2008ApJ...689..721M}, $7.748\pm0.282$~Mpc~\citep{2022ApJ...932...15A}.  \citet{2021ApJ...906..125J} use the maser distance to set their zero point scale. Small differences can arise due to different tip finding methodologies and zero point scaling. \citet{2022ApJ...932...15A} have demonstrated compatibility between TRGB distances found by our team as represented by entries in the CMDs/TRGB catalog within the Extragalactic Distance Database\footnote{\url{https://edd.ifa.hawaii.edu}} and those determined by the Carnegie-Chicago Hubble Program~\citep{2021ApJ...919...16F}.
Recently, \citet{2026arXiv260311160C} provided a new JWST-based zero-point and color calibration for the TRGB distance scale.

This good agreement presumes that proper attention is given to the selection of fields.  With large galaxies, observations should be restricted to halos where the crowding is not severe, internal reddening is not an issue, and old populations are dominated by stars with low metallicities. Small galaxies typically fit comfortably within an entire Advanced Camera for Surveys (ACS) field.  Metallicities in dwarf galaxies are usually low.  In these cases, it is only important to excise regions with ongoing or recent star formation (although those regions are retained for the studies of stellar populations to be discussed further along).

%Fig 6
\begin{figure*}
\centering
\includegraphics[width=0.48\textwidth]{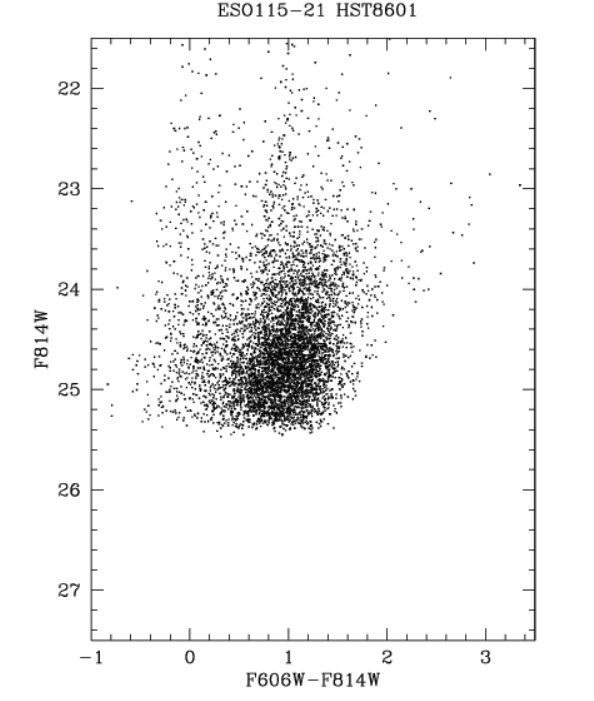}
\includegraphics[width=0.48\textwidth]{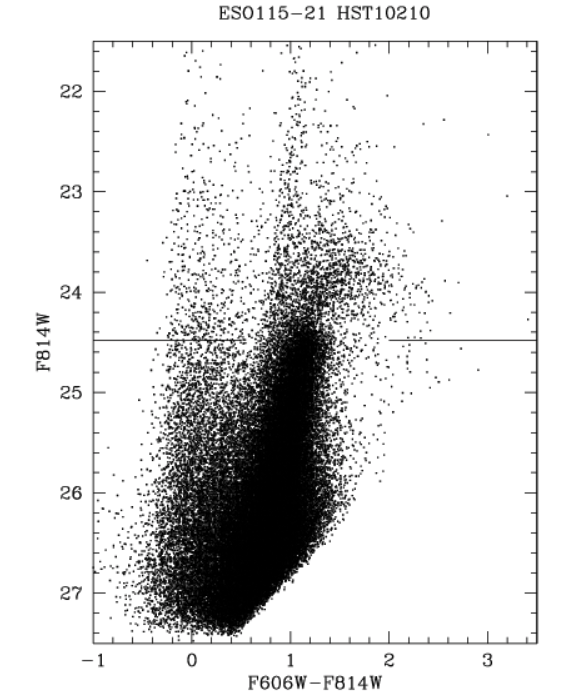}
\caption{
Comparison of color magnitude diagrams from single orbit observations of ESO\,115--21 with HST cameras WFPC2 (left) and ACS (right).  The horizontal line in the ACS plot locates the TRGB.
}
\label{fig:e115-21}
\end{figure*}

There are multiple reasons for ugly.  One reason can be a sparsity of stars near the top of the RGB as can occur in the very small galaxies being found nearby.  An example is MADCASH-2 \citep{2021ApJ...909..211C}.  Such cases fail Madore's rule \citep{1995AJ....109.1645M} that one should have at least 100 stars in the magnitude interval below the RGB tip for a reliable measurement.

Also, TRGB determinations have increased error as the tip approaches the photometric limit with increasing distance. Figure~\ref{fig:e115-21} shows the remarkable difference between observations with the HST earlier generation Wide Field Planetary Camera 2 (WFPC2) and ACS.  Each pairs of F606W and F814W images required to build the color magnitude diagrams (CMD) were acquired in single HST orbits.  For this target, determined to lie at 5.08~Mpc, the ACS product is extremely well defined while the WFPC2 product would lead to an ambiguous TRGB measurement.  
Measurements with WFPC2 become unreliable at distances beyond 5~Mpc. Similarly, the uncertainty of single-orbit ACS observations increases dramatically at distances greater than 10~Mpc, where the TRGB falls below $\mathrm{F814W} = 26$~mag. This limitation of single-orbit HST observations motivates the volume restriction adopted in this study.

Then there are the bad.  There are occasional examples in the literature of marked differences in TRGB distance determinations.  There is the case of NGC\,4038/39 (the Antennae) that \citet{2008ApJ...678..179S} placed at $13.3\pm1.0$~Mpc and \citet{2008AJ....136.1482S} placed at $20.0\pm1.6$~Mpc with the same HST data.
%Schweizer et al.
\citet{2008AJ....136.1482S} asserted (and we agree) that
%Saviane et al.
\citet{2008ApJ...678..179S} misinterpreted the onset of the asymptotic giant branch (AGB) for the tip of the RGB. Entertain the CMD in Figure~\ref{fig:e115-21}.  A substantial population of AGB stars rise to $\sim 1$ mag above the TRGB.  At 20~Mpc, the TRGB would lie at F814W$\sim 27.5$ mag.  Even with multiple HST orbits, detection at this level is difficult and confusion with the onset of the AGB at a magnitude brighter is likely.  Now the greater distance is confirmed with a James Webb Space Telescope (JWST) observation giving $d = 21.3$~Mpc \citep{2025arXiv250311769H}. Another such case has occurred with the galaxy Coma\,P.  \citet{2019AJ....157...76B} report a distance of $5.50\pm0.40$~Mpc while \citet{2018ApJ...861L...6A} report $10.9\pm1.0$~Mpc.  Again the issue is whether the AGB is being mistakenly taken for the RGB. At the shorter distance, Coma\,P would have an extraordinary peculiar velocity.  At the longer distance the peculiar velocity is reasonable but the AGB population is unusually large compared with the RGB population.  There would be the implication that a substantial fraction of its stars were born in the last few Gyr \citep{2011AJ....141..106J} and the ancient population is small.

Going forward with JWST there will be increasing interest in TRGB applications in the infrared.  Care will have to be taken because confusion between the RGB and AGB may be aggravated.  Our team now prefers a distance for Maffei~2 of $5.73\pm0.40$~Mpc from HST F110W and F160W observations \citep{2019ApJ...872L...4A} in preference to our earlier $3.52\pm0.20$~Mpc \citep{2014AJ....148....7W}. Again the issue was whether the AGB was being confused for the RGB.

%%%%%%%%%%%%%%%%%%%%%%%%%%%%%%%%%%%%%%%
%%%%%%%%%%%%%%%%%%%%%%%%%%%%%%%%%%%%%%%

\subsection{Nearby Large Scale Peculiar Velocities}
\label{sec:vpec}

Two fundamental properties of peculiar velocities outside domains of collapse have become clear with observations informing accurate distances within the local volume.  One is that there is coherence in flows within filamentary structures.  Discounting motions internal to halos, \citet{2009MNRAS.393.1265K} found line-of-sight rms peculiar motions are only 25~\kms\ nearby in the Local Sheet.
The scatter turns out to be even smaller, about 15~\kms{}, at the boundary of the Local Group~\citep{2025A&A...698A.178M}.
Such a low value is direct evidence of a low value for the cosmic matter density.

The second now well established property of flows is that there can be abrupt discontinuities between adjacent structures.  The two parallel planes, the Local Sheet and the Leo Spur, separated by $\sim 6$~Mpc, are converging in co-moving coordinates at 260~\kms\ \citep{2015ApJ...805..144K, 2019ApJ...880...52A}.

There is little ambiguity in these two assertions. Distance determinations based on TRGB measurements with HST have 5\% accuracy so, within 10~Mpc, peculiar velocities derived using Eq.~\ref{eq:vpec} can be parsed from uniform expansion at the level of 40~\kms\ at 10 ~Mpc and 20~\kms\ at 5~Mpc.  It is only nearby, and with precision distance estimators, that peculiar velocities can stand out clearly from Hubble flow velocities for individual galaxies. With $\sim 500$ measures, patterns are unmistakable.

%Fig 7
\begin{figure}[]
\centering
\includegraphics[width=0.99\linewidth]{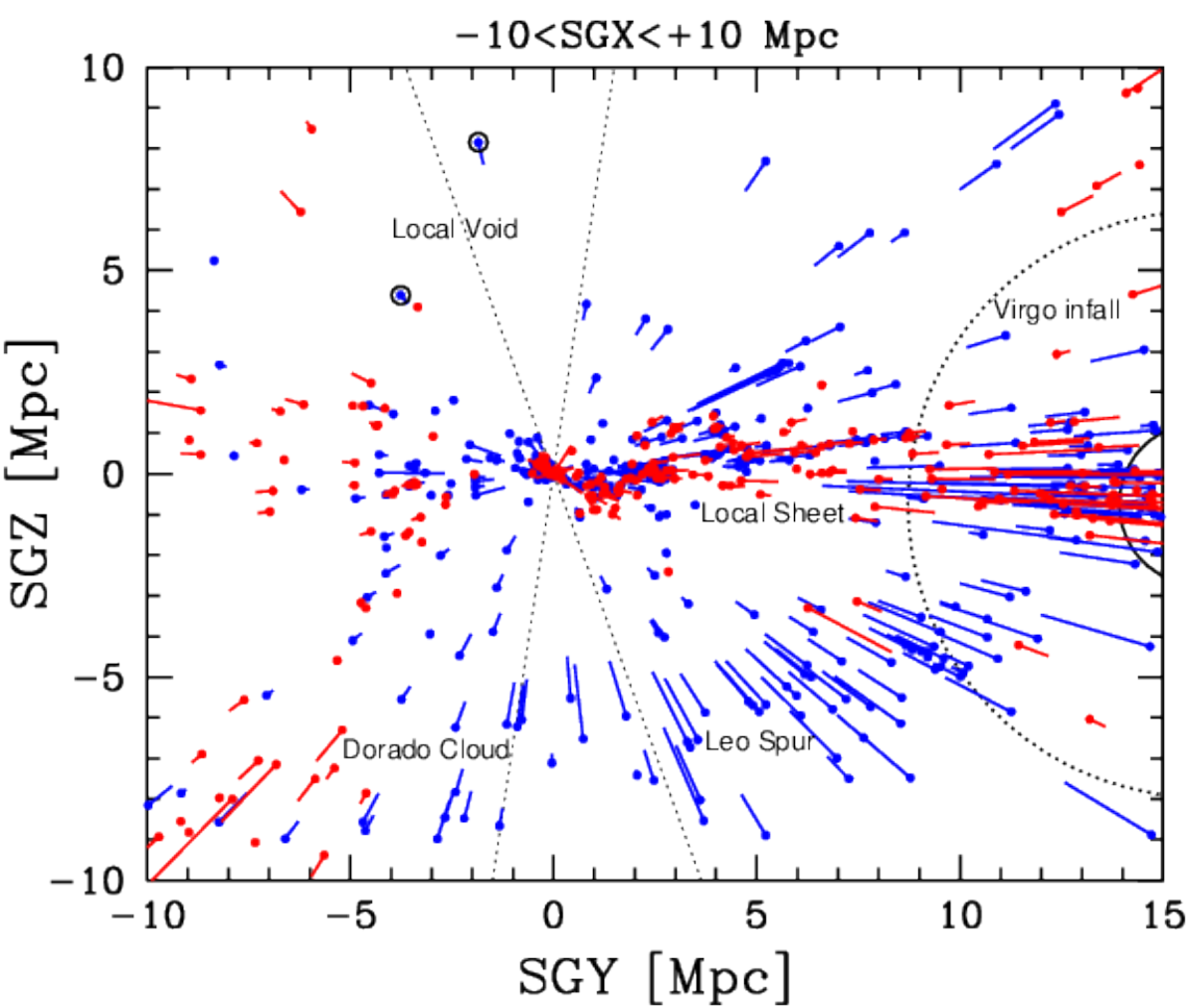}
\caption{
$SGY$--$SGZ$ projection of peculiar velocity vectors in the Local Group frame.  Blue arrows: peculiar velocities  toward us; red arrows: peculiar velocities away from us.  The most obvious trend is the concentration of negative peculiar velocities for galaxies at negative $SGZ$, except in the lower left of the plot where galaxies feel the influence of the Fornax Cluster.  Within the Local Sheet at $SGZ \simeq 0$ the shear in velocities as a function of distance from the Virgo Cluster is marked by predominance of red to blue to red vectors proceeding from left to right in the figure.  The most important cosmographic features are labeled.  Figure modified from \citet{2019ApJ...880...52A}.
}
\label{fig:yzflow}
\end{figure}

The convergence of the Local Sheet and Leo Spur is the most remarkable phenomenon.  Figure~\ref{fig:yzflow} shows with clarity the coherence of peculiar velocities toward us across the domain of the Leo-Dorado-Antlia filament (Leo Spur).  In ignorance of structure on larger scales, one could only speculate on the reason for this motion.  However we are {\it not} ignorant of the structure on relevant scales.  Galaxy orbits have been calculated with Numerical Action Methods (NAM), incorporating the mass distribution inferred from $K$-band luminosities across a volume of radius 100 Mpc and constrained by distance measurements across that volume \citep{2017ApJ...850..207S, 2022ApJ...927..168S}.  {\it All across the entire reach of Laniakea Supercluster, galaxies at positive $SGZ$ are moving toward more negative values}. We are seeing the expansion of the Local Void.  Galaxies at $SGZ \sim 0$ are parts of a wall around the void and are experiencing a downward motion in the cosmic microwave background (CMB) frame.  Galaxies below this equatorial band in the supergalactic coordinate system are not experiencing the expansion of this void.  The apparent motions toward us of galaxies in the Leo Spur are a reflex of our motion toward them.  While the general pattern of void expansion is seen across the 100~Mpc volume, it is in our 10~Mpc region that effects are seen in compelling detail.

Within the Local Void, there are currently only three lonely dwarf galaxies near enough to be easily resolved with HST, hence candidates for TRGB measurements. Two shown as black dots in Figure~\ref{fig:yzflow}, KK\,246 = ESO\,461--036 \citep{2011AJ....141..204K} and ALFAZOA\,J1952+1428 \citep{2011ApJ...739L..26M}, both have negative peculiar velocities; i.e., both are chasing us downward in $SGZ$ following the expansion of the void \citep{2017ApJ...835...78R}.
The third, KKR\,18, is discussed in Section~\ref{sec:young} and in \citet{2026arXiv260904437T}.

%Fig 8
\begin{figure}[]
\centering
\includegraphics[width=0.99\linewidth]{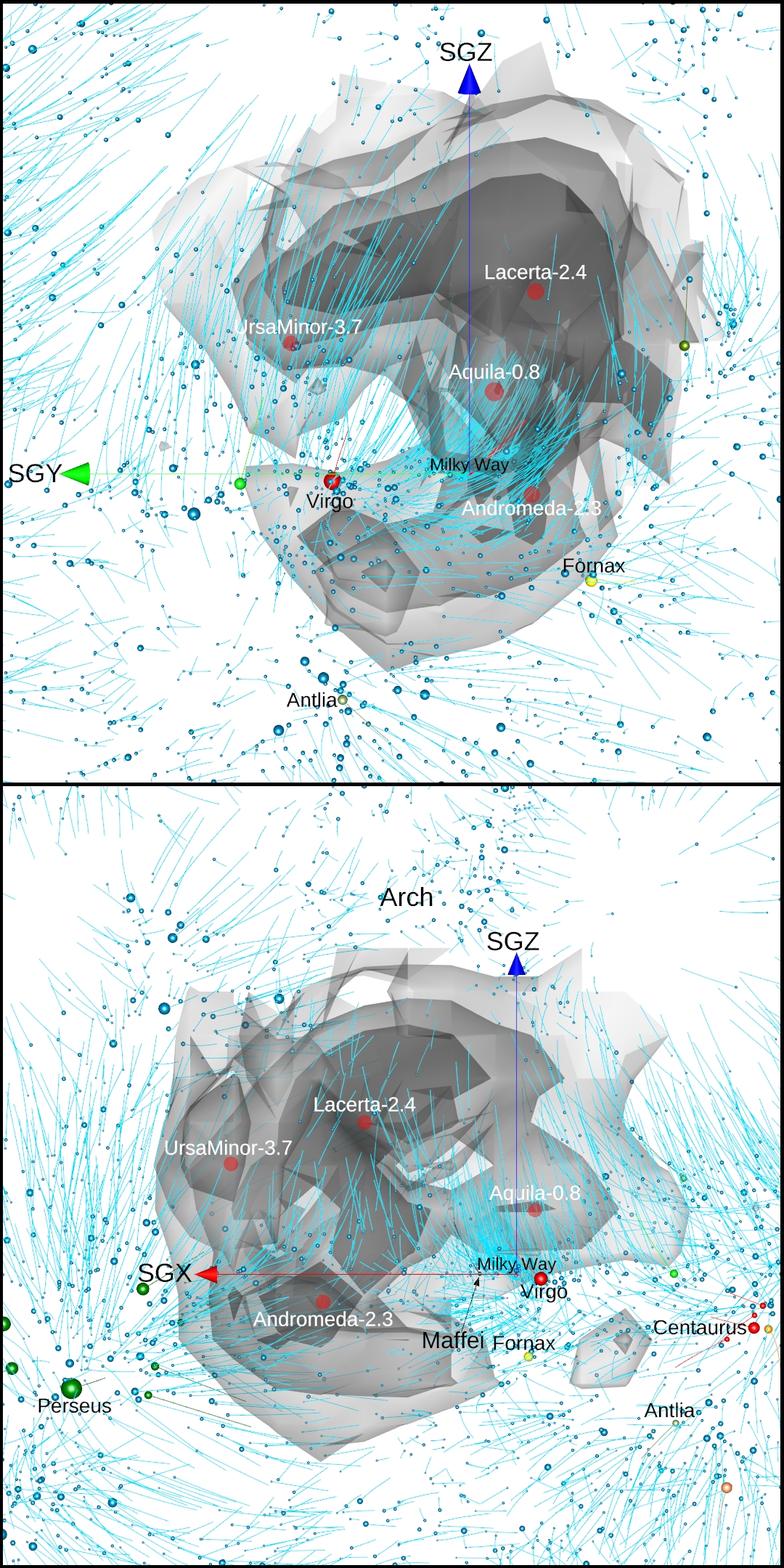}
\caption{
Density contours of the abyss of the Local Void, with numerical action orbits overlain, viewed in the top panel from $-SGX$ and in the bottom panel from $+SGY$. The red, green, blue arrows (directed toward $+SGX,~+SGY,~+SGZ$ respectively) have lengths 3,500~\kms.  Prominent clusters and the Milky Way have black labels.  Local density minima in the Local Void are labeled in white.
There is an interactive version of these figures in \citet{2019ApJ...880...24T}.
}
\label{fig:localvoid}
\end{figure}

The push from the Local Void is not just downward in $SGZ$.  This void wraps around us, particularly toward positive $SGX$.  Recall in Figure~\ref{fig:laniakea} that we are at the edge of the Laniakea Supercluster and, as \citet{1986ApJ...306L..55H} pointed out, there is a deep void between our location and the Perseus-Pisces structure.  \citet{2019ApJ...880...24T} discussed the topology of the Local Void.   Figure~\ref{fig:localvoid} illustrates how the Local Group is nestled within its embrace.  The orbits that galaxies within the 10~Mpc region have followed over the age of the universe are given by the Numerical Action Methods model of \citet{2022ApJ...927..168S}.  The orbits of galaxies since $z=4$ until today for our 10~Mpc sample are shown in Figure~\ref{fig:localvoid}.  To be clear, these orbits are responding to influences drawn from a full radial domain of 100~Mpc.  However local effects cause gradients.  Looking at the lower panel, while essentially all the galaxies right of the void center are moving to the right, toward negative $SGX$, the displacements are greater for those to the left of the Milky Way; i.e., deeper within the void.  The situation with the Maffei galaxies is most notable.  The Maffei Group has a peculiar velocity of $-128\pm33$~\kms\ essentially along the $SGX$ axis toward us, a manifestation of the wrap-around of the Local Void to occupy the space between us and the Perseus-Pisces structure \citep{2019ApJ...872L...4A}.

%\begin{figure}[]
%\centering
%\includegraphics[width=0.99\linewidth]{figures/10Mpc_orbits.png}
%\caption{
%PLACEHOLDER: Look in from -SGY; only orbits for galaxies within %the 10 Mpc volume sample.
%}
%\label{fig:nam_orbits}
%\end{figure}

%Fig 9
\begin{figure}
\centering
\includegraphics[width=0.82\linewidth]{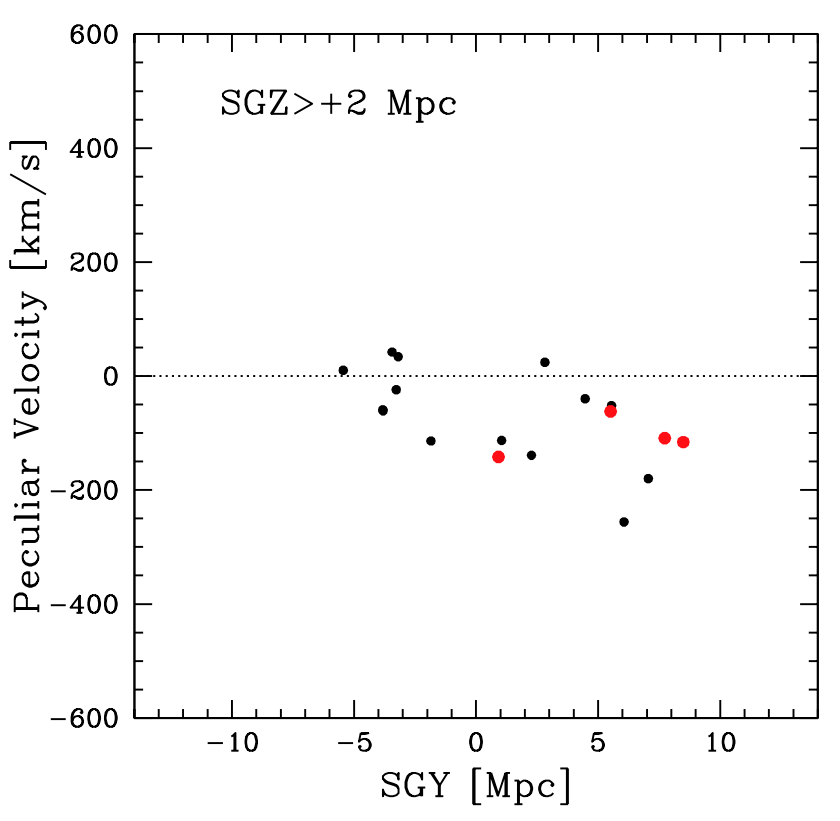}
\includegraphics[width=0.82\linewidth]{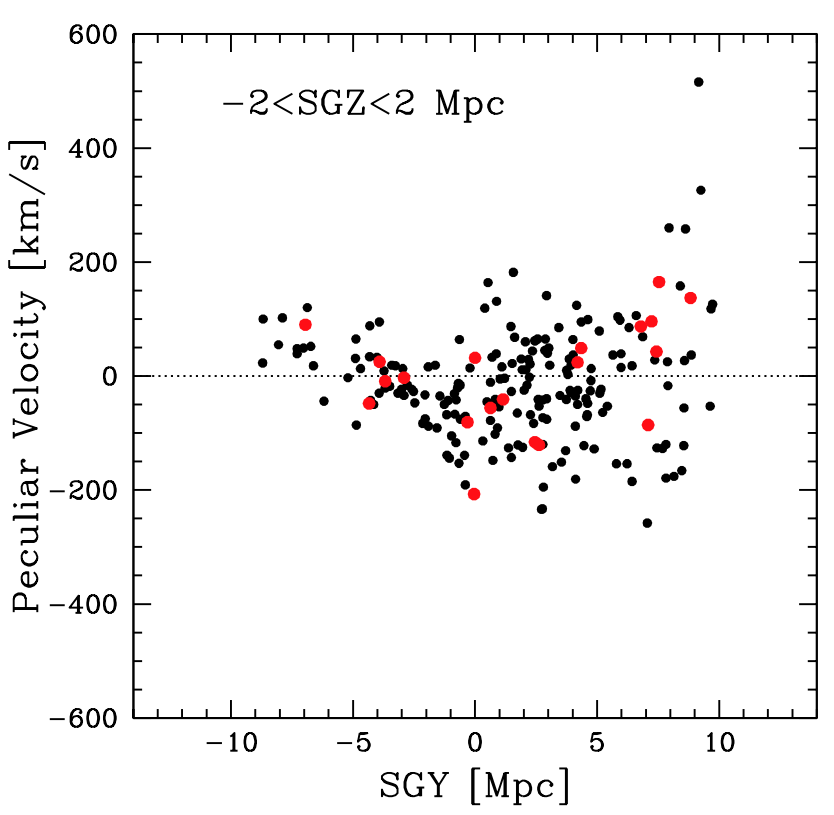}
\includegraphics[width=0.82\linewidth]{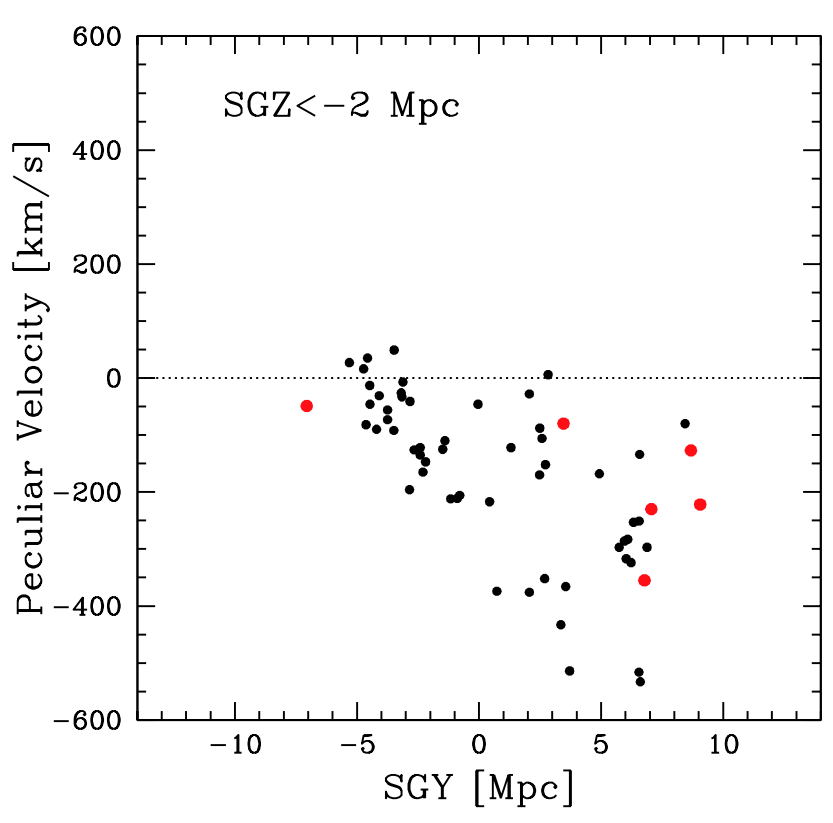}
\caption{
Gradients in the peculiar velocities of groups toward Virgo Cluster in 3 $SGZ$ intervals.  Groups more luminous that $L_K = 10^{10}$~$\Lsun$ are labeled in red.
}
\label{fig:vpec_gradY}
\end{figure}

The local gradient caused by the attraction of the Virgo Cluster is seen in Figure~\ref{fig:vpec_gradY} separated into three $SGZ$ slices.  In this figure and the following, the points that are plotted represent groups, as defined in \S \ref{sec:groups}, in order to suppress the dispersion in velocities within groups.  The middle panel shows the gradient across the Local Sheet.  From our perspective at $SGY=0$, galaxies at positive $SGY$ are increasingly moving away from us (attaining higher positive peculiar velocities) as they near the Virgo Cluster, flairing sharply upward at the surface marking cluster infall (see where infall toward the Virgo Cluster becomes important in Figures~\ref{fig:filam} and \ref{fig:yzflow}).  The trend is maintained at negative $SGY$ where the trend of increasing positive peculiar velocities with distance from us is a signature of the relatively greater attraction that we experience toward Virgo.

A somewhat similar ``U'' shaped run of peculiar velocities is seen within the data of the Leo Spur in the lower panel, with the bottom of the ``U'' displaced a few Mpc to positive $SGY$ because of projection effects.  The Leo groups at $SGY \sim 10$~Mpc are close to the Virgo turnaround surface.  The most extreme negative velocities below $-500$~\kms\ are properties of small dwarf galaxies loosely associated with the Leo Spur.  In the  lower panel of Figure~\ref{fig:LV2} these entities lie below and left of the Leo Group label and are seen to have substantial motions upward toward the Virgo Cluster.  Their upward velocities and our downward velocity are additive.

In the Pegasus Spur structure above the Local Sheet, addressed in the top panel, overall peculiar velocities are negative but trending positive toward negative $SGY$, south of the Milky Way.

%Fig 10
\begin{figure}
\centering
\includegraphics[width=0.82\linewidth]{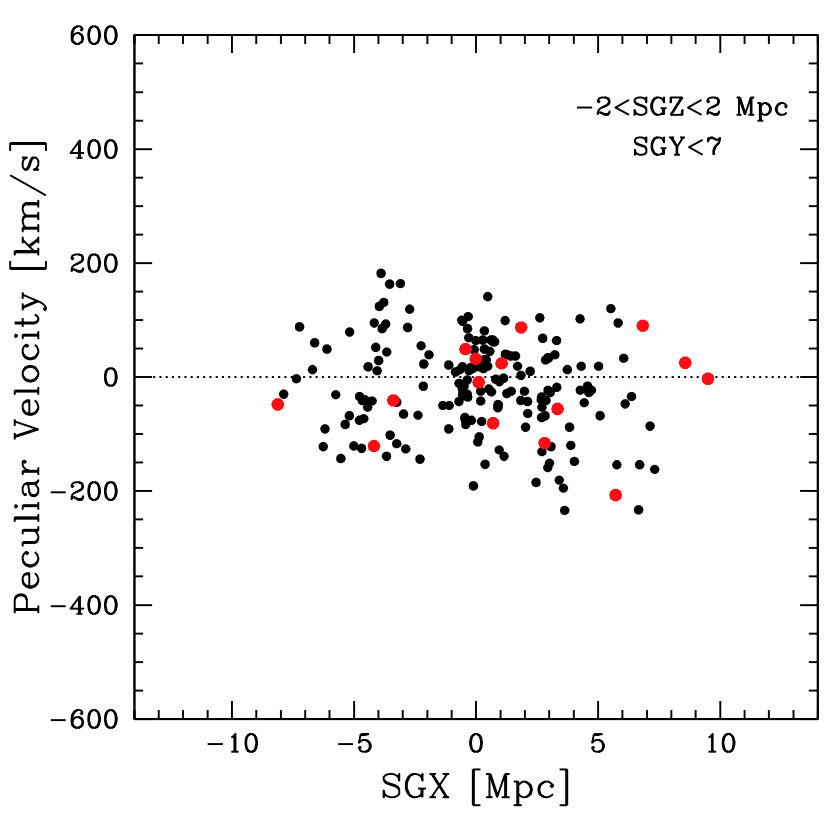}
\caption{
Gradient across the Local Sheet in group peculiar velocities with $SGX$ displacement.  Cut at $SGY=7$~Mpc to avoid the strong positive gradient at the Virgo infall interface.  Most luminous groups are labeled red as in Fig~\ref{fig:vpec_gradY}.
}
\label{fig:vpec_gradX}
\end{figure}

Figure~\ref{fig:vpec_gradX} reveals a slight gradient in the orthogonal $SGX$ direction.  Here, only galaxy groups within the plane of the Local Sheet are displayed to avoid the confusion that would be caused by $SGZ$ projections.  Also, galaxies at $SGY>7$~Mpc are excluded to avoid the flair toward positive peculiar velocities as the Virgo Cluster is approached.  The gradient is modest because the causes are less local: the pull of the distant Great Attractor and the repulsion caused by the void between Laniakea and Perseus-Pisces superclusters.

\begin{figure*}[]
\centering
\includegraphics[width=0.99\linewidth]{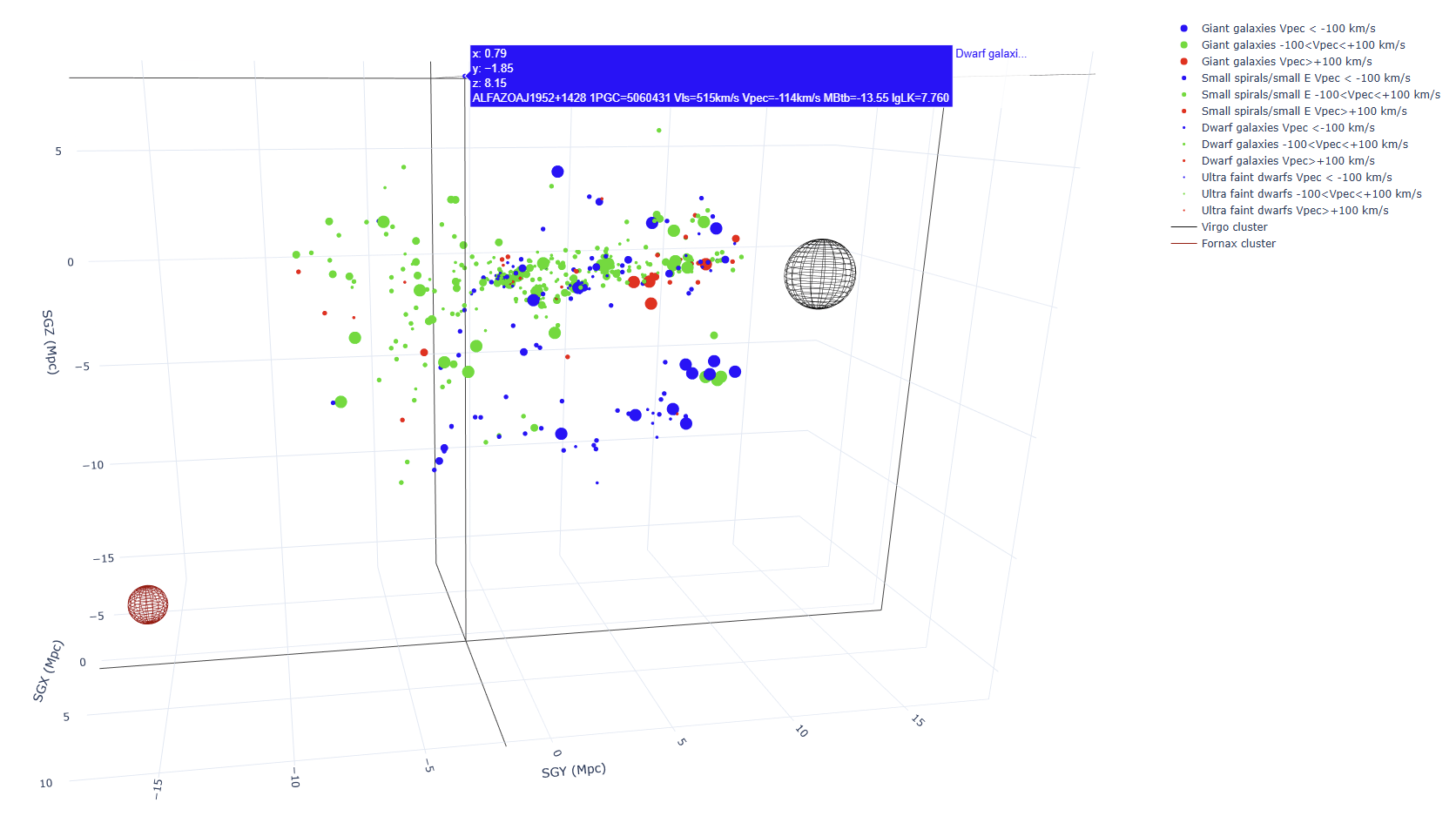}
\caption{An \href{https://irfu.cea.fr/Projets/COAST/10Mpc_Groups.html}{online interactive 3D visualization} of the 10 Mpc Groups Catalog. This visualization can also be found  \href{https://doi.org/10.5281/zenodo.23726804}{here}. The view can be rotated, zoomed in and out and translated using proper mouse or touch actions on classical displays and touch screens. Hover on the objects on display to get their positions and a selection of metadata: name, identifier of the group's dominant member, redshift velocity, peculiar velocity, magnitude and luminosity. Double-click or single-click on the objects' categories listed in the right-hand side column to isolate or hide them away. Two spheroid meshes are used to materialize the positions and approximate extents of the Virgo and Fornax clusters.
}
\label{fig:10Mpc_Groups_3D}
\end{figure*}

The overall trends in local peculiar velocities can be seen in the interactive Figure~\ref{fig:10Mpc_Groups_3D}.  Here, the colors of symbols are coded to peculiar velocities, with values for those in blue less than $-100$~\kms, those in red greater than 100~\kms, and those in green within $\pm 100$~\kms.  Cases lying in the equatorial Local Sheet are mostly green except as the Virgo Cluster is approached where large positive peculiar velocities are found.  Negative peculiar velocities are associated with galaxies below the Local Sheet and the relatively few found above the Local Sheet.  Selecting on objects in the interactive display gives individual parameter values.

%Fig 11

As an aside, it is noted that a small number of objects at distances up to $\sim 20$~Mpc have been the targets of TRGB studies based on HST observations.  Multiple orbits have been required but the efforts have been justified by the special natures of the galaxies, say, because they hosted Type Ia supernovae or lie in the important Virgo or Fornax clusters.  Such galaxies lie outside the structures of our primary interest.  Given the coherence in velocities within structures, just one excellent distance gives a strong clue to the kinematic behaviour of the associated structure.  Accordingly, NGC\,4038, at a distance of 21.3~Mpc and within a group with $\langle V_\mathrm{LS}\rangle=1438$~\kms\ \citep{2017ApJ...843...16K}, has a group peculiar velocity of $-160$~\kms\ that implies a relative motion toward us of the structure at negative $SGZ$ called the Crater Cloud in the Nearby Galaxies Atlas and Catalog.  Likewise, NGC\,5907 at 17.3~Mpc and in a group with $\langle V_\mathrm{LS}\rangle=949$~\kms\ whence $V_\mathrm{pec}=-348$~\kms\ for the group, suggests a large downward motion of the structure at positive $SGZ$ called the Draco Cloud.
These peculiar velocities converging on the supergalactic equator are quite in accord with the motions modeled in the numerical action study of \citet{2022ApJ...927..168S}.

As another aside, it might be considered curious that most large peculiar velocities are toward us.  The evident physical reason is a compression toward the supergalactic mid-plane that we live in.  From an observational perspective, though, galaxies with low velocities compared to Hubble flow expectations attract greater attention than galaxies with high velocities compared to Hubble flow.  It must be asked if there are structures nearby that are poorly surveyed because their observed velocities are large.  Such is the case in the environs of the Virgo Cluster, where galaxies to the foreground falling into the cluster attain high velocities away from us.  An example of difficulties is with galaxies associated with the Sombrero (NGC\,4594) Group.  Here, galaxies lie at roughly the tangent point of the zero velocity surface about the Virgo Cluster, with the consequence that systems over a considerable range of distances have roughly the same observed velocity from our perspective. Deeper TRGB surveys will be required to clarify the situation in this extremely interesting region.  Otherwise, the Numerical Action Model study by \citet{2017ApJ...850..207S} based on a basket of distance methodologies over a large range revealed only the Pegasus Cloud among nearby structures to have significant positive peculiar velocities.  Evidently this structure, as it extends at negative $SGY$, is nearing the stall point in velocity flows between the Laniakea and Perseus-Pisces attractors \citep{2017ApJ...845...55P}.  We see a hint of this trend in the peculiar velocities of the adjacent Pegasus Spur in the top panel of Figure~\ref{fig:vpec_gradY}.

%%%%%%%%%%%%%%%%%%%%%%%%%%%%%%%%%%
%%%%%%%%%%%%%%%%%%%%%%%%%%%%%%%%%%

\section{Groups}
\label{sec:groups}

The propensity of nearby galaxies to lie in groups was described by \citet{1975gaun.book..557D}.  There have been many more recent formulations of galaxy groups.  We keep attention on those with a relatively nearby focus \citep{1982ApJ...257..423H, 1987ApJ...321..280T, 1993ApJS...85....1N, 2011MNRAS.412.2498M}.  There has been recognition that the tendency for galaxies to group extends to the regime of dwarfs \citep{2006AJ....132..729T, 2011EAS....48..289M, 2012AstBu..67..135M} which is certainly relevant to the current study. The analysis carried out here builds on the work of \citet{2017ApJ...843...16K} and \citet{2021AN....342..999K}. An interactive 3D visualization of the groups sample considered here is presented in Figure~\ref{fig:10Mpc_Groups_3D}. The tremendous opportunity we have for improvement is the availability of distances with sufficient accuracy to unambiguously assign systems to one of three regimes: (i) within the second turnaround radius of a collapsed halo (either as the primary or a secondary component), (ii) within the first turnaround radius of infall into a collapsed halo, or (iii) isolated.

The conceptual idea of the grouping scheme employed by \citet{2017ApJ...843...16K} was described by \citet{2015AJ....149...54T, 2015AJ....149..171T} and is similar to the method of \citet{2017MNRAS.470.2982L}.  It invokes theoretically motivated scaling relationships between halo mass and the observables velocity dispersion and halo dimensions:
\begin{equation}
    \sigma_p \propto r_{2t} \propto M_v^{1/3}
\label{eq:scaling}
\end{equation}
where $M_v$ is the (virial) mass of the halo, $\sigma_p$ is the line-of-sight (peculiar) velocity dispersion of galaxies within the halo, and $r_{2t}$ is the three-dimensional radius of second turnaround about the halo.  The second turnaround radius has also come to be called the splashback radius \citep{2014ApJ...789....1D, 2014JCAP...11..019A} and follows from theory of self-similar spherical gravitational collapse \citep{1984ApJ...281....1F, 1985ApJS...58...39B}.  The novel aspect of our application is in carrying the scaling relations from rich clusters continuously to the halo masses of dwarf galaxies.

%Fig 12
\begin{figure}
\centering
\includegraphics[width=0.99\linewidth]{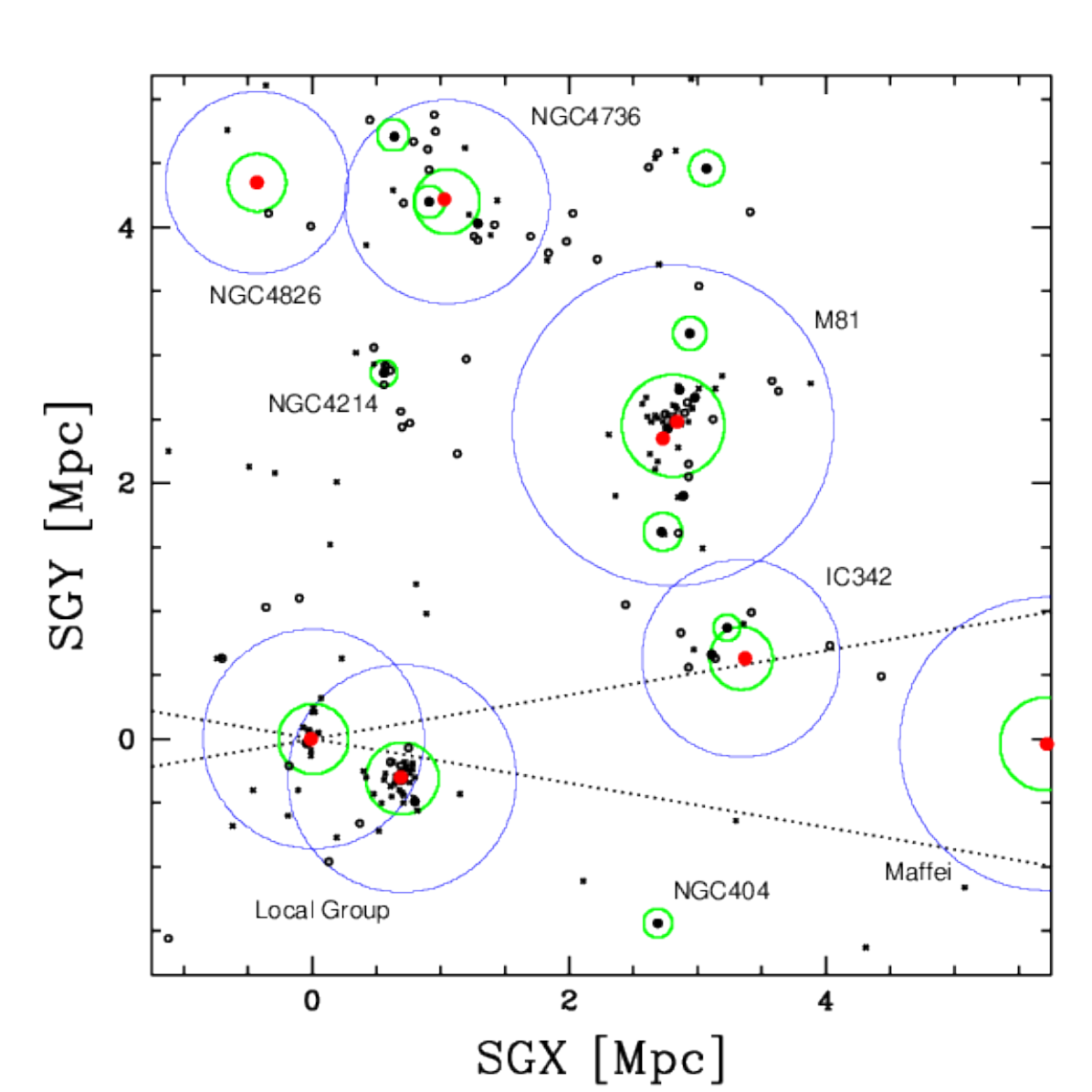}
\includegraphics[width=0.99\linewidth]{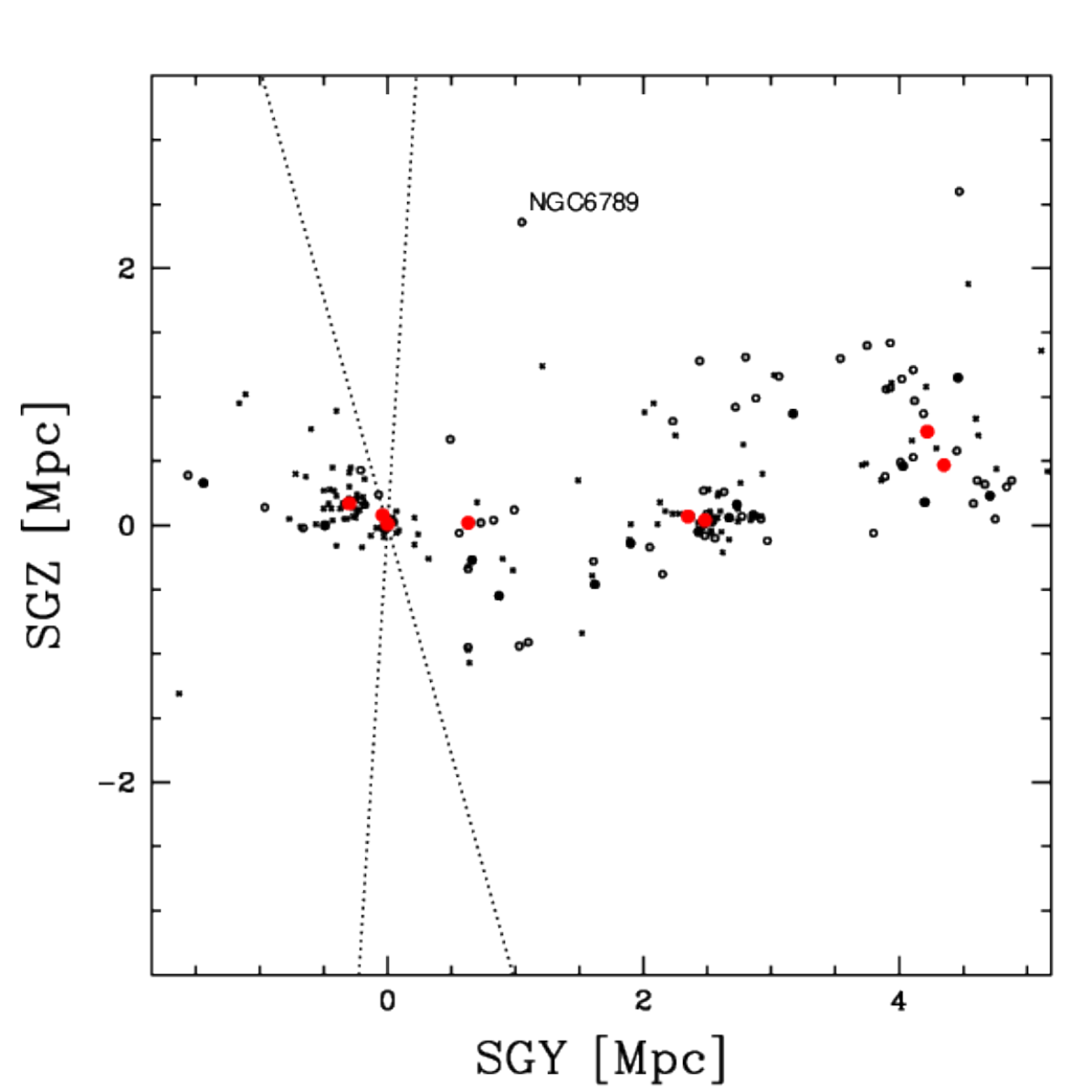}
\caption{
Locations of nearby galaxies within a cube 7~Mpc on a side.  Top panel, supergalactic $SGX$--$SGY$ projection.  Red filled circles identify galaxies with log$L_K>10$; black filled circles identify galaxies with log$L_K$ between 9 and 10; open black circles identify galaxies with log$L_k<9$ but $M_B$ brighter than $-13$; black crosses identify galaxies fainter than $M_B=-13$.  Green circles enclose the projected second turnaround domain for cases with log$L_K > 9$ while blue circles enclose the projected first turnaround domain for cases with log$L_K >10$. The dotted wedge is the projection of $b\pm10^{\circ}$.  The same sample is shown in the lower panel in an $SGY$--$SGZ$ projection but without the green and blue circles.
}
\label{fig:xy_loc}
\end{figure}

A zoom of the distribution of galaxies in our immediate neighborhood is shown in two projections in Figure~\ref{fig:xy_loc}.  It is seen in the $SGY$--$SGZ$ projection that essentially all these nearest galaxies lie within 2~Mpc of the supergalactic equator; i.e., lie within the Local Sheet.  As a result, the display in the top panel is not confused by depth effects.  With only 5\% distance uncertainties, the concentrations of galaxies to the proximities of the most luminous systems is apparent (the heavily obscured Maffei Group has not been properly inventoried). In the case of the Local Group, the first turnaround domains around the separate Milky Way and Andromeda halos defined by the projected first turnaround blue circles are strongly overlapping.  By implication, the Milky Way and Andromeda halos are within a common infall zone, consistent with the knowledge that they are destined to merge \citep{2012ApJ...753....9V}.  There is ambiguity whether the M\,81 and IC\,342 halos or the NGC\,4736 and NGC\,4826 halos will merge.

Figures that zoom in on other sections of the 10~Mpc region are gathered and shown in Appendix~\ref{sec:appendixA}.

%%%%%%%%%%%%%%%%%%%%%%%%%%%%%%%%%%%%%%%
%%%%%%%%%%%%%%%%%%%%%%%%%%%%%%%%%%%%%%%
\section{Mid Infrared Photometry and Stellar Masses}
\label{sec:photometry}
%%%%%%%%%%%%%%%%%%%%%%%%%%%%%%%%%%%%%%%

Flux from galaxies at ultraviolet, optical, and far infrared wavelengths provide information on star formation history, moderated by attenuation by dust \citep{2008ApJS..178..247K, 2009ApJ...703..517D, 2011ApJS..192....6L, 2017ApJ...850...68C}.
By contrast, flux at mid infrared bands is dominated by the accumulated stellar content manifested in the Rayleigh-Jeans tail of the stellar spectral energy distribution (SED), relatively free of dust reddening.
These mid infrared fluxes can be directly interpreted as stellar masses \citep{2012AJ....143..139E, 2014MNRAS.445..899C, 2014ApJ...782...90C},.

Photometry at 3--4~$\mu$m of nearby galaxies has been carried out from space with Spitzer Space Telescope \citep{2004ApJS..154....1W} and WISE, the Wide-field Infrared Explorer \citep{2010AJ....140.1868W} and the supplement NEOWISE \citep{2014ApJ...792...30M}.
The Spitzer Local Volume Legacy (LVL) survey \citep{2009ApJ...703..517D} observed 258 galaxies within 11~Mpc with filters at 3.4~$\mu$m and 4.6~$\mu$m.
A larger and somewhat deeper program \citep{2026arXiv260605780T} extracted cutouts from the all sky WISE and NEOWISE observations, providing photometry for 567 galaxies at 3.6~$\mu$m (W1) and 558 galaxies at 4.5~$\mu$m (W2).
The fluxes in the W1 band can be translated into stellar mass \citep{2014MNRAS.445..899C}
\begin{equation}
    \Upsilon_{\ast}^{W1} = 0.5 M_{\odot}/L_{\odot}^{W1}.
\label{eq:upsilon}
\end{equation}
Stellar mass is then given by
\begin{equation}
    M_{\ast} = \Upsilon_{\ast}^{W1} L^{W1}
\label{eq:stellarmass}
\end{equation}
where the absolute magnitude of the Sun at W1 is 5.91~mag AB.

%%%%%%%%%%%%%%%%%%%%%%%%%%%%%%%%%%%%%%%
%%%%%%%%%%%%%%%%%%%%%%%%%%%%%%%%%%%%%%%

\section{Halo luminosity and Mass Functions}
\label{sec:halofn}
%%%%%%%%%%%%%%%%%%%%%%%%%%%%%%%%%%%%%%

%Fig 13
\begin{figure}
\centering
\includegraphics[width=0.99\linewidth]{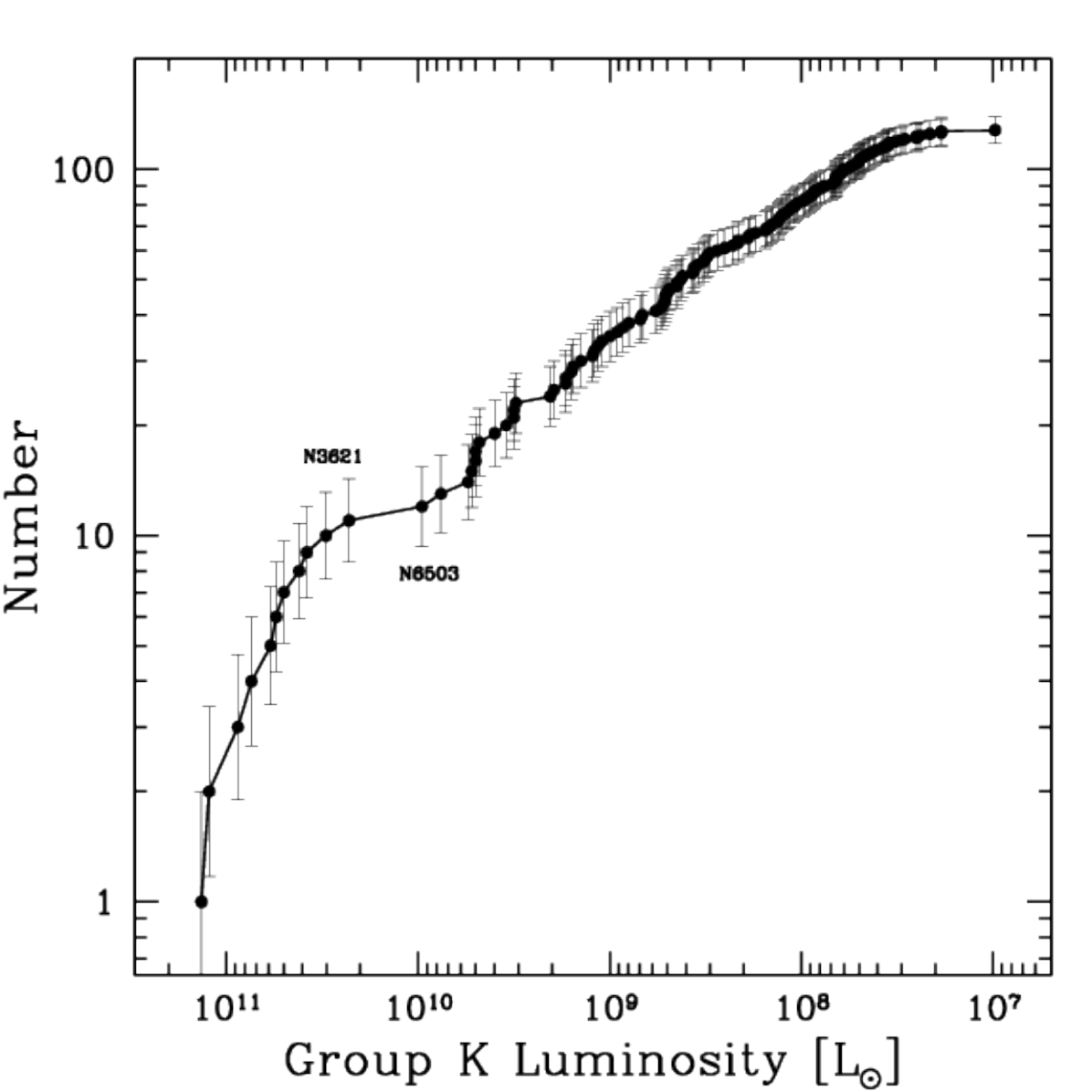}
\includegraphics[width=0.99\linewidth]{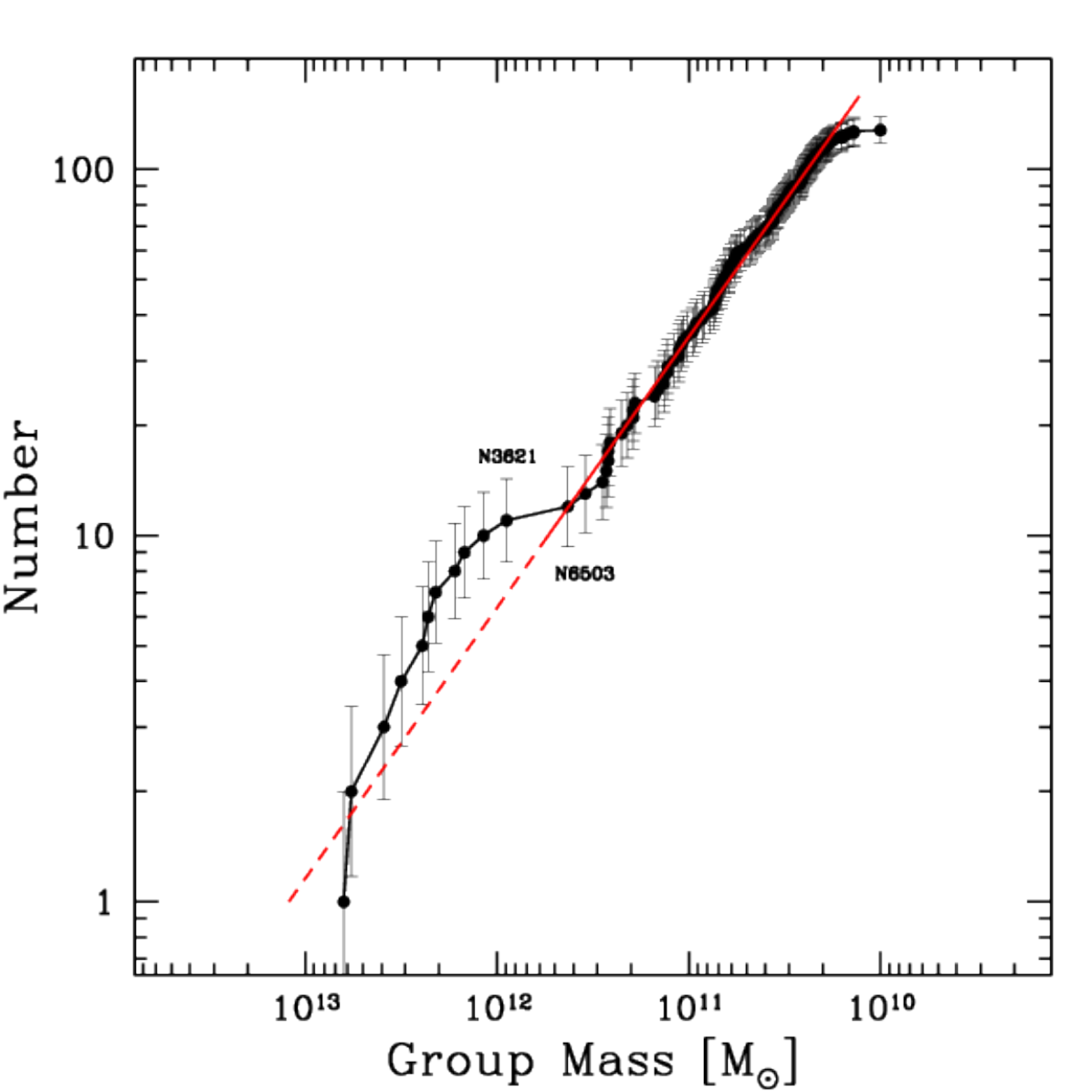}
\caption{
The top panel gives the cumulative $K$ band luminosity function for 128 groups within 7 Mpc and brighter than $M_B=-13$, while the bottom panel shows the mass function for the same groups.  The solid red line is a fit to the points with log mass between 10.1 and 11.9 and the dashed line is an extension of the linear fit to higher masses.  Two groups are identified that sit on each side of curious gaps in the functions.
}
\label{fig:LandMfn}
\end{figure}

%Fig 14
\begin{figure}
\centering
\includegraphics[width=0.99\linewidth]{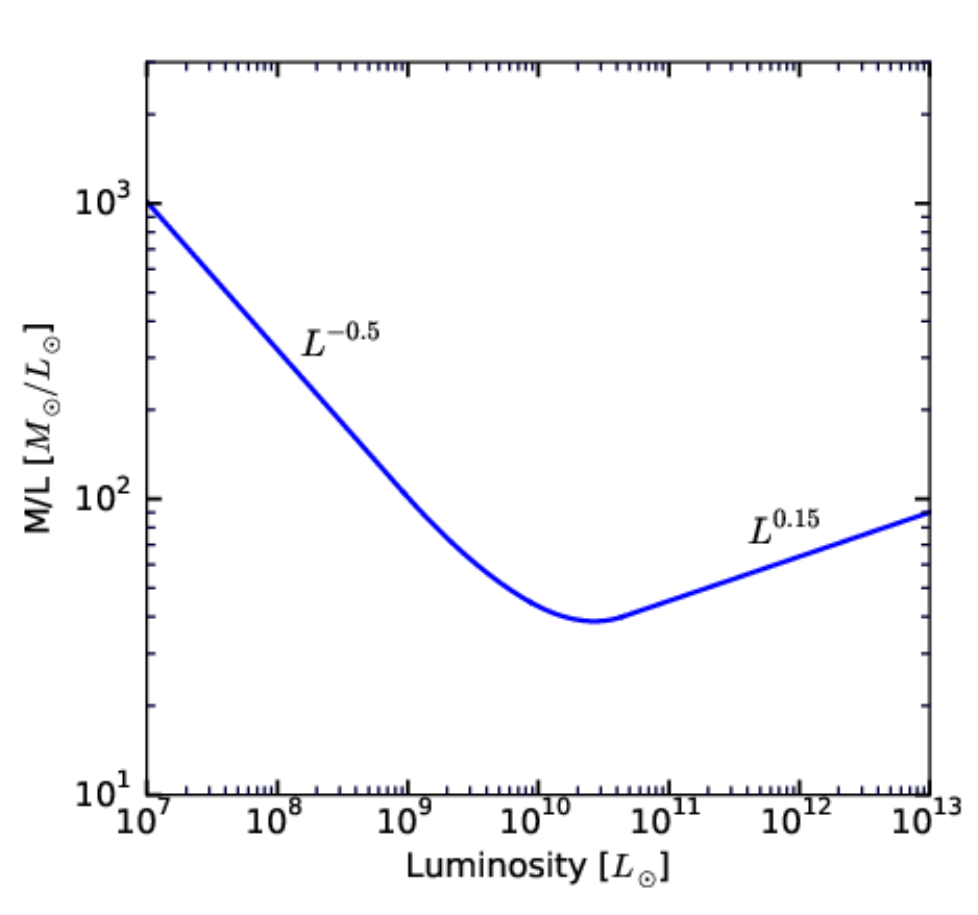}
\caption{
Assumed relationship between group $K$ luminosities and masses.
}
\label{fig:MoL}
\end{figure}

%Fig 15
\begin{figure}
\centering
\includegraphics[width=0.99\linewidth]{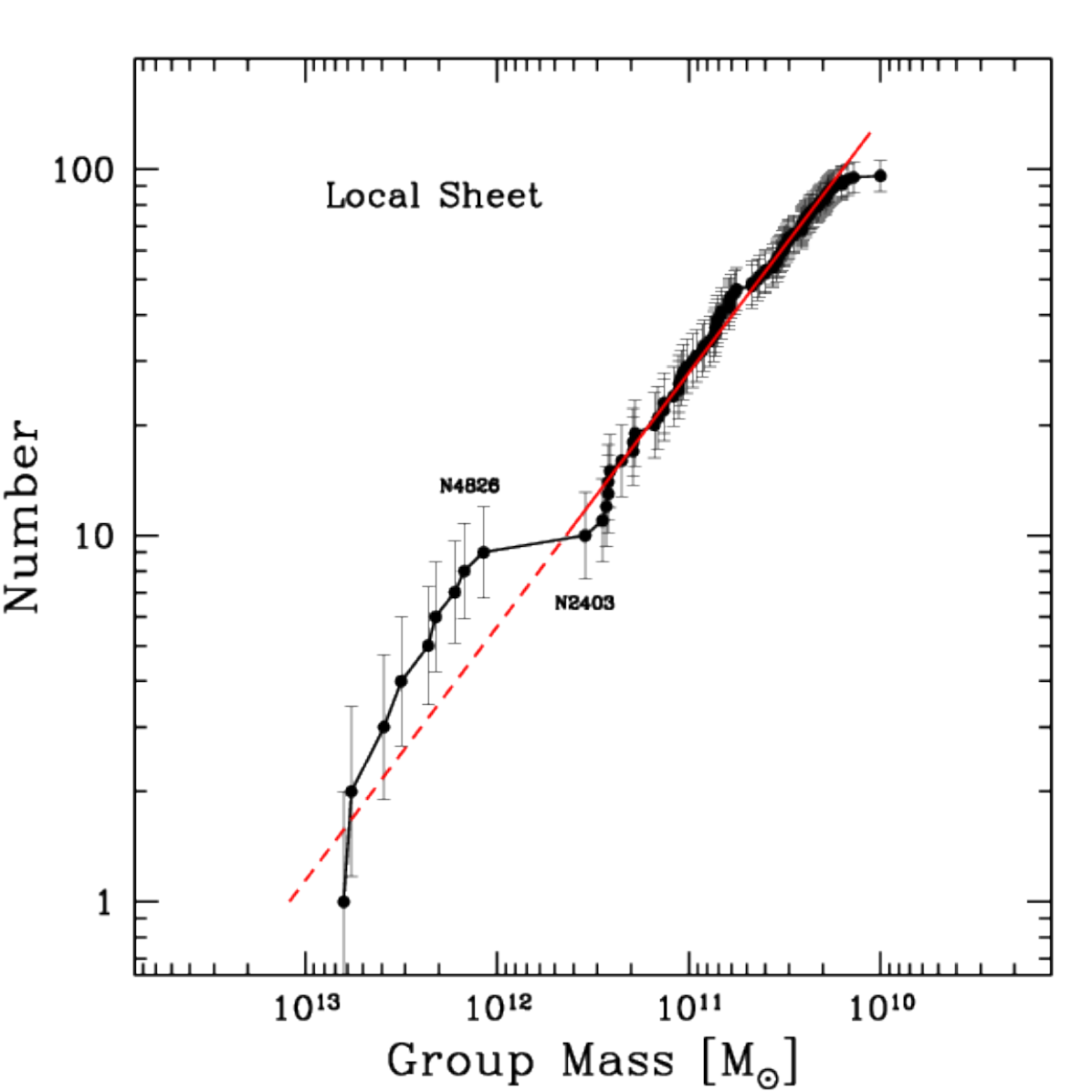}
\includegraphics[width=0.99\linewidth]{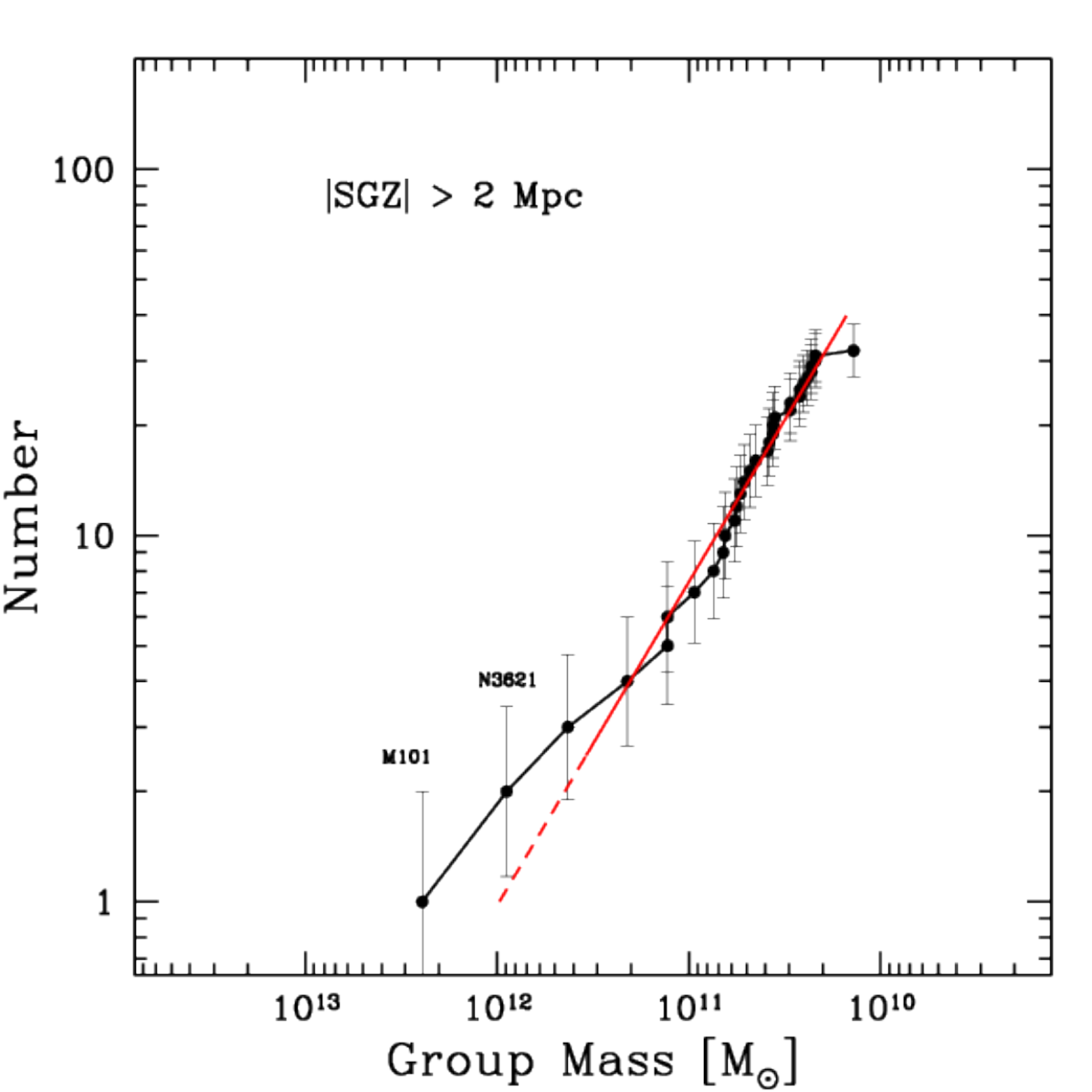}
\caption{
The cumulative mass function split into the contributions within $\pm2$~Mpc of the supergalactic equator (top panel) and contributions further than these limits from the supergalactic equator (bottom panel).
}
\label{fig:2Mfn}
\end{figure}

With the restrictions group distances between 1 and 7~Mpc, absolute Galactic latitudes greater than $10^{\circ}$ and individual galaxy luminosities brighter than $M_B = -13$, we have 180 galaxies in 128 halos, with only 3 galaxies lacking precision distances.  For each galaxy, a luminosity at the 2MASS $K_s$ band is acquired, either directly from either the Extended Source Catalog or the Large Galaxy Atlas \citep{2000AJ....119.2498J, 2003AJ....125..525J} or inferred from $B$ magnitudes and galaxy types following \citet{2003AJ....125..525J} and \citet{2005AstL...31..299K}.  The luminosities associated with halos are the sums over all members of a halo.  Absolute luminosities assume the group distance.

A cumulative halo luminosity function for our 7~Mpc sample is seen in the top panel of Figure~\ref{fig:LandMfn}.  The cumulative halo mass function is shown in the bottom panel for the same sample assuming the translation between light and mass illustrated in Figure~\ref{fig:MoL} that was used by \citet{2017ApJ...843...16K}.  The number of halos are found to follow a tight power law at masses below $10^{12}~\Msun$, with a significant excess above an extension of the power law in the decade of higher masses.  There is a jump in the number$-$luminosity and number$-$mass relations at $\sim 10^{10}~\Lsun$ and $\sim 10^{12}~\Msun$, with the halos associated with NGC\,3621 and NGC\,6503 on the high and low sides respectively.  At lower masses than that associated with NGC\,6503, the cumulative numbers of halos $N$ depend on mass $M$ as $N \propto M^{-0.74\pm0.01}$.

At face value, there is an excess of halos in the half decade of masses above $10^{12}~\Msun$ with respect to the extension of the power law describing numbers at lower masses.  Details depend on the assumed linkage between luminosities and masses but a feature with the same implication is seen in the halo luminosity function of the upper panel of Figure~\ref{fig:LandMfn}.  The statistics are poor in this volume restricted to 7~Mpc, but a similar feature in the halo mass function was seen in data assembled in a larger volume by \citet{2017ApJ...843...16K}.  The excess of giant to dwarf galaxies within 8~Mpc in the Local Sheet was remarked by \citet{2020MNRAS.494.2600N} and cited as unexpected within a $\Lambda$CDM cosmology.

Is this feature of the halo mass function restricted to the Local Sheet?  The sample shown in Figure~\ref{fig:LandMfn} is split in Figure~\ref{fig:2Mfn} into the 75\% of galaxies lying in the Local Sheet, top panel, and the 25\% that are greater than $\pm2$~Mpc from the central plane, lower panel.  Not surprisingly, the cumulative mass function of the dominant Local Sheet component is similar to that of the ensemble.  The growth in number toward lower masses is somewhat shallower, with $N \propto M^{-0.69\pm0.02}$.  The gap in the mass function at $\sim 10^{12}~\Msun$ is widened compared to the ensemble since the gap in the ensemble is bounded by two galaxies off the Local Sheet.  Among galaxies in the local Sheet, the gap is bounded by the halos of NGC\,4826 on the high side an NGC\,2403 on the low side.  The gap amounts to a factor of 4 in mass.

Looking to the bottom panel of Figure~\ref{fig:2Mfn}, only two galaxies lie above the break at $\sim10^{12}~\Msun$: M\,101 and NGC\,3621.  These galaxies lie in the Local Sheet fringes, second panels of Figure~\ref{fig:filam}, at $SGZ=+2.56$~Mpc and $-3.34$~Mpc, respectively.   The slope of the cumulative mass function faintward of these two is steeper than found in the Local Sheet, with $N \propto M^{-0.88\pm0.03}$.  The hint of an excess above the break in this lower panel is not significant, but the steeper slope in counts of fainter galaxies in regions off the Local Sheet compared with in the Local Sheet has statistical significance.

%%%%%%%%%%%%%%%%%%%%%%%%%%%%%%%%%%%%%%%
%%%%%%%%%%%%%%%%%%%%%%%%%%%%%%%%%%%%%%%

\section{Galaxy Mass in Stars}
\label{sec:pops}
%%%%%%%%%%%%%%%%%%%%%%%%%%%%%%%%%%%%%%%

%Fig 16
\begin{figure}
\centering
\includegraphics[width=0.99\linewidth]{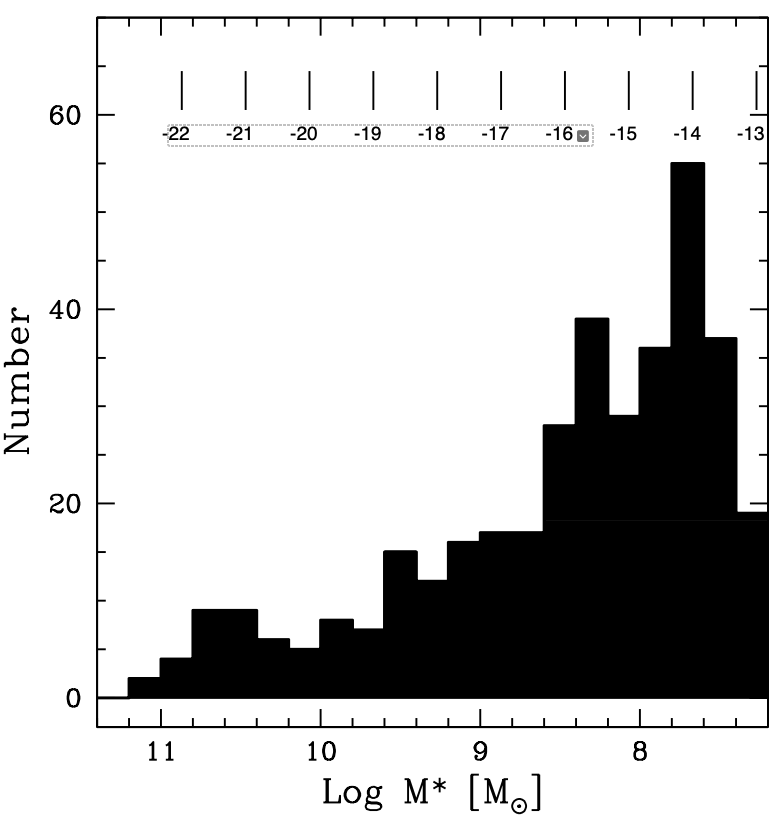}
\caption{
Number of galaxies in log stellar mass bins for the sample of 370 galaxies with $1<d<10$~Mpc, $|b|>10^{\circ}$, and $M_{W1}<-13$~mag. Correspondence between $M_{W1}$ and log~$M^{\ast}$ is shown across the top of the figure.
}
\label{fig:N_Mstar}
\end{figure}

As discussed in Section~\ref{sec:photometry}, there is a close connection between mid-infrared luminosities and stellar masses.
Given the linkage of Eq.~\ref{eq:stellarmass}, every galaxy in our sample with W1 photometry can be assigned a mass.
Figure~\ref{fig:N_Mstar} is a histogram of the numbers of galaxies with these estimated stellar masses in log stellar mass intervals.
$M_{W1}$ equivalences are given across the top of the figure.

\begin{figure}
\centering
\includegraphics[width=0.99\linewidth]{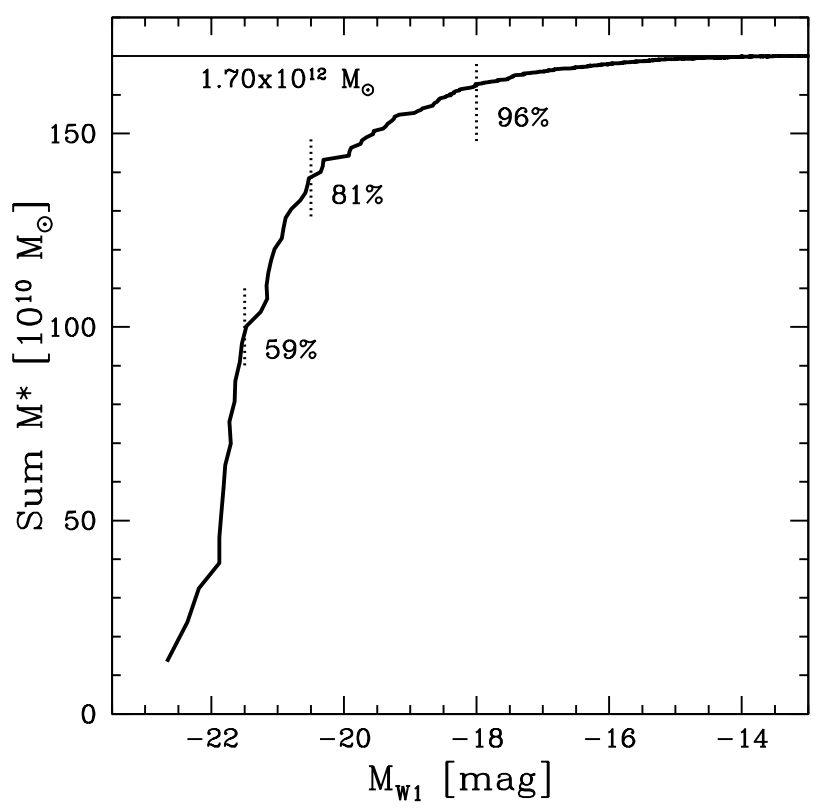}
\caption{
Sum of stellar mass as a function of $M_{W1}$ for the sample of 370 galaxies with $1<d<10$~Mpc, $|b|>10^{\circ}$, and $M_{W1}<-13$~mag.
}
\label{fig:sumMstarW1}
\end{figure}

\begin{figure}
\centering
\includegraphics[width=0.99\linewidth]{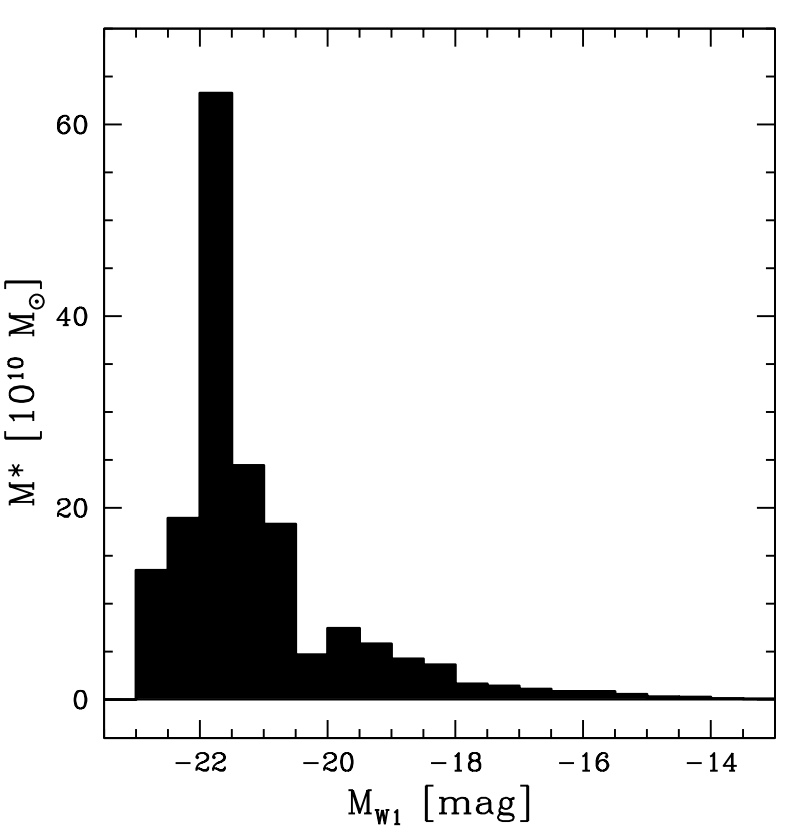}
\caption{
Stellar mass in intervals of $M_{W1}$ for the sample of Figure~\ref{fig:sumMstarW1}.
}
\label{fig:MstarMW1}
\end{figure}

Figures~\ref{fig:sumMstarW1} and \ref{fig:MstarMW1} present a very different view of the same material.
The first of these two figures gives the sum of stellar mass starting at the most massive galaxy (NGC\,5128) and successively adding the stellar masses of additional galaxies ordered by decreasing mass.
In total, the sample sums to $1.70\times10^{12}~\Msun$.
Over 80\% of the mass in stars is contained within the galaxies brighter than $M_{W1}=-20.5$ and 96\% is contained by $M_{W1}=-18$~mag.

The summation of masses in bins of $M_{W1}$ is shown in Figure~\ref{fig:MstarMW1}.
The difference from Figure~\ref{fig:N_Mstar} is dramatic.
Most of the stellar mass is within the few brightest galaxies.

%Fig 17

%Fig 18

\begin{figure}
\centering
\includegraphics[width=0.99\linewidth]{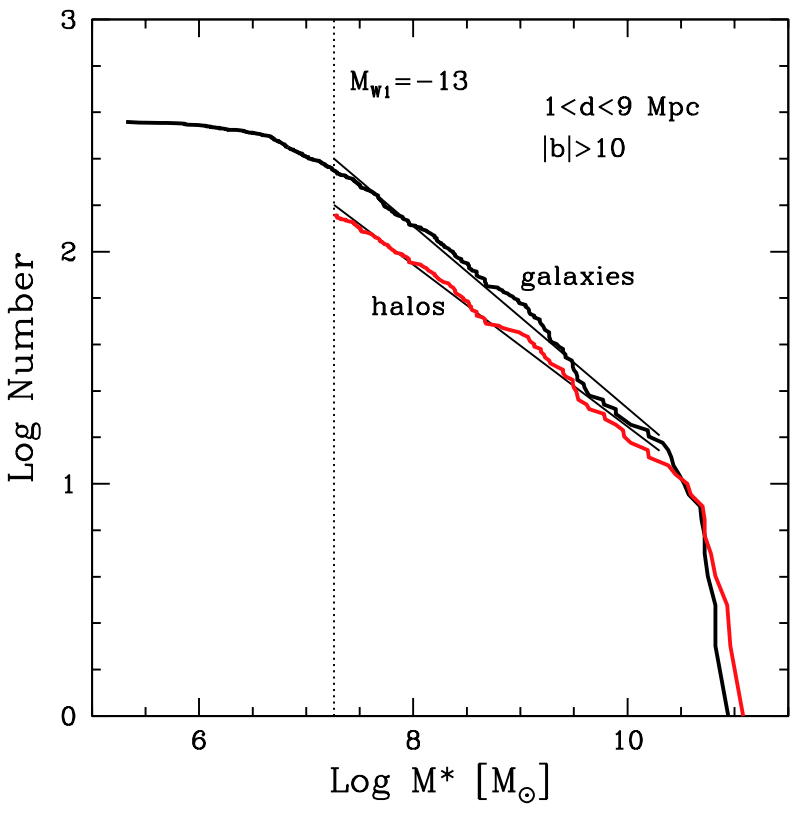}
\caption{
Growth in the number of galaxies with decreasing stellar mass for the sample within $1-9$~Mpc. The black line shows the function for individual galaxies and the red line shows the function for group halos.  The vertical dotted line is at the completeness limit of $M_{W1}=-13$~mag.  Straight lines are fit over the ranges of power law growth.
}
\label{fig:halo_massfn}
\end{figure}

The sample is slightly restricted in Figure~\ref{fig:halo_massfn} to distances $1<d<9$~Mpc because of incompleteness and distance uncertainties beyond 9~Mpc.
The black line labeled {\it galaxies} gives the stellar mass function for the sample.
Incompletion sets in at $M_{W1}=-13~\simeq~ {\rm log}M^{\ast}=7.25~\Msun$.
The rapid rise in the brightest 0.3 logarithmic mass interval is notable (involving eight galaxies).
Toward lower masses, there is a power law between stellar mass and number, with $N \propto {M^{\ast}}^{-0.39}$.

%Fig 19

%Fig 20
\begin{figure}
\centering
\includegraphics[width=0.99\linewidth]{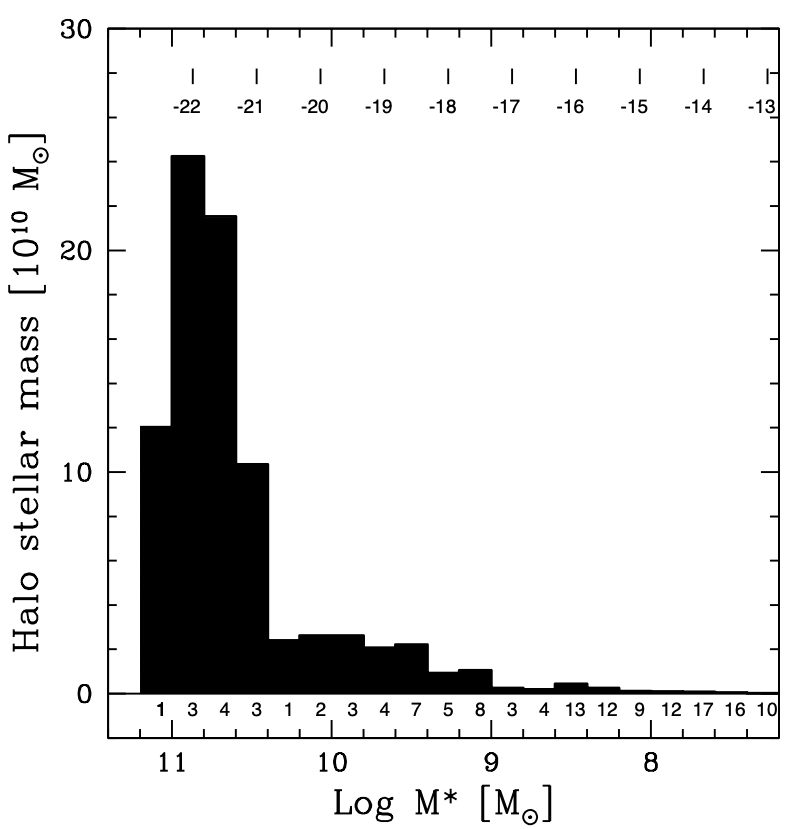}
\caption{
Stellar mass in group halos are given in intervals of log $M^{\ast}$. Numbers of halos in each bin are recorded beneath the histogram. Corresponding values of $M_{W1}$ are given across the top of the figure.
}
\label{fig:haloMstarMW1}
\end{figure}

The 10 galaxies with stellar masses above $\rm{log}~M^{\ast} = 10.5$ are all the primary galaxies within their halos.
The next three most massive systems are secondaries: M\,82 to M\,81, NGC\,5195 to M\,51, and NGC\,4945 to Cen\,A = NGC\,5128.
The stellar populations of these and other secondaries can be added to primary populations to give the total stellar masses within virializing halos.
The red line in Figure~\ref{fig:halo_massfn} labeled {\it halos} shows the stellar mass function for the distinct halos in the sample, extending down to the completeness limit.
The power law fit is $N \propto {M^{\ast}}^{-0.45}$, shallower than that for the full sample of individuals galaxies.

Figure~\ref{fig:haloMstarMW1} is a histogram of the cumulative stellar masses in halo log stellar mass intervals.
The concentration of stellar mass to the most dominant halos is even more extreme than the concentration to the individual dominant galaxies seen in Figure~\ref{fig:MstarMW1}.
Half the mass is within the five dominant halos (associated with NGC\,5128, NGC\,5194, NGC\,3031, NGC\,5055, and NGC\,253) and 3/4 of the mass is accounted for with the next five (NGC\,4258, NGC\,6946, NGC\,5236, NGC\,5457, and IC\,342).
The difference between the most and least massive of these ten halos is only a factor three.
Going down a further factor of three there are only four additional halos.

%Fig 21
\begin{figure}
\centering
    \includegraphics[width=0.99\linewidth]{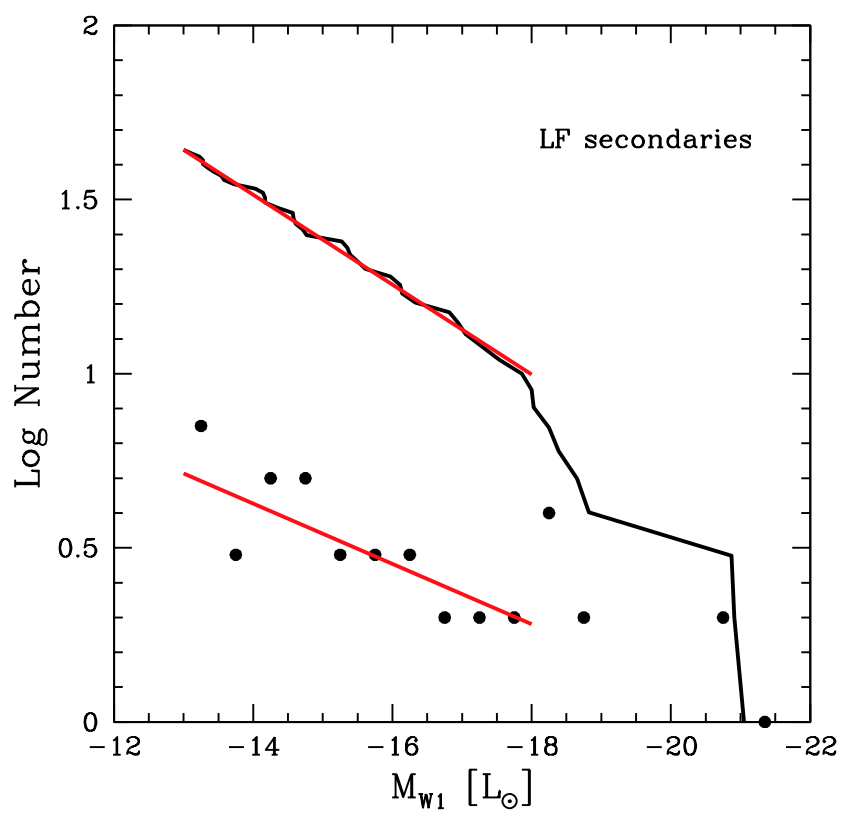}
\caption{
Luminosity function in the W1 band for 44 secondaries brighter than $M_{W1} = -13$ in the halos of the 10 dominant galaxies at $1<d<9$~Mpc. The continuous black line gives the cumulative distribution and the dots show the individual counts in 0.5~mag bins of $M_{W1}$.  The functions can be represented by power laws over the domain from $-18$ to $-13$~mag as shown by the red lines.
}
\label{fig:LFsecondaries}
\end{figure}

The stellar mass functions within groups can be translated into more familiar luminosity functions.
Figure~\ref{fig:LFsecondaries} shows both the binned and cumulative luminosity functions for the secondary galaxies in the halos of the dominant 10 groups at $1<d<9$~Mpc.
There is the familiar deficiency immediately faintward of the primary galaxies.  Then below $M_{W1}=-18$ the distribution increases toward fainter galaxies as described by the Schechter function \citep{1976ApJ...203..297S} faint end power law value $\alpha = -1.27 \pm0.06$.
This value compares with faint end power law values for groups that range $-1.1<\alpha<-1.4$ at optical $R$ band \citep{2009MNRAS.398..722T}.
%The shallower slope for the halo relation tells us that dwarf galaxies are more numerous as secondaries in halos than primaries in isolated halos.

The most distinctive feature of this nearby sample is the dominance of the few major galaxies, distinguished by the gap that sets them apart from the next wrung of systems.
This evidence from stellar masses supports the inferences from masses dominated by dark matter presented in Section~\ref{sec:halofn} where a corresponding gap was found below group mass $\sim 10^{12}~\Msun$.
Statistics are insufficient with this nearby sample but \citet{2017ApJ...843...16K} found a similar feature in a volume extending to 3500~\kms.
It can be asked if this dominance of massive galaxies combined with the shallow stellar mass functions are at odds with standard $\Lambda$CDM theory.
\citet{2020MNRAS.494.2600N} compared the ratio of giants to dwarfs locally with expectations from simulations and found the high value to be a $2.5\sigma$ excursion from $\Lambda$CDM.  This comparison is made more extreme with the observation that most of stellar mass is within a small number of halos within a factor 3 above $3~10^{10}~\Msun$ and there is an evident deficiency in the factor 3 interval below this mass.

%%%%%%%%%%%%%%%%%%%%%%%%%%%%%%%%%%%%%%%
\section{Galaxy Metallicities}
\label{sec:metals}
%%%%%%%%%%%%%%%%%%%%%%%%%%%%%%%%%%%%%%%

Successive episodes of star formation lead to the build-up of heavy elements in a galaxy.
There is the complex interplay driven by star formation efficiency but modulated by influx of pristine interstellar gas and losses from energetic events.
It has long been known that metallicities are lower in dwarf galaxies than in giant galaxies \citep{1986ApJ...303...39D}.
The documented decrease in metallicity with decreasing stellar mass down to log$M^{\ast}\sim8.5$ \citep{2004ApJ...613..898T} has been followed to smaller systems \citep{2006ApJ...647..970L, 2010ApJ...711..671K, 2011ApJ...727...79K} down to log$M^{\ast}\sim4$ and $[Fe/H]\sim-2.5$.

Metallicities can be evaluated through spectroscopy, the fitting of isochrones to the CMD of resolved stellar populations, or the related characteristic of the color of the upper RGB.
The initial calibrations established with Galactic globular clusters \citep{1988AJ.....96...92A, 1990AJ....100..162D} were extended to galaxies \citep{1993AJ....106..964L} over the color range $1.22<V-I<1.74$ corresponding to the metallicity range $-2.17<[Fe/H]<-0.7$.

%Fig 22
\begin{figure}
\centering
    \includegraphics[width=0.99\linewidth]{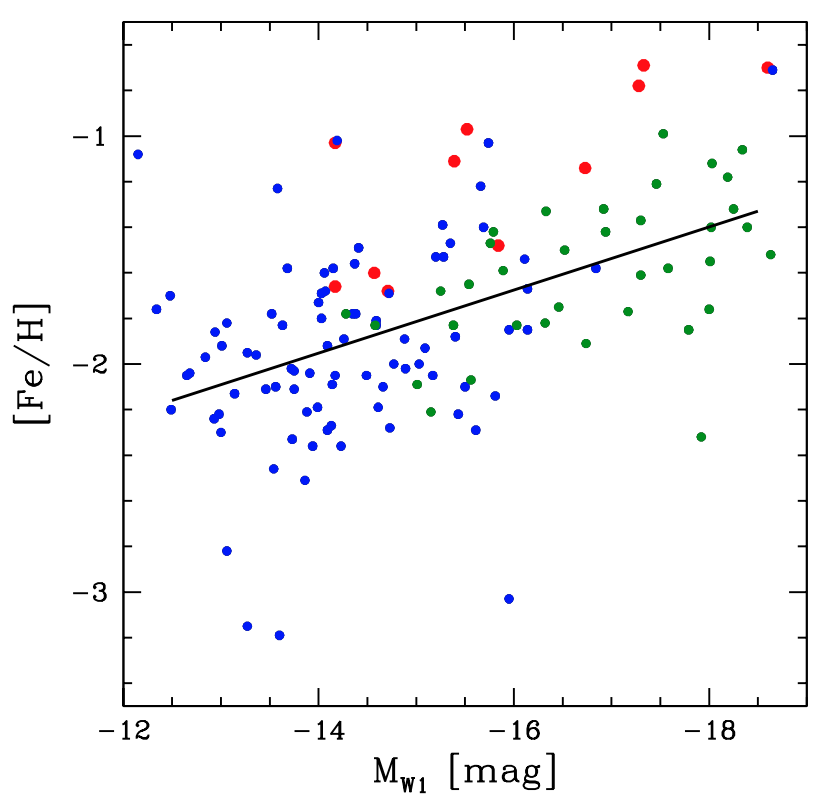}
\caption{
Metal index $[Fe/H]$ dependence on absolute W1 magnitude, $M_{W1}$. Early type galaxies (mostly dSph) in red, irregulars typed 9, 10 in blue, and types 1-7 in green. Fitted line excludes early types and outliers.
}
\label{fig:MZ}
\end{figure}

The RGB is a prominent (and necessary) feature for the determination of TRGB distances and as a secondary product provides an estimate of galaxy metallicity.
A characteristic color is measured at $M_I = -3.5$, 0.5 mag below the TRGB \citep{2008MNRAS.384.1544S, 2026arXiv260418183M}.
Taking metallicity values from this latter paper, Fig.~\ref{fig:MZ} is a plot of the index $[Fe/H]$ over the range in $M_{W1}$ between $-18.5$ and $-12.5$ (more luminous galaxies are excluded because the camera placement for a TRGB measurement often excluded regions expected to contain high metallicity stars).
There are evident separations between morphological types in the figure.
Those in red, the gas poor systems dominated by older stars, mostly dwarf spheroidals (dSph), have systematically higher $[Fe/H]$ values at a given absolute W1 magnitude.
Along the main trend line of gas bearing systems, galaxies typed Sdm and earlier (T$\le8$) and generally brighter than $M_{W1} = -15.5$ shown in green have statistically higher metallicities than irregulars typed Sm and Im (T values 9 and 10) shown in blue.
The least squares fit straight line (errors in $[Fe/H]$) is described by the relation:
\begin{equation}
    [Fe/H] = -3.90 - 0.14 M_{W1}
    \label{Eq:MZ}
\end{equation}

%\textcolor{red}{The straight line in Fig 22 lies $\sim0.7$ dex below the relation shown by Tassis, Gnedin, Kravtsov 2012, ApJ, 745, 68. Is there an explanation? (To check how they get $M_{star}$.)}

Returning to the offset of the early-type systems in Fig.~\ref{fig:MZ}, the effect has been noted by \citet{2003AJ....125.1926G}.
Various explanations have been explored by \citet{2008MNRAS.384.1544S}.
It is possible that the offset may not lie with metallicity. There is a degeneracy in the colors of RGB stars between metallicity and age, with colors shifted redward with both increasing metallicity and increasing age \citep{1994ApJS...95..107W, 2003AJ....125.1261D}.
With continuing star formation, the average age of stars on the RGB will be younger in dIrr compared with dSph.

If the offset is predominantly a metallicity effect, it could be that the continuing star formation in dIrr is fed by continuing accretion of intragalactic gas that dilutes the metal budget.
Alternatively, perhaps the offset lies in the histories of the dSph.
Perhaps they had been more massive but lost stellar mass in the events that removed their gas.
Adding to the puzzle, comparative star formation histories of nearby dwarfs indicate similar developments between gas rich and poor systems, only diverging in the last few Gyr \citep{2011ApJ...739....5W}.

%%%%%%%%%%%%%%%%%%%%%%%%%%%%%%%%%%%%%%%
\section{Galaxy Morphologies with Environment}
\label{sec:environment}
%%%%%%%%%%%%%%%%%%%%%%%%%%%%%%%%%%%%%%%

Morphological classifications based on appearance in low resolution images such as provided in HyperLEDA \citep{2014A&A...570A..13M} provide a rough evaluation of the stellar content of galaxies.  The presence or absence of neutral hydrogen monitors the opportunity for star formation.  Most critically, the resolution of stars at the bright end of the CMD from HST observations convincingly reveals the occurrences of recent star formation referenced to the importance of the old population \citep{2009ApJS..183...67D, 2015ApJS..218...29M}.
The resolution faint limit reached for a given galaxy depends on it distance and the exposure duration.  With an ACS observation over a single HST orbit, observations reach F814W~$\sim 27$~mag corresponding to $M_I \sim -3$ at 10~Mpc.  For galaxies within 10~Mpc, star formation with ages 100~Myr or less will be detected in a populated main sequence.

It has long been known from studies of clusters that galaxy morphologies vary with environment \citep{1974ApJ...194....1O, 1980ApJ...236..351D}.
This characteristic extends to the smallest galaxies \citep{2011EAS....48..315G, 2022ApJ...933...47C}.
What follows are statistics for the highly complete sample brighter than $M_B \simeq M_{W1} =-13$, $|b|>10^{\circ}$, and $1<d<9$~Mpc.

All but one of the ten dominant galaxies within the sample volume are gas-rich spiral galaxies with abundant star formation. The exception is NGC\,5128 (Centaurus\,A), a system experiencing an interaction.
Of the further 135 dominant galaxies in halos (contributing to the red curve in Figure~\ref{fig:halo_massfn}) there are only four systems overwhelmingly dominated by old stars yet none can be called red and dead.
The brightest, NGC\,404, with S0 morphology, has an outer HI ring and UV emission in it  from young stars \citep{2010ApJ...714L.171T} as well as spectral emission from the nuclear region where there is an evident black hole \citep{1995ApJS...98..477H}.
Another S0 galaxy, NGC\,59, has emission lines from what \citet{2015MNRAS.450.1338S} call a 1.5\% frosting of stars younger than 70 Myr.
KUG\,1234+299 = AGC\,220847 could be called a transition galaxy with an HI detection \citep{2018ApJ...861...49H} and a small population on the upper main sequence.
Similarly, ESO\,384--16 has limited evidence of stars on the upper main sequence and emission lines \citep{2009MNRAS.399..683J}.

% \begin{figure}
% \centering
% \includegraphics[width=0.5\linewidth]{figures/ESO6-01.png}
% \includegraphics[width=0.46\linewidth]{figures/ESO6-1_CMD.png}
% \caption{
% Left: an image of ESO\,006--001 formed from HST F814W and F606W exposures. Right: the CMD of the resolved stars in this image.
% }
% \label{fig:ESO6-1}
% \end{figure}

%Fig 23
\begin{figure}
\centering
\includegraphics[width=1\linewidth]{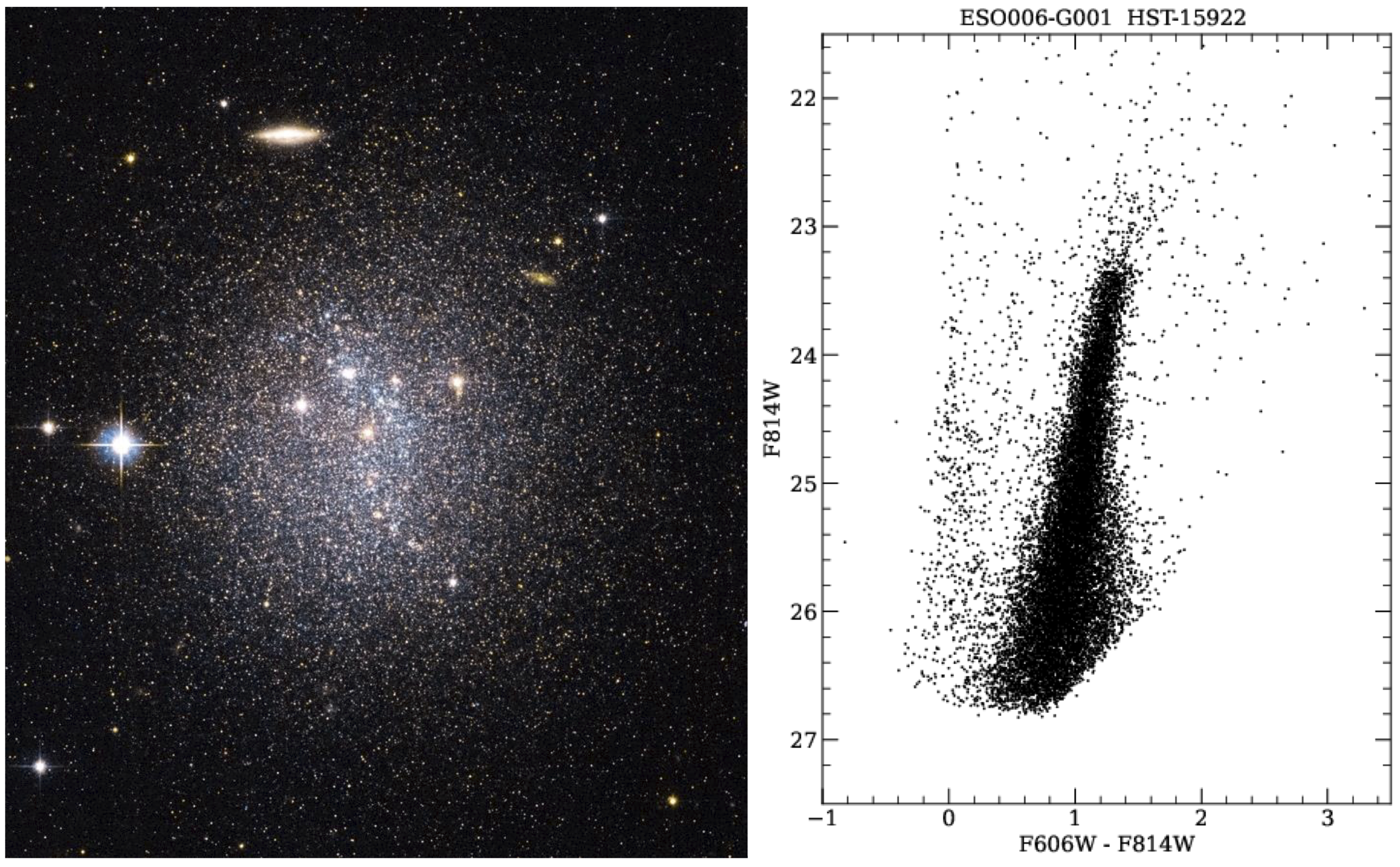}
\caption{
Left: an image of ESO\,006--001 formed from HST F814W and F606W exposures. Right: the CMD of the resolved stars in this image.
}
\label{fig:ESO6-1}
\end{figure}

ESO\,006--001 \citep{2023ApJ...943..139M}, with $2\times10^7~\Msun$, is an interesting case shown in Figure~\ref{fig:ESO6-1}.  Overall, it is estimated that 90\% of the stars formed $12-14$~Gyr ago. Yet, the main sequence is populated to above the level of the TRGB indicating modest star formation within the last 100 Myr. No HI is found superimposed on the galaxy but there is an HI cloud projected 9~kpc removed.
ESO\,006--001 and, indeed, all the galaxies discussed in the previous paragraph, can be called transition systems, where the bulk of stars are old but a small amount of star formation persists.  \citet{2015MNRAS.450.4207B} found that this pattern of delayed star formation is common in isolated dwarf galaxies.

In sum, of 145 primary galaxies brighter than $M_{W1}=-13$ in the sample within 9~Mpc, only the peculiar NGC\,5128 = Centaurus\,A and four small galaxies can be called early-types and all of these have strong evidence or hints of young populations. The situation is different with the secondaries in halos.
Of the 48 secondaries brighter than $M_{W1}=-13$, 15 (32\%) are early types.
Two are well known: M\,82 in interaction with M\,81 and NGC\,5195 in interaction with M\,51.
If the Local Group halos of M\,31 and the Milky Way are included, then four gas-rich and 5 gas-poor systems are added, such that 42\% of the unobscured galaxies within 9~Mpc are early types.
The characteristic in common with these 20 early-type galaxies are CMD with well developed RGB but no, or essentially no, hint of stars on the main sequence.  Likewise, they are almost all devoid of HI.
NGC\,5102 and NGC\,5237 are exceptions that are detected in HI (and perhaps UGC\,7356 although the HI might be associated with the primary NGC\,4258).
Most of these galaxies can be classified as dwarf spheroidals (dSph) although a few might qualify as transitional between spheroidal and irregular: e.g., NGC\,4627, NGC\,5237, and DDO\,82.

These statistics support what is generally known.
Star formation in secondary galaxies is frequently depleted. Various mechanisms have been proposed that lead to the loss of gas necessary for the maintenance of star formation.
An inevitable consequence of a smaller galaxy falling into a more massive halo is strangulation \citep{1984ApJ...281...56S, 2015Natur.521..192P}.
Any gas in outer reservoirs held by a small galaxy in isolation finds itself outside the Roche limit of the galaxy once entering the more massive halo.
The transition truncating star formation can be more abrupt through localized events \citep{2022MNRAS.511.1544F}.
Galaxies can interact with each other in ways that are called harassment \citep{1998ApJ...495..139M, 2015A&A...576A.103B}.
Gas can be removed by ram-pressure stripping \citep{1972ApJ...176....1G}. This latter mechanism is particularly viable in massive clusters but must affect dwarfs passing through the disks of spirals.
Then there are mechanisms internal to galaxies that can cause the loss of gas such as energetic events in the nucleus or supernovae.
However these phenomena occur equally in isolated galaxies and halo secondaries so do not explain differences between systems in the two environments.

The free-fall timescale in a halo is roughly a Gyr.
Small galaxies get absorbed into primary galaxies as testified in our Milky Way by the abundance of stellar streams \citep{2022ApJ...928...30L}.
New galaxies enter halos as previous arrivals spend time in the inhospitable halo environment and likely ultimately disappear.
If the transitioning fraction of about 1/3 of secondaries found in our sample is representative, there is the implication that the absorption of small galaxies into the primary occurs within only a couple of free-fall timescales.

%%%%%%%%%%%%%%%%%%%%
\section{Young Galaxies?}
\label{sec:young}
%%%%%%%%%%%%%%%%%%%%

A ubiquitous feature of the CMD of galaxies is a prominent RGB.  A good example is seen in Fig.~\ref{fig:ESO6-1}. 
Almost all galaxies imaged with sufficient sensitivity to reach fainter than $M_I = -4$ with HST resolve an abundance of stars on the red giant branch of the stellar evolution of low mass stars, consistent with the proposition that star formation typically began in protogalaxies before the epoch of reionization at $z \sim 6$.
In the cases of the very nearest ones that have been imaged to depths reaching the main sequence turnoff, there is evidence that a substantial fraction of their stars (40\%-90\%) were formed 13 Gyr ago \citep{2014ApJ...789..148W, 2017ApJ...837..102S}, consistent with the theory that baryons could collapse and form stars in halos in the early cold universe but were dispersed in low mass halos with reionization \citep{2002ApJ...572L..23S}.

%Fig 24
\begin{figure}
\centering
\includegraphics[width=.89\linewidth]{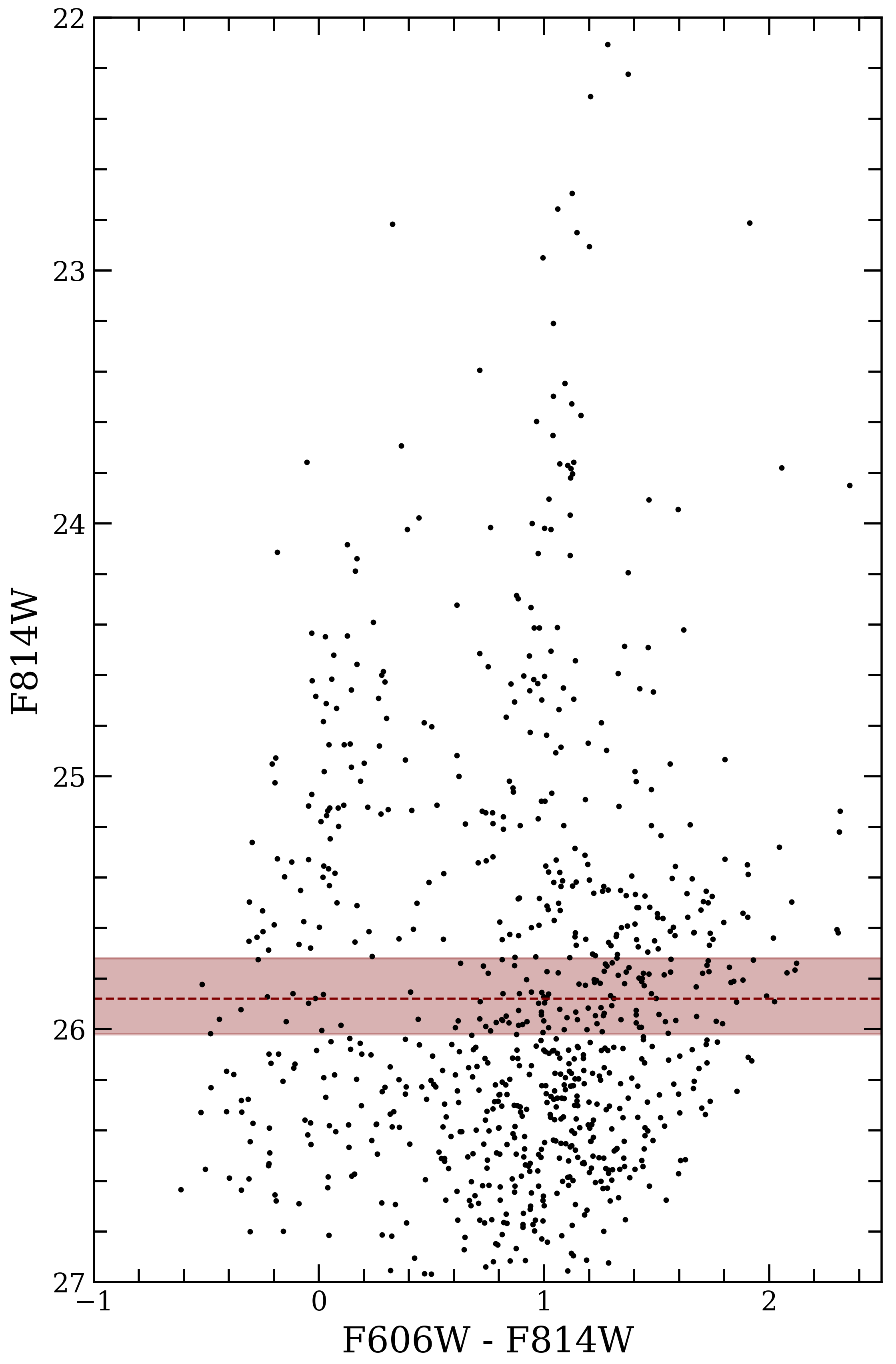}
\includegraphics[width=.89\linewidth]{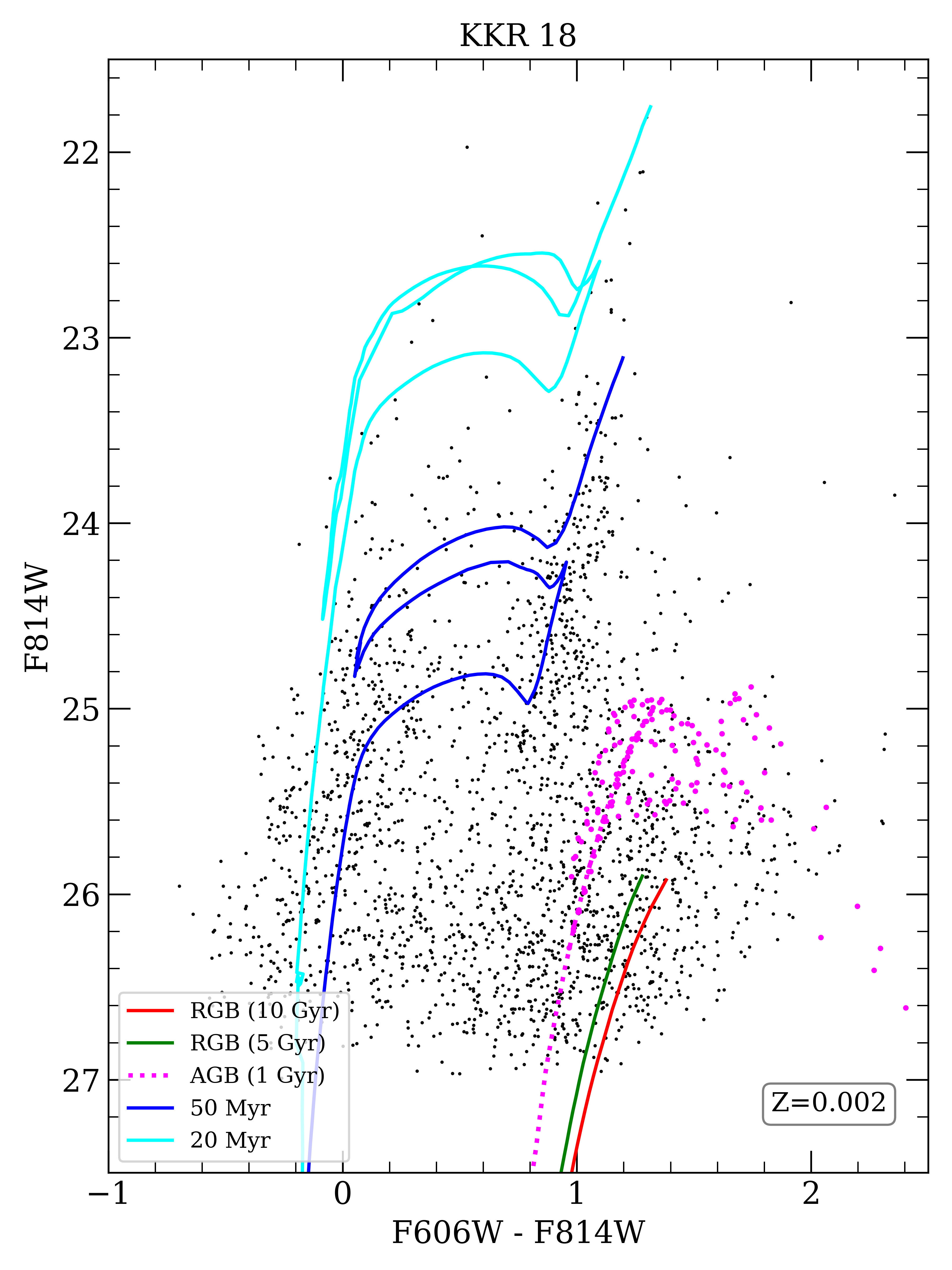}
\caption{
Top: CMD for outer field of KKR\,18
with TRGB maximum likelihood fit at $F814W \simeq 25.9$.
Bottom: isochrones imposed on full field CMD of KKR\,18.
}
\label{fig:z106-29}
\end{figure}

Although a prominent RGB is the norm there may be exceptions.  Two examples in the literature lie beyond our 10~Mpc boundary: I~Zw~18 \citep{1972ApJ...173...25S} and Leoncino \citep{2020ApJ...891..181M}.
Two interesting cases lie within 10~Mpc. Peekaboo Galaxy, so-named because it hides in the glare of a 10th magnitude star, lies at a distance of 7~Mpc \citep{2023MNRAS.518.5893K}.
Very approximately, its absolute magnitude is $M_B\sim-10$, much fainter than our completeness cutoff.
The other nearby galaxy, KKR\,18, has been observed in HST GO-18070 \citep{2025hst..prop18070T} and has a TRGB distance determination of $d=9.7\pm0.8$~Mpc.

CMD for KKR\,18 are seen in Fig.~\ref{fig:z106-29}.  For the CMD seen in the top panel, the center of the galaxy dominated by young stars is excluded in order to maximize the RGB contrast. The RGB feature is discerned but it is weak and the TRGB fit is tenuous.  
The CMD for the entire field is shown in the bottom panel with selected isochrones at Z=0.002 set at the distance 9.7~Mpc.
The great majority of the features in the CMD can be attributed to stars with ages less than $\sim 1$~Gyr. The RGB can be attributed to stars with ages of up to $\sim 5$~Gyr. Older stars cannot be excluded but they could not be abundant given the weakness of the RGB feature.
The possibility within the $\L$CDM paradigm of late forming stars in dwarfs  has been discussed by \citet{2021ApJ...921L...9B}.

KKR\,18 is a dwarf galaxy with $M_B\simeq-14$ lying in isolation in the Local Void. Its position is located in the two panels of Fig.~\ref{fig:xy_peg}.  A full discussion of this interesting galaxy is the topic of a companion paper \citep{2026arXiv260904437T}.  There is the take-away that likely the overwhelming fraction of stars in KKR\,18 have been born within the last few Gyr.  This property is singular among 240 galaxies with HST CMDs at distances between one and ten Mpc and brighter than $M_B = -13$.

%%%%%%%%%%%%%%%%%%%%
\section{Summary}
\label{sec:summary}
%%%%%%%%%%%%%%%%%%%%

It is a reasonable assumption that the place where we live in the Universe is not unusual.
Yet, if the survey region were extended by only 1~Mpc the census would include five additional substantial groups: about  NGC\,6744; the complex involving NGC\,891, NGC\,925, and NGC\,1023; the M\,104 (Sombrero) region that is complicated by its proximity to the Virgo Cluster Southern Extension infall cloud; and the two big groups in Leo around NGC\,3379 (M\,105) and NGC\,3627.
The NGC\,3379 Group warrants particular attention as the nearest knot of early-type galaxies, an environment that is not explored within 10~Mpc.

Our survey region lies in close proximity to the Virgo Cluster but only mildly feels its influence.
In the standard $\Lambda$CDM cosmology, though, ultimately galaxies currently as far as 12~Mpc from the Virgo core will fall into the cluster, hence, including the NGC\,4258, M\,51, and NGC\,5055 groups and probably even M\,101 \citep{2017ApJ...850..207S}.
We will avoid that fate but are destined to merge with M\,31 \citep{2012ApJ...753....8V}.
The halos of Centaurus\,A and M\,83 will surely merge and there will likely be a merger between the halos of M\,81 and IC\,342.

The concentration of stellar and group mass in relatively few galaxies is an outstanding characteristic of the 10~Mpc volume.
%(an extension of the ``council of giants'' phenomenon \citep{2014MNRAS.440..405M}).
The feature is set apart by the apparent deficiency of galaxies in an interval faintward of the dominant systems, evident as gaps in both the inferred halo masses at $10^{12}~\Msun$ from  2MASS $K_s$ fluxes shown in Figure~\ref{fig:LandMfn} and the stellar mass estimates at $10^{10}~\Msun$ from WISE W1 fluxes shown in Figures~\ref{fig:MstarMW1} to \ref{fig:LFsecondaries}.
The statistical evidence is poor in this small volume but there is a similar gap feature at $10^{12}~\Msun$ in the group mass function found with a sample extending to 3,500~\kms\ by \citet{2017ApJ...843...16K}, their Figure~17.
Comparisons with simulations suggest a tension with $\Lambda$CDM cosmology \citep{2020MNRAS.494.2600N} but further tests are warranted. 

Almost all galaxies in our volume-complete sample have dominant RGB consistent with the ubiquity of the prominence of early star formation.  The galaxy KKR\,18 is an exception, $0.4\%$ of the complete sample, that may not have a significant old population.  The Peekaboo dwarf \citep{2023MNRAS.518.5893K}, a decade less luminous than our sample faint limit may be a second such galaxy within 10~Mpc. 

The single most outstanding kinematic feature of the 10~Mpc region is the $\sim230$~\kms\ motion toward the south supergalactic pole of all galaxies on and north of the supergalactic equator \citep{2017ApJ...835...78R}.  This motion, with the Virgo Cluster as a participant, is roughly orthogonal to the motion toward the apex of the CMB dipole.
Galaxies south of the supergalactic equator are shielded from this flow pattern.

This motion is attributed to Local Void expansion.
The phenomenon extends well beyond the 10~Mpc region.
It was found from Numerical Action orbit modeling that there is a downward flow toward the supergalactic equator over the full domain of the Laniakea structure \citep{2022ApJ...927..168S} extending over 8,000~\kms.
Overlaying this entire domain is the Local Void blending into the Hercules Void \citep{2019ApJ...880...24T}.

Our story goes full circle. Large scales inform small scales inform large scales.
There is clarity in the local information that brings comprehension to statistical information with large samples.

\bigskip
\noindent
Aknowledgement.
Support of this study has come from the Space Telescope Science Institute associated with Hubble Space Telescope awards GO-9162, 9771, 10210, 10235, 10905, 12546, 12877, 12878, 13442, 14636, 15922, and 18070.
Support for the WISE observations were provided by NASA grant No. 88NSSC18K0424.
IDK, DIM and LNM are supported by the Russian Science Foundation grant \textnumero~24--12--00277.

%%%%%%%%%%%%%%%%%%%%
\appendix

\section{Galaxy Groups in the Local Volume}
\label{sec:appendixA}
%%%%%%%%%%%%%%%%%%%%

%Fig 25
\begin{figure}
\centering
\includegraphics[width=0.43\textwidth]{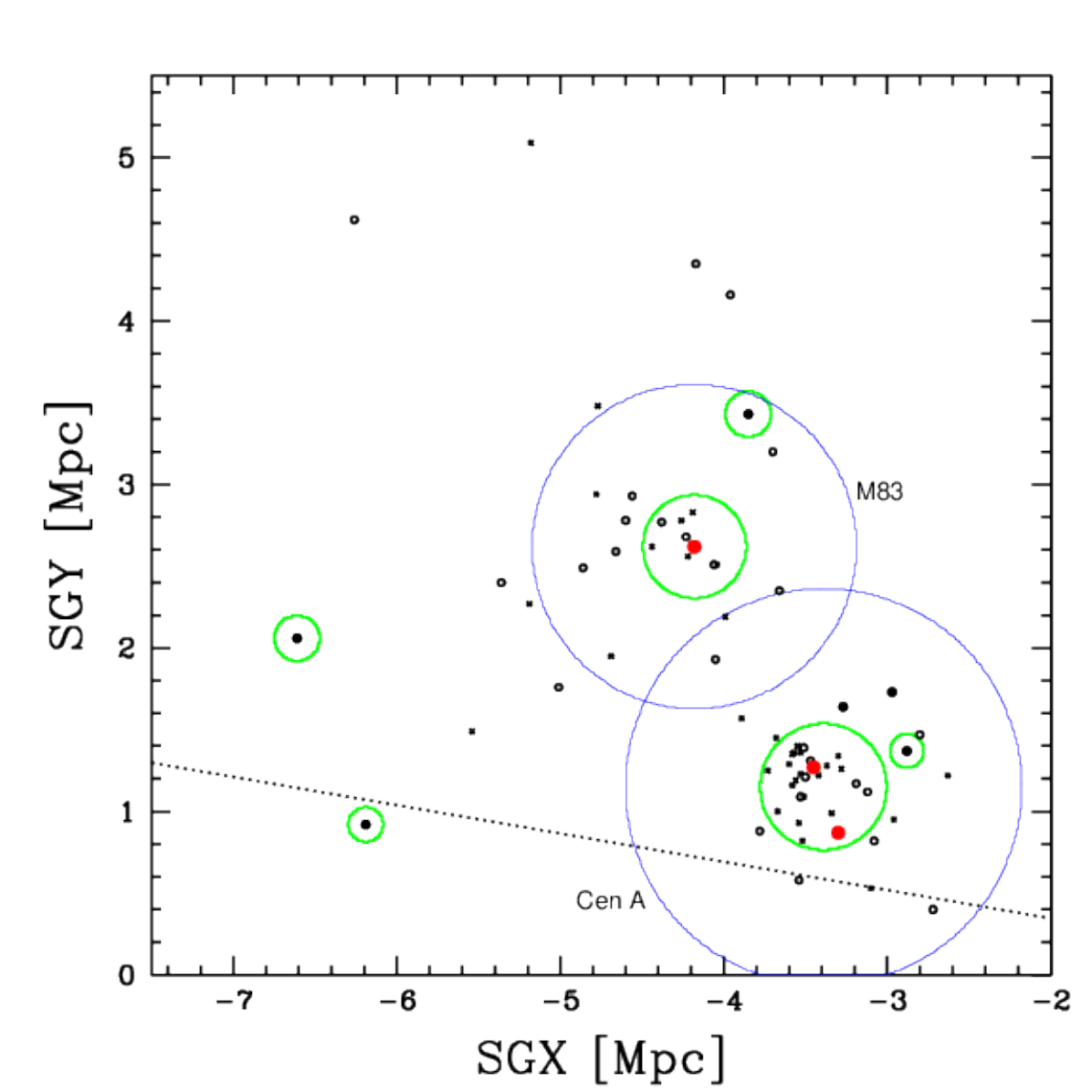}
\includegraphics[width=0.43\textwidth]{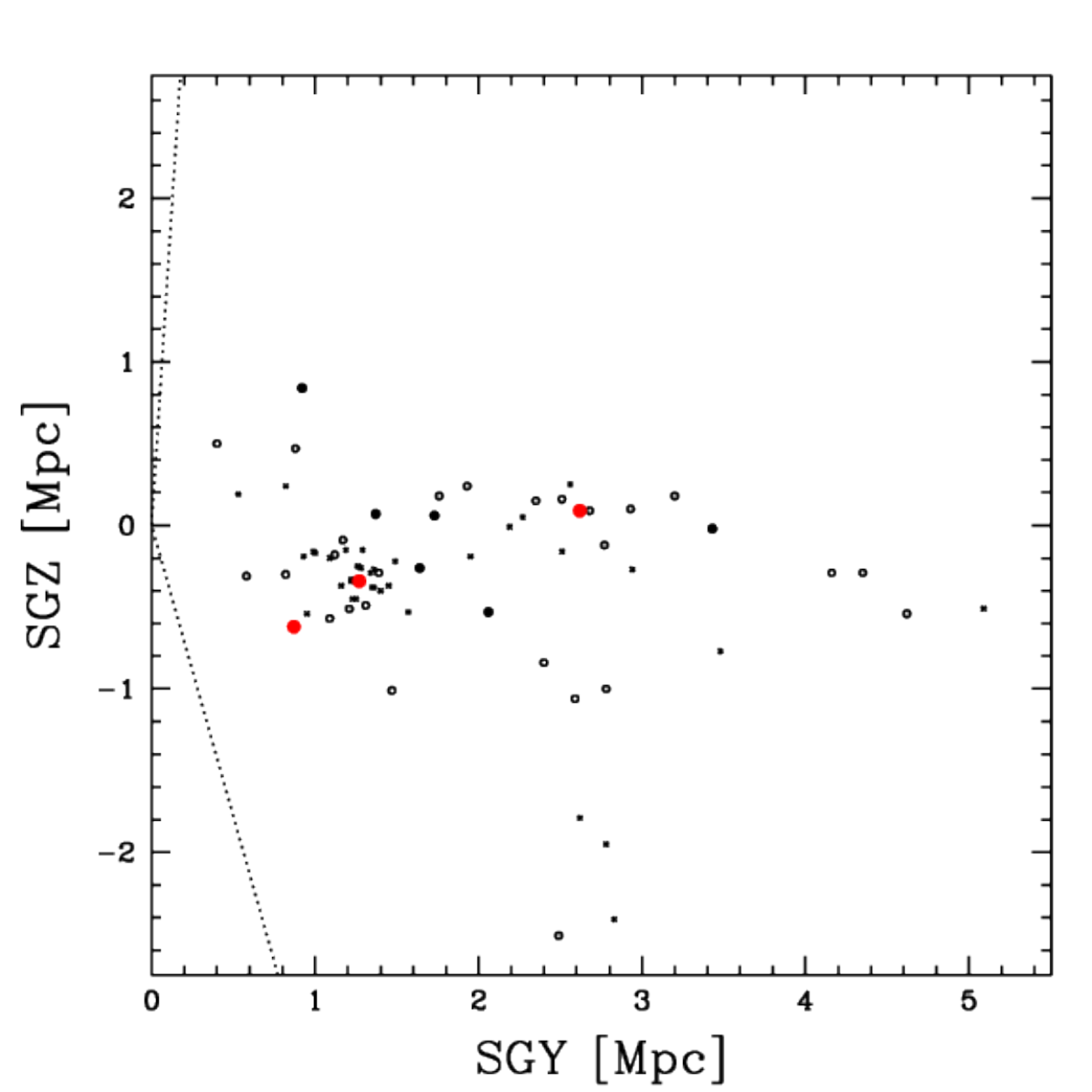}
\caption{
Galaxies in a box 5.5~Mpc on a side in the vicinity of the Centaurus\,A Group.  The format is the same as described in the legend of Figure~\ref{fig:xy_loc}.    Dotted lines indicate the $|b|=10^{\circ}$ proximity of the Galactic equator.
}
\label{fig:xy_cen}
\end{figure}

Centaurus\,A is the dominant galaxy in the most substantial nearby group.  As seen in Figure~\ref{fig:xy_cen}, there is a halo around M\,83 that is distinct.  The infall domains around these two halos strongly overlap.  The numerical action orbit reconstruction by \citet{2022ApJ...927..168S} demonstrate that these two halos are destined to merge.
It is seen in the $SGY$--$SGZ$ projection that almost all the galaxies in this box are within $\pm1$~Mpc of the supergalactic equatorial plane.

%Fig 26
\begin{figure}
\centering
\includegraphics[width=0.43\textwidth]{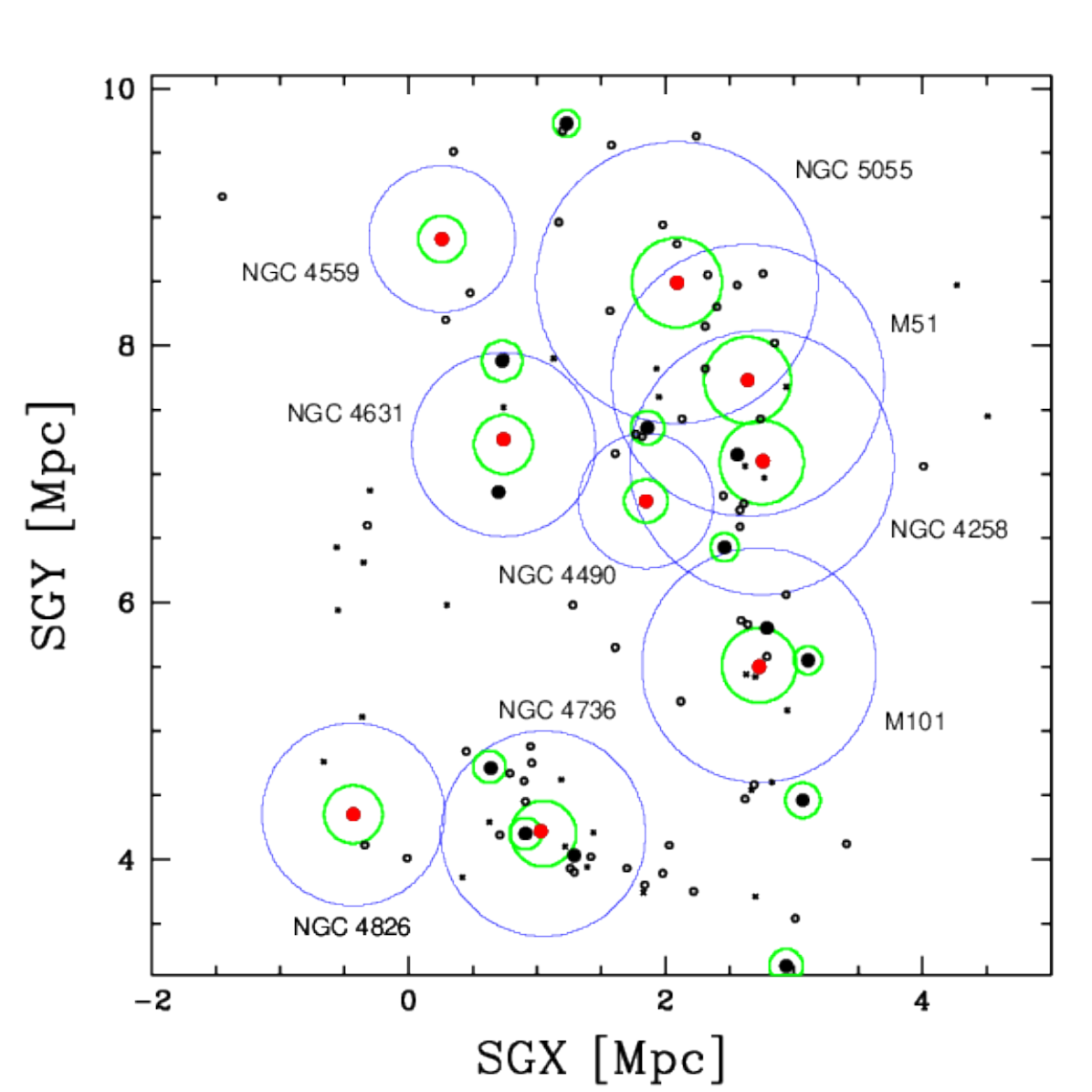}
\includegraphics[width=0.43\textwidth]{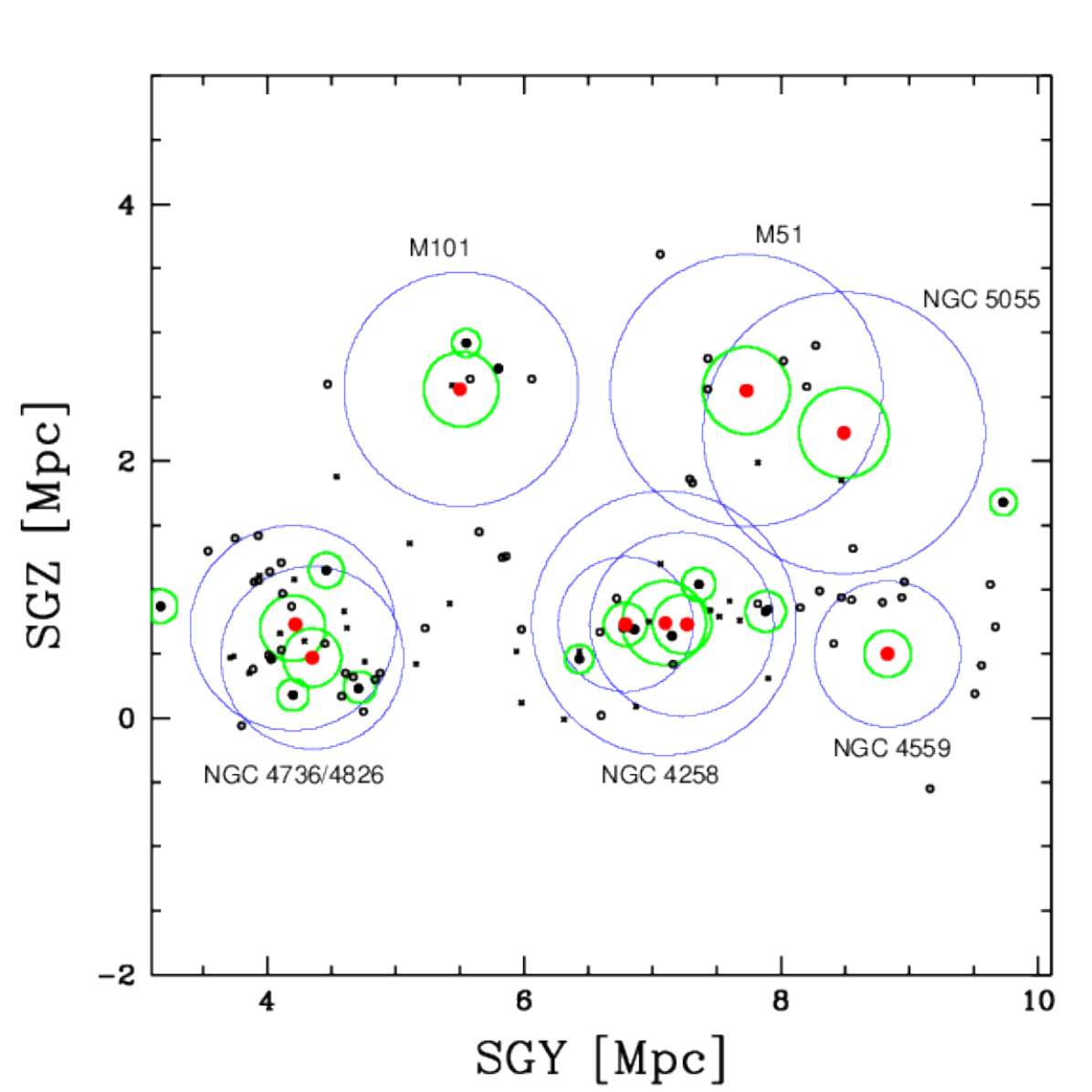}
\caption{
A box 7~Mpc on a side showing the distribution of galaxies in the Canes Venatici region of the Local Sheet. The format is the same as described in the legend of Figure~\ref{fig:xy_cvn}.  There is projection confusion in the $SGX$--$SGY$ view of the upper panel that is resolved with the $SGY$--$SGZ$ view of the lower panel.
}
\label{fig:xy_cvn}
\end{figure}

The following Figure~\ref{fig:xy_cvn} illustrates the complex of groups in the densest part of the Local Sheet.  It is seen in the lower panel that a filament runs between M\,101 and NGC\,5055 about 2~Mpc \textit{above} the galaxies in the Local Sheet.  This minor filament is seen in the upper panel to lie at $SGX\sim2$--2.5~Mpc, over top of the NGC\,4258 and NGC\,4490 groups that lie within the Local Sheet.  This minor filament is the singular Local Sheet fringe feature at positive $SGZ$ seen in Figure~\ref{fig:filam}.

In this box, there is substantial overlap of the infall regions with two of the major halos: between M\,51 and NGC\,5055 and between NGC\,4258 and NGC\,4490.  Several lesser galaxies are entrained within the collapsing region around NGC\,4258.

%Fig 27
\begin{figure}
\centering
\includegraphics[width=0.43\textwidth]{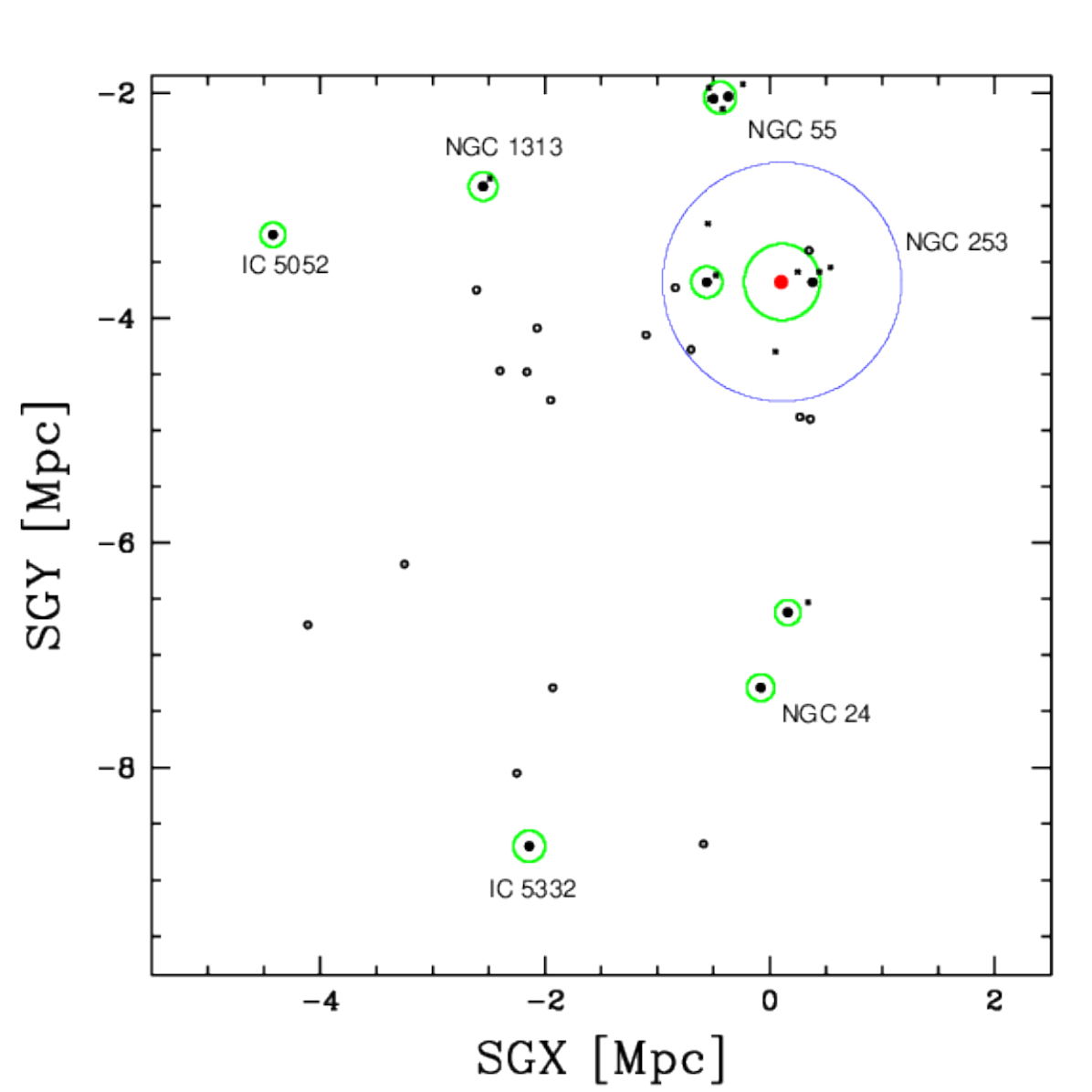}
\includegraphics[width=0.43\textwidth]{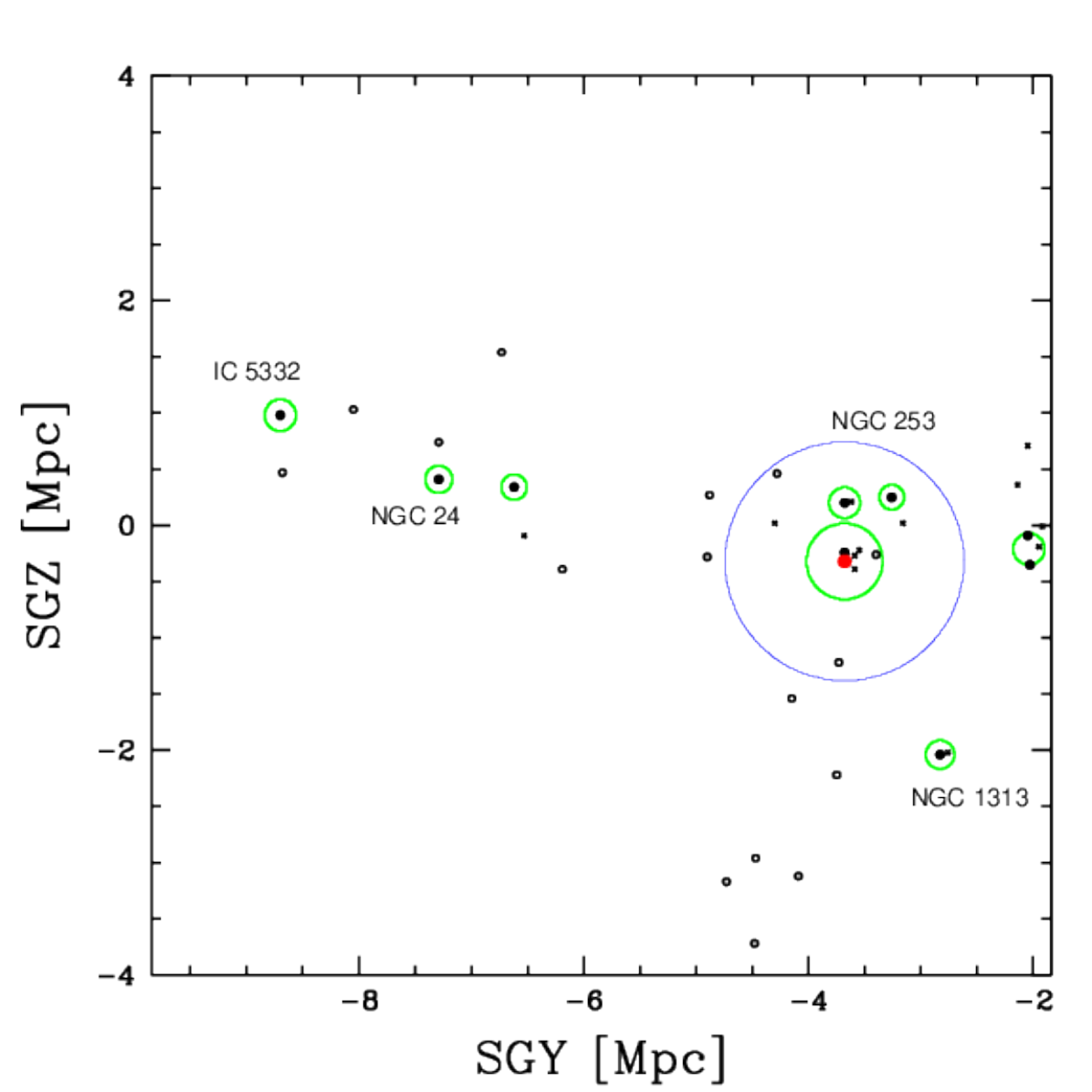}
\caption{
This box of 8~Mpc on a side contains galaxies south of the Galactic plane ($SGY<0$) and at negative $SGX$. Same format as the previous plots in the Appendix.
}
\label{fig:xy_scl}
\end{figure}

%Fig 28
\begin{figure}
\centering
\includegraphics[width=0.43\textwidth]{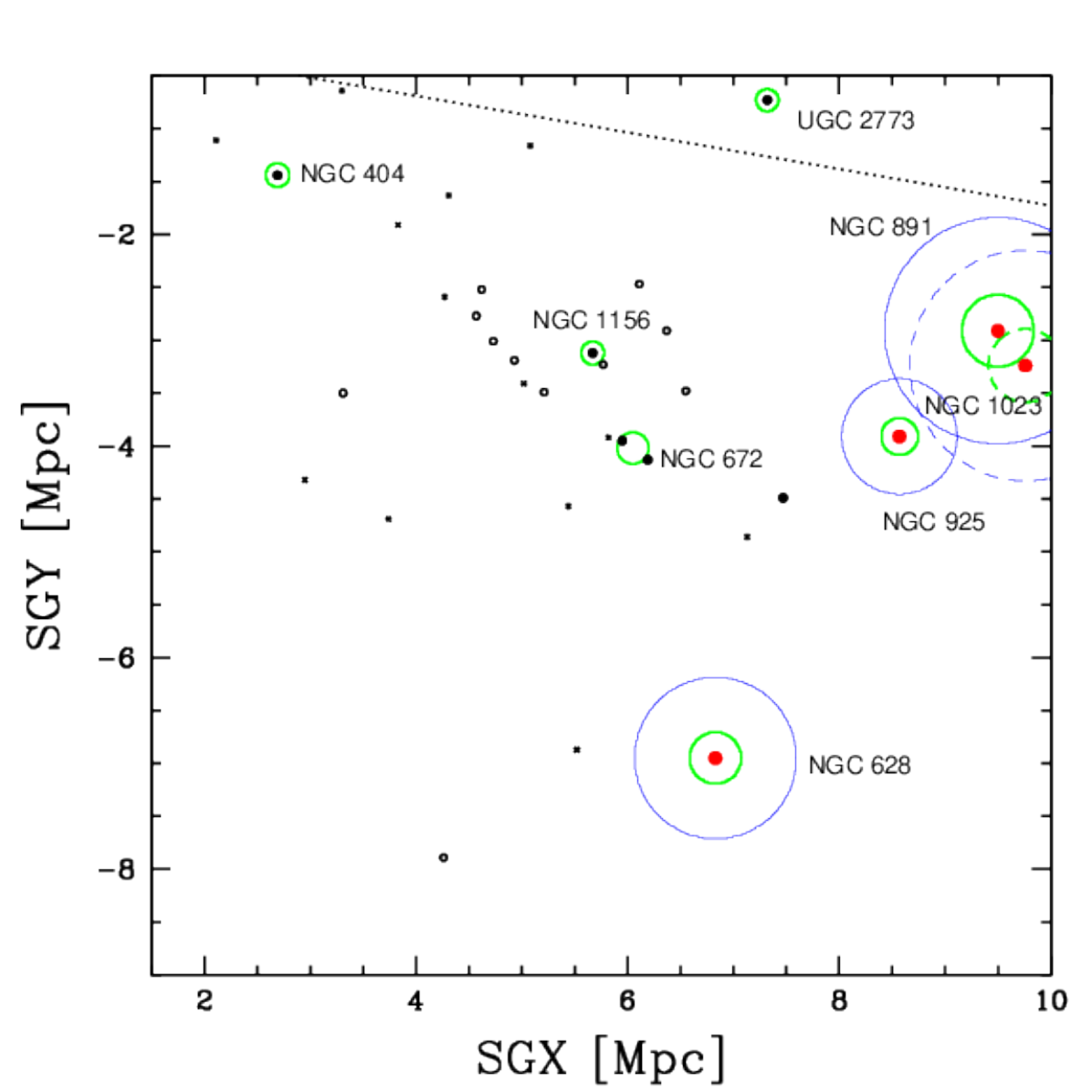}
\includegraphics[width=0.43\textwidth]{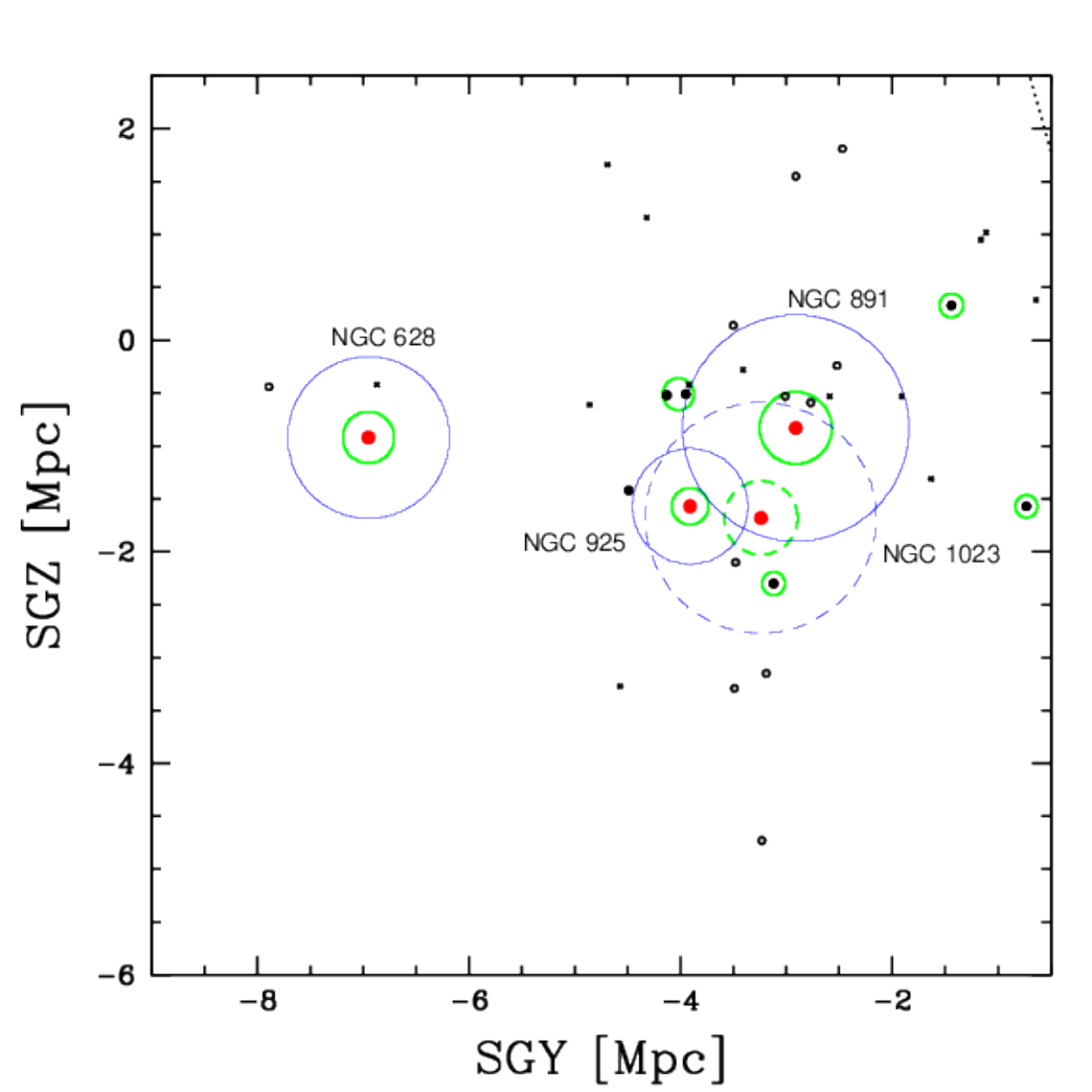}
\caption{
This box of 8.5 Mpc on a side lies south of the Galactic plane at positive $SGX$.  The centroid of the NGC\,1023 Group lies slightly beyond 10~Mpc so the second and first turnaround radii are shown as dashed circles. Otherwise, same format as the previous plots.
}
\label{fig:xy_1023}
\end{figure}

Figure~\ref{fig:xy_scl} and Figure~\ref{fig:xy_1023} display the distribution of galaxies in the much lower density sky south of the Milky Way.  The concentration to the supergalactic equatorial plane is still present but the population is sparse.  There are over seven times as many halos more luminous than log$L_K=10$ per unit of volume in the northern region shown in Figure~\ref{fig:xy_cvn} than in these regions at $SGY < 0$.

At the edge of our 10~Mpc sample, the NGC\,891, NGC\,925, and NGC\,1023 infall zones overlap although their cores within their respective second turnaround radii are discrete.  These three entities are likely destined to merge.  Galaxies in this complex scatter across our 10~Mpc boundary leading to incompletion in our inventory.

Of minor note, it is seen in Figure~\ref{fig:xy_scl} that a straggle of dwarf galaxies, with NGC\,1313 the largest, lie below $SGZ=-2$~Mpc.  These objects are the negative $SGY$ part of the wispy fringe feature at positive $SGX$ seen in the second panel of Figure~\ref{fig:filam}.  Elements north of the Milky Way are seen at negative $SGZ$ in Figure~\ref{fig:xy_cen} although the only substantial galaxy in this structure, NGC\,3621, lies slightly to more negative $SGZ$ than the boundary of this box.
The remaining fringe objects of Figure~\ref{fig:filam} show up in Figure~\ref{fig:xy_1023} lying at negative values of $SGZ$ below NGC\,1156.

%Fig 29
\begin{figure}
\centering
\includegraphics[width=0.43\textwidth]{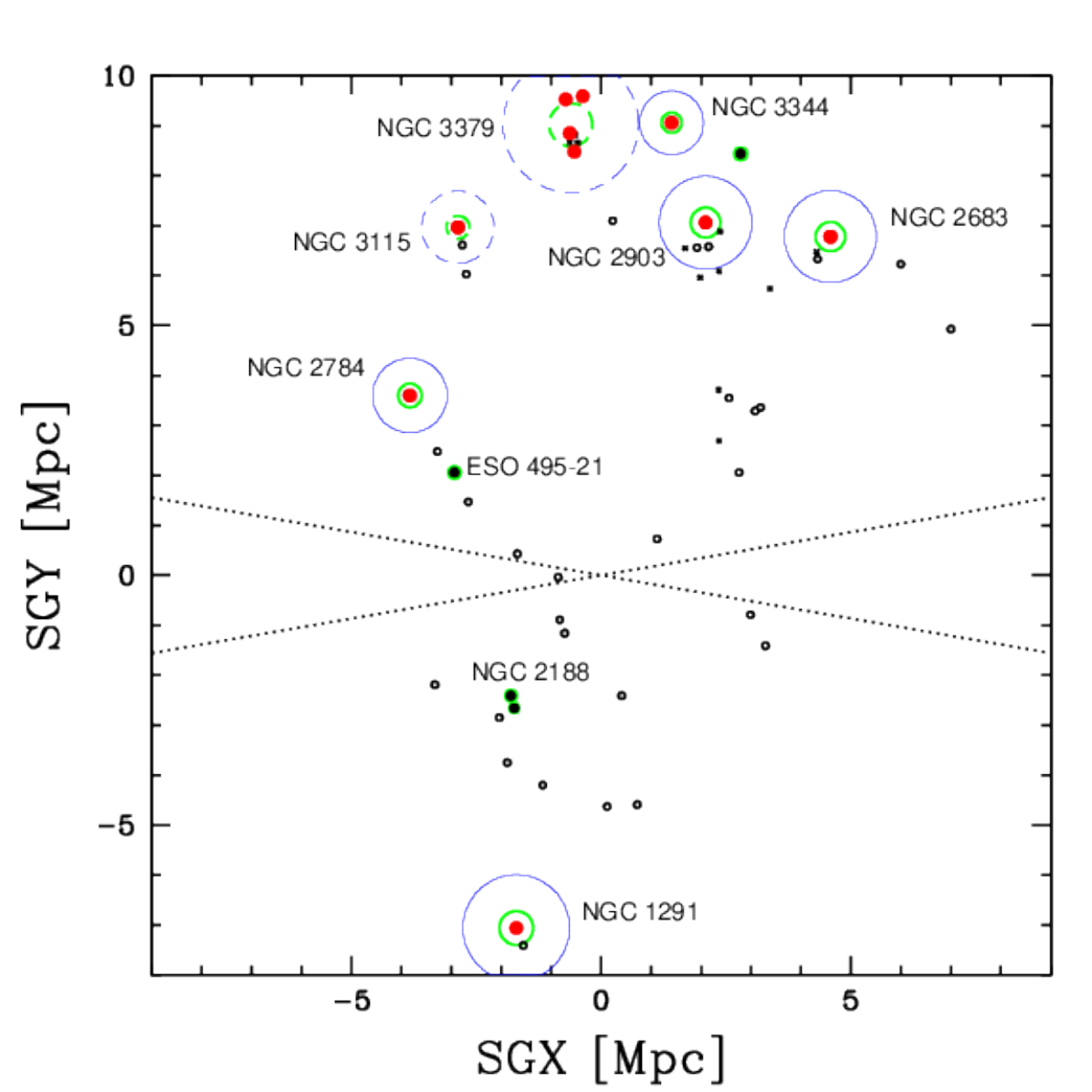}
\caption{
The Leo Spur-Antlia Cloud-Dorado Cloud ("Leo Spur") filaments lying at negative $SGZ$ with respect to the Local Sheet.  In this case, the $SGX$--$SGY$ projection is restricted to $-5>SGZ>-10$~Mpc across the entire scene, elevated to $-2>SGZ<-10$~Mpc if $SGY>5$~Mpc to accommodate the rise in $SGZ$ values in the region of the Leo groups as seen in the third lower panel of Figure~\ref{fig:filam}. The centroids of the NGC\,3115 and NGC\,3379 (Leo) groups lie slightly beyond 10~Mpc so the turnaround domains about them are shown by dashed circles.
}
\label{fig:xy_spur}
\end{figure}

%Fig 30
\begin{figure}
\centering
\includegraphics[width=0.43\textwidth]{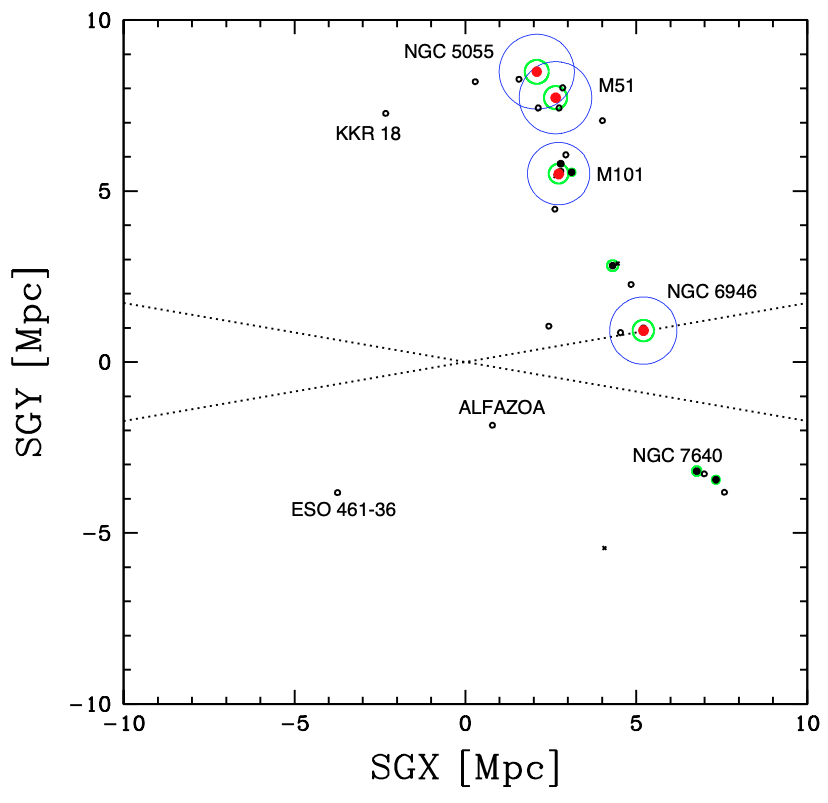}
\includegraphics[width=0.43\textwidth]{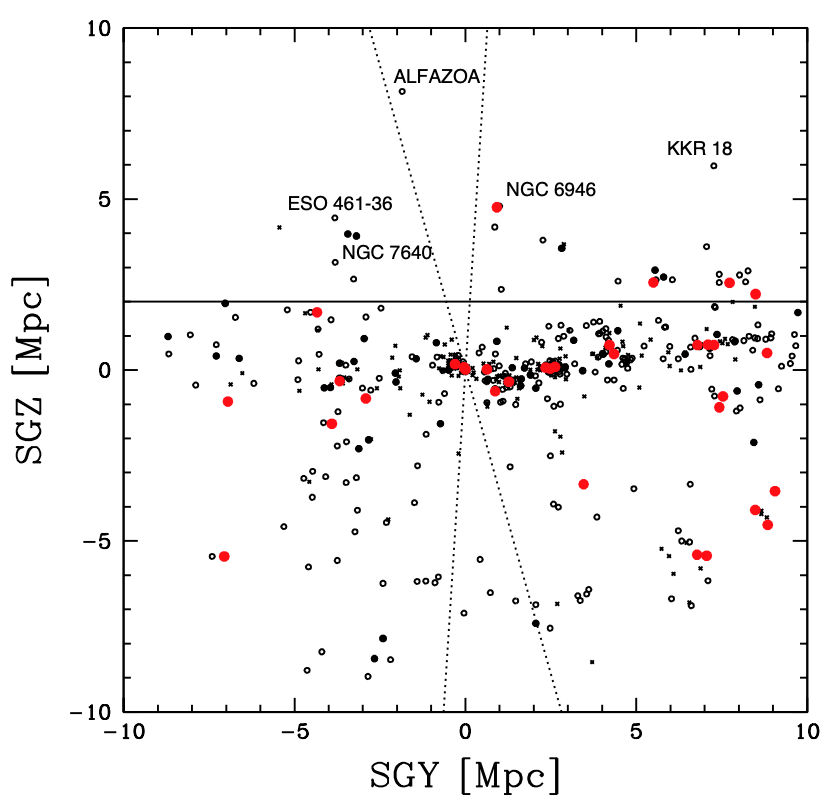}
\caption{
The left panel shows a projection of all galaxies within $2<SGZ<10$~Mpc, a volume dominated by the Local Void but permeated by the Pegasus Spur.  The right panel is an orthogonal projection with a solid line at $SGZ=2$~Mpc delineating the lower boundary of the scene in the left panel.
}
\label{fig:xy_peg}
\end{figure}

In Figure~\ref{fig:xy_spur} attention is turned to the volume below the Local Sheet and its minor fringes.  The substantial NGC\,1291 Group, as the Fornax Cluster is approached, is at the edge of our study and not properly sampled.  Otherwise the major entities are near the Virgo Cluster turnaround surface.  The Leo Group with the dominant member NGC\,3379 and the lesser group around NGC\,3115 lie at and across our 10~Mpc boundary and are incompletely sampled.  In the Nearby Galaxies Atlas, the main strand of the Leo Spur runs out of our study zone toward positive $SGX$ at $SGY\sim8$~Mpc (see footnote 1 in \S\ref{sec:10mpc}).  It is only a tenuous tendril that passes under the Milky Way to connect with the somewhat more substantial strand running through NGC\,2188 (Dorado Cloud) and NGC\,2784 (Antlia Cloud).  The galaxies in these features crossing behind the Milky Way are important for their revelatory kinematics (see Figure~\ref{fig:yzflow}) but almost all are inconsequential dwarfs,

One of the most outstanding features of the 10~Mpc volume is the Local Void, given attention in Figure~\ref{fig:xy_peg}.  The region being considered lies above the horizontal line at $SGZ=2$~Mpc in the right panel.  The Pegasus Spur, as identified in the Nearby Galaxies Atlas, continues toward negative $SGY$ and positive $SGX$ beyond our region of interest and we see here only a piece of the structure.  It would be quite reasonable to associate the filament fragment running from M\,101 to NGC\,5055 to this same feature (rather than viewing this fragment as a fringe of the Local Sheet).  There is arbitrariness in aspects of delineating structures.

The Pegasus Spur intrudes on the Local Void where, otherwise in our volume, only three dwarf galaxies are known: ESO\,461--36, KKR\,18 and ALFAZOA\,J1952+1428 (the latter shortened to ALFAZOA in the figure).

%%%%%%%%%%%%%%%%%%%%
\section{Group members}
\label{sec:appendixB}
%%%%%%%%%%%%%%%%%%%%

Information on the samples at 1 to 9 Mpc, $|b|>10^{\circ}$, and $M_{W1}<-13$ is gathered in two tables.
Table~\ref{tab:galaxy_catalog_10mpc} pertains to 221 individual galaxies and Table~\ref{tab:group_galaxy_properties} pertains to 146 halos, potentially containing multiple galaxies.
Galaxies are given names from HyperLEDA: Principal Galaxy Catalog (PGC) identifications \citep{1996A&A...311...12P}.  The dominant galaxy in a group carries the designation 1PGC.

In Table~\ref{tab:galaxy_catalog_10mpc}, in the column labeled $PS$, a flag 1 identifies the dominant (primary) galaxy in a halo, while flag 2 identifies secondaries in the halo.
The entries in the column labeled $Ty$ are usually the type given in RC3, the Third Reference Catalog  \citep{1991trcb.book.....D}, but replacements were made in cases discussed in Section~\ref{sec:environment} that have types poorly described in RC3.
Here, type $-8$ is used for transition cases between dwarf irregular and dwarf spheroidal.
The WISE W1 magnitudes in Table~\ref{tab:galaxy_catalog_10mpc} are derived by \citet{2026arXiv260605780T} and translated to absolute values from the distances given in the column labeled $d$ (from precision TRGB measurements if the flag in column $t$ is 1).
The absolute $B$ (blue) magnitudes are from \citet{2009ApJ...703..517D} assuming the same distances.
The stellar masses follow from W1 luminosities as given in Eq.~\ref{eq:stellarmass}.

In Table~\ref{tab:group_galaxy_properties}, information is appropriately added or summed over members of a halo, with absolute values determined from the averaged group distances.
The group masses (in units of $10^{12}~\Msun$) and $R_{2t}$, second turn around radii approximating the Virial radius \citep{2015AJ....149...54T, 2017ApJ...843...16K} account for dark matter as discussed in Section~\ref{sec:halofn}.
The stellar mass values given in logarithmic units are summed over all members of the halo but, as discussed in Section~\ref{sec:pops}, inevitably almost all the stellar mass is in the dominant galaxy.

%%%%%%%%%%%%%%%%%%%%%%%%%%%%%%%%%%%%%%%%%%%%%%
% 10 Mpc tables ...
%%%%%%%%%%%%%%%%%%%%%%%%%%%%%%%%%%%%%%%%%%%%%%

\clearpage
\begin{turnpage}

\begin{table*}
\centering
\caption{Galaxy properties for the local-volume sample ($1<d<9$~Mpc, $|b|>10^\circ$, $M_{W1}<-13$~mag).}
\label{tab:galaxy_catalog_10mpc}

\begingroup
\scriptsize
\setlength{\tabcolsep}{3pt}
\renewcommand{\arraystretch}{0.95}

%==================== PART A ====================
\begin{tabular*}{\textwidth}{@{\extracolsep{\fill}} c c l r r r r c r r r r c}
\hline\hline
PGC & 1PGC & RA/Dec (J2000) & $l$ & $b$ & sgl & sgb & PS & Ty & $V_h$ & $V_{ls}$ & $d$ & $t$ \\
~   & ~    & (hhmmss.s$\pm$ddmmss.s) & (deg) & (deg) & (deg) & (deg) & ~ & ~ &
(km s$^{-1}$) & (km s$^{-1}$) & (Mpc) & ~ \\
(1) & (2)  & (3) & (4) & (5) & (6) & (7) & (8) & (9) & (10) & (11) & (12) & (13) \\
\hline
388  & 388  & 000531.9-280554.0 & 24.8340  & -79.7884 & 266.0981 & 3.1192  & 1 & 5.4  & 732 & 753 & 8.51 & 1 \\
591  & 591  & 000744.1+405233.2 & 114.0867 & -21.2440 & 334.9193 & 19.0438 & 1 & 9.0  & 296 & 588 & 7.97 & 1 \\
701  & 701  & 000956.3-245749.8 & 43.6860  & -80.4343 & 269.3872 & 3.2262  & 1 & 5.1  & 554 & 595 & 7.13 & 1 \\
930  & 930  & 001403.9-231055.5 & 55.9027  & -80.6717 & 271.3799 & 2.9060  & 1 & 7.7  & 466 & 515 & 6.48 & 1 \\
1014 & 1014 & 001453.6-391147.7 & 332.8820 & -75.7309 & 256.2492 & -2.3594 & 1 & 8.8  & 129 & 95  & 2.06 & 1 \\
1034 & 1034 & 001525.1-212640.4 & 65.7098  & -80.0220 & 273.1300 & 3.1559  & 1 & -2.9 & 365 & 432 & 4.78 & 1 \\
1777 & 1777 & 002904.1+432550.9 & 118.6330 & -19.2541 & 338.0042 & 15.3737 & 1 & 10.0 & 188 & 465 & 6.67 & 1 \\
2544 & 2544 & 004230.1+403433.0 & 121.0985 & -22.2640 & 335.4861 & 12.4752 & 1 & 10.0 & 242 & 504 & 7.01 & 1 \\
2578 & 2578 & 004303.6-221451.2 & 101.1867 & -84.7619 & 274.2325 & -3.2132 & 1 & 9.9  & 360 & 400 & 4.81 & 1 \\
2758 & 2789 & 004708.6-204537.6 & 113.9479 & -83.5572 & 275.9225 & -3.7299 & 2 & 6.9  & 156 & 198 & 3.6 & 1 \\
\nodata \\
\hline
\end{tabular*}

\vspace{6pt}

%==================== Divider ====================
\begin{tabular*}{\textwidth}{@{\extracolsep{\fill}} c}
\end{tabular*}

\vspace{4pt}

%==================== PART B ====================
\begin{tabular*}{\textwidth}{@{\extracolsep{\fill}} c r r r r r r r r l}
\hline\hline
PGC & sgx & sgy & sgz & $A_g$ & $W1$ & $M_B$ & $M_{W1}$ & $M_\star$ & Name \\
~   & (Mpc) & (Mpc) & (Mpc) & (mag) & (mag) & (mag) & (mag) & $\log M_\odot$ & ~ \\
(1) & (14) & (46115) & (16) & (17) & (18) & (19) & (20) & (21) & (22) \\
\hline
388  & -0.58 & -8.48 & 0.46  & 0.07 & 15.698 & -14.59 & -13.95 & 7.65 & ESO409-015 \\
591  &  6.83 & -3.19 & 2.60  & 0.35 & 15.439 & -13.90 & -14.08 & 7.70 & UGC00064 \\
701  & -0.08 & -7.12 & 0.40  & 0.08 & 11.303 & -17.27 & -17.97 & 9.25 & NGC0024 \\
930  &  0.16 & -6.47 & 0.33  & 0.09 & 10.765 & -17.75 & -18.29 & 9.38 & NGC0045 \\
1014 & -0.49 & -2.00 & -0.09 & 0.06 &  8.046 & -18.24 & -18.52 & 9.47 & NGC0055 \\
1034 &  0.26 & -4.77 & 0.26  & 0.09 & 12.611 & -15.36 & -15.79 & 8.38 & NGC0059 \\
1777 &  5.97 & -2.41 & 1.77  & 0.33 & 15.277 & -13.15 & -13.86 & 7.61 & UGC00288 \\
2544 &  6.22 & -2.84 & 1.51  & 0.38 & 15.832 & -13.01 & -13.41 & 7.43 & Andromeda-4 \\
2578 &  0.35 & -4.79 & -0.27 & 0.07 & 14.336 & -14.01 & -14.07 & 7.69 & IC1574 \\
2758 &  0.37 & -3.60 & -0.23 & 0.08 &  9.031 & -18.21 & -18.77 & 9.57 & NGC0247 \\
\nodata \\
\hline
\end{tabular*}

\vspace{4pt}
\noindent{\scriptsize\textit{Note.} The complete version of this catalog is available at \url{https://edd.ifa.hawaii.edu}.}

\endgroup
\end{table*}

\end{turnpage}
\clearpage

%%%%%%%%%%%%%%%%%%%%%%%%%%%%%%%%%%%%%%%%%%%%%%
% Group / galaxy summary table (single table)
%%%%%%%%%%%%%%%%%%%%%%%%%%%%%%%%%%%%%%%%%%%%%%

\clearpage
\begin{table*}
\centering
\caption{Halo (group) properties for the local-volume sample ($1<d<9$~Mpc, $|b|>10^\circ$, $M_{W1}<-13$~mag).}
\label{tab:group_galaxy_properties}

\begingroup
\scriptsize
\setlength{\tabcolsep}{3pt}
\renewcommand{\arraystretch}{0.95}

\begin{tabular*}{\textwidth}{@{\extracolsep{\fill}} r r r r r r r r r r l}
\hline\hline
\colhead{1PGC} & \colhead{$V_{ls}$} & \colhead{$d$} & \colhead{sgx} & \colhead{sgy} & \colhead{sgz} &
\colhead{$\log L_K$} & \colhead{$Mass_{12}$} & \colhead{$R_{2t}$} & \colhead{$M_\star$} & \colhead{Name} \\
~ & (km s$^{-1}$) & (Mpc) & (Mpc) & (Mpc) & (Mpc) &
($L_\odot$) & ($10^{12}\,M_\odot$) & (Mpc) & ($\log M_\odot$) & ~ \\
(1) & (2) & (3) & (4) & (5) & (6) & (7) & (8) & (9) & (10) & (11) \\
\hline
46957 & 230 & 3.53 & -3.31 &  1.12 & -0.42 & 10.866 & 5.5182 & 0.462 & 11.06 & NGC5128 \\
47404 & 533 & 8.36 &  2.58 &  7.55 &  2.49 & 10.917 & 3.6716 & 0.404 & 10.94 & NGC5194 \\
28630 & 164 & 3.64 &  2.75 &  2.39 &  0.05 & 10.927 & 6.0935 & 0.477 & 10.91 & NGC3031 \\
46153 & 561 & 8.79 &  2.04 &  8.29 &  2.17 & 10.962 & 4.0941 & 0.418 & 10.80 & NGC5055 \\
2789  & 269 & 3.61 &  0.11 & -3.60 & -0.31 & 10.908 & 3.7229 & 0.404 & 10.76 & NGC0253 \\
65001 & 392 & 6.96 &  5.09 &  0.90 &  4.65 & 10.837 & 2.9616 & 0.375 & 10.70 & NGC6946 \\
39600 & 487 & 7.46 &  2.69 &  6.93 &  0.72 & 10.859 & 3.3736 & 0.392 & 10.70 & NGC4258 \\
48082 & 249 & 4.82 & -4.08 &  2.56 &  0.09 & 10.838 & 3.0309 & 0.378 & 10.68 & NGC5236 \\
50063 & 438 & 6.51 &  2.67 &  5.38 &  2.50 & 10.738 & 2.3535 & 0.348 & 10.57 & NGC5457 \\
13826 & 199 & 3.32 &  3.26 &  0.62 &  0.00 & 10.523 & 1.4273 & 0.294 & 10.54 & IC0342 \\
\nodata \\
\hline
\end{tabular*}
\vspace{4pt}
\noindent{\scriptsize\textit{Note.} The complete version of this catalog is available at \url{https://edd.ifa.hawaii.edu}.}

\endgroup
\end{table*}
\clearpage

\bibliography{paper}

\begin{thebibliography}{}
\expandafter\ifx\csname natexlab\endcsname\relax\def\natexlab#1{#1}\fi
\providecommand{\url}[1]{\href{#1}{#1}}
\providecommand{\dodoi}[1]{doi:~\href{http://doi.org/#1}{\nolinkurl{#1}}}
\providecommand{\doeprint}[1]{\href{http://ascl.net/#1}{\nolinkurl{http://ascl.net/#1}}}
\providecommand{\doarXiv}[1]{\href{https://arxiv.org/abs/#1}{\nolinkurl{https://arxiv.org/abs/#1}}}

\bibitem[{{Adhikari} {et~al.}(2014){Adhikari}, {Dalal}, \&
  {Chamberlain}}]{2014JCAP...11..019A}
{Adhikari}, S., {Dalal}, N., \& {Chamberlain}, R.~T. 2014, \jcap, 2014, 019,
  \dodoi{10.1088/1475-7516/2014/11/019}

\bibitem[{{Anand} {et~al.}(2018){Anand}, {Tully}, {Karachentsev}, {Makarov},
  {Makarova}, {Rizzi}, \& {Shaya}}]{2018ApJ...861L...6A}
{Anand}, G.~S., {Tully}, R.~B., {Karachentsev}, I.~D., {et~al.} 2018, \apjl,
  861, L6, \dodoi{10.3847/2041-8213/aacc2b}

\bibitem[{{Anand} {et~al.}(2019{\natexlab{a}}){Anand}, {Tully}, {Rizzi}, \&
  {Karachentsev}}]{2019ApJ...872L...4A}
{Anand}, G.~S., {Tully}, R.~B., {Rizzi}, L., \& {Karachentsev}, I.~D.
  2019{\natexlab{a}}, \apjl, 872, L4, \dodoi{10.3847/2041-8213/aafee6}

\bibitem[{{Anand} {et~al.}(2022){Anand}, {Tully}, {Rizzi}, {Riess}, \&
  {Yuan}}]{2022ApJ...932...15A}
{Anand}, G.~S., {Tully}, R.~B., {Rizzi}, L., {Riess}, A.~G., \& {Yuan}, W.
  2022, \apj, 932, 15, \dodoi{10.3847/1538-4357/ac68df}

\bibitem[{{Anand} {et~al.}(2019{\natexlab{b}}){Anand}, {Tully}, {Rizzi},
  {Shaya}, \& {Karachentsev}}]{2019ApJ...880...52A}
{Anand}, G.~S., {Tully}, R.~B., {Rizzi}, L., {Shaya}, E.~J., \& {Karachentsev},
  I.~D. 2019{\natexlab{b}}, \apj, 880, 52, \dodoi{10.3847/1538-4357/ab24e5}

\bibitem[{{Anand} {et~al.}(2021){Anand}, {Rizzi}, {Tully}, {Shaya},
  {Karachentsev}, {Makarov}, {Makarova}, {Wu}, {Dolphin}, \&
  {Kourkchi}}]{2021AJ....162...80A}
{Anand}, G.~S., {Rizzi}, L., {Tully}, R.~B., {et~al.} 2021, \aj, 162, 80,
  \dodoi{10.3847/1538-3881/ac0440}

\bibitem[{{Armandroff} \& {Zinn}(1988)}]{1988AJ.....96...92A}
{Armandroff}, T.~E., \& {Zinn}, R. 1988, \aj, 96, 92, \dodoi{10.1086/114792}

\bibitem[{{Barnes} {et~al.}(2001){Barnes}, {Staveley-Smith}, {de Blok},
  {Oosterloo}, {Stewart}, {Wright}, {Banks}, {Bhathal}, {Boyce}, {Calabretta},
  {Disney}, {Drinkwater}, {Ekers}, {Freeman}, {Gibson}, {Green}, {Haynes}, {te
  Lintel Hekkert}, {Henning}, {Jerjen}, {Juraszek}, {Kesteven}, {Kilborn},
  {Knezek}, {Koribalski}, {Kraan-Korteweg}, {Malin}, {Marquarding}, {Minchin},
  {Mould}, {Price}, {Putman}, {Ryder}, {Sadler}, {Schr{\"o}der}, {Stootman},
  {Webster}, {Wilson}, \& {Ye}}]{2001MNRAS.322..486B}
{Barnes}, D.~G., {Staveley-Smith}, L., {de Blok}, W.~J.~G., {et~al.} 2001,
  \mnras, 322, 486, \dodoi{10.1046/j.1365-8711.2001.04102.x}

\bibitem[{{Benitez-Llambay} \& {Fumagalli}(2021)}]{2021ApJ...921L...9B}
{Benitez-Llambay}, A., \& {Fumagalli}, M. 2021, \apjl, 921, L9,
  \dodoi{10.3847/2041-8213/ac3006}

\bibitem[{{Ben{\'\i}tez-Llambay} {et~al.}(2015){Ben{\'\i}tez-Llambay},
  {Navarro}, {Abadi}, {Gottl{\"o}ber}, {Yepes}, {Hoffman}, \&
  {Steinmetz}}]{2015MNRAS.450.4207B}
{Ben{\'\i}tez-Llambay}, A., {Navarro}, J.~F., {Abadi}, M.~G., {et~al.} 2015,
  \mnras, 450, 4207, \dodoi{10.1093/mnras/stv925}

\bibitem[{{Bertschinger}(1985)}]{1985ApJS...58...39B}
{Bertschinger}, E. 1985, \apjs, 58, 39, \dodoi{10.1086/191028}

\bibitem[{{Bialas} {et~al.}(2015){Bialas}, {Lisker}, {Olczak}, {Spurzem}, \&
  {Kotulla}}]{2015A&A...576A.103B}
{Bialas}, D., {Lisker}, T., {Olczak}, C., {Spurzem}, R., \& {Kotulla}, R. 2015,
  \aap, 576, A103, \dodoi{10.1051/0004-6361/201425235}

\bibitem[{{Brunker} {et~al.}(2019){Brunker}, {McQuinn}, {Salzer}, {Cannon},
  {Janowiecki}, {Leisman}, {Rhode}, {Adams}, {Ball}, {Dolphin}, {Giovanelli},
  \& {Haynes}}]{2019AJ....157...76B}
{Brunker}, S.~W., {McQuinn}, K. B.~W., {Salzer}, J.~J., {et~al.} 2019, \aj,
  157, 76, \dodoi{10.3847/1538-3881/aafb39}

\bibitem[{{Carlin} {et~al.}(2021){Carlin}, {Mutlu-Pakdil}, {Crnojevi{\'c}},
  {Garling}, {Karunakaran}, {Peter}, {Tollerud}, {Forbes}, {Hargis}, {Lim},
  {Romanowsky}, {Sand}, {Spekkens}, \& {Strader}}]{2021ApJ...909..211C}
{Carlin}, J.~L., {Mutlu-Pakdil}, B., {Crnojevi{\'c}}, D., {et~al.} 2021, \apj,
  909, 211, \dodoi{10.3847/1538-4357/abe040}

\bibitem[{{Carlsten} {et~al.}(2020){Carlsten}, {Greco}, {Beaton}, \&
  {Greene}}]{2020ApJ...891..144C}
{Carlsten}, S.~G., {Greco}, J.~P., {Beaton}, R.~L., \& {Greene}, J.~E. 2020,
  \apj, 891, 144, \dodoi{10.3847/1538-4357/ab7758}

\bibitem[{{Carlsten} {et~al.}(2022){Carlsten}, {Greene}, {Beaton}, {Danieli},
  \& {Greco}}]{2022ApJ...933...47C}
{Carlsten}, S.~G., {Greene}, J.~E., {Beaton}, R.~L., {Danieli}, S., \& {Greco},
  J.~P. 2022, \apj, 933, 47, \dodoi{10.3847/1538-4357/ac6fd7}

\bibitem[{{Castignani} {et~al.}(2022){Castignani}, {Combes}, {Jablonka},
  {Finn}, {Rudnick}, {Vulcani}, {Desai}, {Zaritsky}, \&
  {Salom{\'e}}}]{2022A&A...657A...9C}
{Castignani}, G., {Combes}, F., {Jablonka}, P., {et~al.} 2022, \aap, 657, A9,
  \dodoi{10.1051/0004-6361/202040141}

\bibitem[{{Chazov} {et~al.}(2026){Chazov}, {Makarov}, {Tully}, {Anand},
  {Makarova}, {Cohen}, {Blakeslee}, {Cantiello}, {Jensen}, \&
  {Raimondo}}]{2026arXiv260311160C}
{Chazov}, M.~I., {Makarov}, D.~I., {Tully}, R.~B., {et~al.} 2026, arXiv
  e-prints, arXiv:2603.11160, \dodoi{10.48550/arXiv.2603.11160}

\bibitem[{{Cluver} {et~al.}(2017){Cluver}, {Jarrett}, {Dale}, {Smith},
  {August}, \& {Brown}}]{2017ApJ...850...68C}
{Cluver}, M.~E., {Jarrett}, T.~H., {Dale}, D.~A., {et~al.} 2017, \apj, 850, 68,
  \dodoi{10.3847/1538-4357/aa92c7}

\bibitem[{{Cluver} {et~al.}(2014){Cluver}, {Jarrett}, {Hopkins}, {Driver},
  {Liske}, {Gunawardhana}, {Taylor}, {Robotham}, {Alpaslan}, {Baldry}, {Brown},
  {Peacock}, {Popescu}, {Tuffs}, {Bauer}, {Bland-Hawthorn}, {Colless},
  {Holwerda}, {Lara-L{\'o}pez}, {Leschinski}, {L{\'o}pez-S{\'a}nchez},
  {Norberg}, {Owers}, {Wang}, \& {Wilkins}}]{2014ApJ...782...90C}
{Cluver}, M.~E., {Jarrett}, T.~H., {Hopkins}, A.~M., {et~al.} 2014, \apj, 782,
  90, \dodoi{10.1088/0004-637X/782/2/90}

\bibitem[{{Cook} {et~al.}(2014){Cook}, {Dale}, {Johnson}, {Van Zee}, {Lee},
  {Kennicutt}, {Calzetti}, {Staudaher}, \& {Engelbracht}}]{2014MNRAS.445..899C}
{Cook}, D.~O., {Dale}, D.~A., {Johnson}, B.~D., {et~al.} 2014, \mnras, 445,
  899, \dodoi{10.1093/mnras/stu1787}

\bibitem[{{Courtois} {et~al.}(2012){Courtois}, {Hoffman}, {Tully}, \&
  {Gottl{\"o}ber}}]{2012ApJ...744...43C}
{Courtois}, H.~M., {Hoffman}, Y., {Tully}, R.~B., \& {Gottl{\"o}ber}, S. 2012,
  \apj, 744, 43, \dodoi{10.1088/0004-637X/744/1/43}

\bibitem[{{Courtois} {et~al.}(2013){Courtois}, {Pomar{\`e}de}, {Tully},
  {Hoffman}, \& {Courtois}}]{2013AJ....146...69C}
{Courtois}, H.~M., {Pomar{\`e}de}, D., {Tully}, R.~B., {Hoffman}, Y., \&
  {Courtois}, D. 2013, \aj, 146, 69, \dodoi{10.1088/0004-6256/146/3/69}

\bibitem[{{Da Costa} \& {Armandroff}(1990)}]{1990AJ....100..162D}
{Da Costa}, G.~S., \& {Armandroff}, T.~E. 1990, \aj, 100, 162,
  \dodoi{10.1086/115500}

\bibitem[{{Dalcanton} {et~al.}(2009){Dalcanton}, {Williams}, {Seth}, {Dolphin},
  {Holtzman}, {Rosema}, {Skillman}, {Cole}, {Girardi}, {Gogarten},
  {Karachentsev}, {Olsen}, {Weisz}, {Christensen}, {Freeman}, {Gilbert},
  {Gallart}, {Harris}, {Hodge}, {de Jong}, {Karachentseva}, {Mateo}, {Stetson},
  {Tavarez}, {Zaritsky}, {Governato}, \& {Quinn}}]{2009ApJS..183...67D}
{Dalcanton}, J.~J., {Williams}, B.~F., {Seth}, A.~C., {et~al.} 2009, \apjs,
  183, 67, \dodoi{10.1088/0067-0049/183/1/67}

\bibitem[{{Dale} {et~al.}(2009){Dale}, {Cohen}, {Johnson}, {Schuster},
  {Calzetti}, {Engelbracht}, {Gil de Paz}, {Kennicutt}, {Lee}, {Begum},
  {Block}, {Dalcanton}, {Funes}, {Gordon}, {Johnson}, {Marble}, {Sakai},
  {Skillman}, {van Zee}, {Walter}, {Weisz}, {Williams}, {Wu}, \&
  {Wu}}]{2009ApJ...703..517D}
{Dale}, D.~A., {Cohen}, S.~A., {Johnson}, L.~C., {et~al.} 2009, \apj, 703, 517,
  \dodoi{10.1088/0004-637X/703/1/517}

\bibitem[{{Davis} \& {Scrimgeour}(2014)}]{2014MNRAS.442.1117D}
{Davis}, T.~M., \& {Scrimgeour}, M.~I. 2014, \mnras, 442, 1117,
  \dodoi{10.1093/mnras/stu920}

\bibitem[{{de Vaucouleurs}(1953)}]{1953AJ.....58...30D}
{de Vaucouleurs}, G. 1953, \aj, 58, 30, \dodoi{10.1086/106805}

\bibitem[{{de Vaucouleurs}(1956)}]{1956VA......2.1584D}
---. 1956, Vistas in Astronomy, 2, 1584, \dodoi{10.1016/0083-6656(56)90087-3}

\bibitem[{{de Vaucouleurs}(1961)}]{1961ApJS....6..213D}
---. 1961, \apjs, 6, 213, \dodoi{10.1086/190064}

\bibitem[{{de Vaucouleurs}(1975)}]{1975gaun.book..557D}
---. 1975, {Nearby Groups of Galaxies}, ed. A.~{Sandage}, M.~{Sandage}, \&
  J.~{Kristian} (the University of Chicago Press), 557

\bibitem[{{de Vaucouleurs} {et~al.}(1991){de Vaucouleurs}, {de Vaucouleurs},
  {Corwin}, {Buta}, {Paturel}, \& {Fouqu{\'e}}}]{1991trcb.book.....D}
{de Vaucouleurs}, G., {de Vaucouleurs}, A., {Corwin}, Jr., H.~G., {et~al.}
  1991, {Third Reference Catalogue of Bright Galaxies} (Volume 1-3, XII, 2069
  pp.~7 figs..~ Springer-Verlag Berlin Heidelberg New York)

\bibitem[{{Dekel} \& {Silk}(1986)}]{1986ApJ...303...39D}
{Dekel}, A., \& {Silk}, J. 1986, \apj, 303, 39, \dodoi{10.1086/164050}

\bibitem[{{Diemer} \& {Kravtsov}(2014)}]{2014ApJ...789....1D}
{Diemer}, B., \& {Kravtsov}, A.~V. 2014, \apj, 789, 1,
  \dodoi{10.1088/0004-637X/789/1/1}

\bibitem[{{Dolphin} {et~al.}(2003){Dolphin}, {Saha}, {Skillman}, {Dohm-Palmer},
  {Tolstoy}, {Cole}, {Gallagher}, {Hoessel}, \& {Mateo}}]{2003AJ....125.1261D}
{Dolphin}, A.~E., {Saha}, A., {Skillman}, E.~D., {et~al.} 2003, \aj, 125, 1261,
  \dodoi{10.1086/346279}

\bibitem[{{Dressler}(1980)}]{1980ApJ...236..351D}
{Dressler}, A. 1980, \apj, 236, 351, \dodoi{10.1086/157753}

\bibitem[{{Dressler} {et~al.}(1987){Dressler}, {Faber}, {Burstein}, {Davies},
  {Lynden-Bell}, {Terlevich}, \& {Wegner}}]{1987ApJ...313L..37D}
{Dressler}, A., {Faber}, S.~M., {Burstein}, D., {et~al.} 1987, \apjl, 313, L37,
  \dodoi{10.1086/184827}

\bibitem[{{Eskew} {et~al.}(2012){Eskew}, {Zaritsky}, \&
  {Meidt}}]{2012AJ....143..139E}
{Eskew}, M., {Zaritsky}, D., \& {Meidt}, S. 2012, \aj, 143, 139,
  \dodoi{10.1088/0004-6256/143/6/139}

\bibitem[{{Fillmore} \& {Goldreich}(1984)}]{1984ApJ...281....1F}
{Fillmore}, J.~A., \& {Goldreich}, P. 1984, \apj, 281, 1,
  \dodoi{10.1086/162070}

\bibitem[{{Fisher} \& {Tully}(1981)}]{1981ApJS...47..139F}
{Fisher}, J.~R., \& {Tully}, R.~B. 1981, \apjs, 47, 139, \dodoi{10.1086/190755}

\bibitem[{{Font} {et~al.}(2022){Font}, {McCarthy}, {Belokurov}, {Brown}, \&
  {Stafford}}]{2022MNRAS.511.1544F}
{Font}, A.~S., {McCarthy}, I.~G., {Belokurov}, V., {Brown}, S.~T., \&
  {Stafford}, S.~G. 2022, \mnras, 511, 1544, \dodoi{10.1093/mnras/stac183}

\bibitem[{{Freedman}(2021)}]{2021ApJ...919...16F}
{Freedman}, W.~L. 2021, \apj, 919, 16, \dodoi{10.3847/1538-4357/ac0e95}

\bibitem[{{Grebel}(2011)}]{2011EAS....48..315G}
{Grebel}, E.~K. 2011, in EAS Publications Series, Vol.~48, EAS Publications
  Series, ed. M.~{Koleva}, P.~{Prugniel}, \& I.~{Vauglin} (EDP), 315--327,
  \dodoi{10.1051/eas/1148074}

\bibitem[{{Grebel} {et~al.}(2003){Grebel}, {Gallagher}, \&
  {Harbeck}}]{2003AJ....125.1926G}
{Grebel}, E.~K., {Gallagher}, III, J.~S., \& {Harbeck}, D. 2003, \aj, 125,
  1926, \dodoi{10.1086/368363}

\bibitem[{{Gunn} \& {Gott}(1972)}]{1972ApJ...176....1G}
{Gunn}, J.~E., \& {Gott}, III, J.~R. 1972, \apj, 176, 1, \dodoi{10.1086/151605}

\bibitem[{{Haynes} \& {Giovanelli}(1986)}]{1986ApJ...306L..55H}
{Haynes}, M.~P., \& {Giovanelli}, R. 1986, \apjl, 306, L55,
  \dodoi{10.1086/184705}

\bibitem[{{Haynes} {et~al.}(2018){Haynes}, {Giovanelli}, {Kent}, {Adams},
  {Balonek}, {Craig}, {Fertig}, {Finn}, {Giovanardi}, {Hallenbeck}, {Hess},
  {Hoffman}, {Huang}, {Jones}, {Koopmann}, {Kornreich}, {Leisman}, {Miller},
  {Moorman}, {O'Connor}, {O'Donoghue}, {Papastergis}, {Troischt}, {Stark}, \&
  {Xiao}}]{2018ApJ...861...49H}
{Haynes}, M.~P., {Giovanelli}, R., {Kent}, B.~R., {et~al.} 2018, \apj, 861, 49,
  \dodoi{10.3847/1538-4357/aac956}

\bibitem[{{Ho} {et~al.}(1995){Ho}, {Filippenko}, \&
  {Sargent}}]{1995ApJS...98..477H}
{Ho}, L.~C., {Filippenko}, A.~V., \& {Sargent}, W.~L. 1995, \apjs, 98, 477,
  \dodoi{10.1086/192170}

\bibitem[{{Hoyt} {et~al.}(2025){Hoyt}, {Jang}, {Freedman}, {Madore}, {Owens},
  \& {Lee}}]{2025arXiv250311769H}
{Hoyt}, T.~J., {Jang}, I.~S., {Freedman}, W.~L., {et~al.} 2025, arXiv e-prints,
  arXiv:2503.11769, \dodoi{10.48550/arXiv.2503.11769}

\bibitem[{{Huchra} \& {Geller}(1982)}]{1982ApJ...257..423H}
{Huchra}, J.~P., \& {Geller}, M.~J. 1982, \apj, 257, 423,
  \dodoi{10.1086/160000}

\bibitem[{{Huchtmeier} {et~al.}(2000{\natexlab{a}}){Huchtmeier},
  {Karachentsev}, \& {Karachentseva}}]{2000A&AS..147..187H}
{Huchtmeier}, W.~K., {Karachentsev}, I.~D., \& {Karachentseva}, V.~E.
  2000{\natexlab{a}}, \aaps, 147, 187, \dodoi{10.1051/aas:2000369}

\bibitem[{{Huchtmeier} {et~al.}(2000{\natexlab{b}}){Huchtmeier},
  {Karachentsev}, {Karachentseva}, \& {Ehle}}]{2000A&AS..141..469H}
{Huchtmeier}, W.~K., {Karachentsev}, I.~D., {Karachentseva}, V.~E., \& {Ehle},
  M. 2000{\natexlab{b}}, \aaps, 141, 469, \dodoi{10.1051/aas:2000324}

\bibitem[{{Jacobs} {et~al.}(2011){Jacobs}, {Tully}, {Rizzi}, {Karachentsev},
  {Chiboucas}, \& {Held}}]{2011AJ....141..106J}
{Jacobs}, B.~A., {Tully}, R.~B., {Rizzi}, L., {et~al.} 2011, \aj, 141, 106,
  \dodoi{10.1088/0004-6256/141/4/106}

\bibitem[{{Jang} {et~al.}(2021){Jang}, {Hoyt}, {Beaton}, {Freedman}, {Madore},
  {Lee}, {Neeley}, {Monson}, {Rich}, \& {Seibert}}]{2021ApJ...906..125J}
{Jang}, I.~S., {Hoyt}, T.~J., {Beaton}, R.~L., {et~al.} 2021, \apj, 906, 125,
  \dodoi{10.3847/1538-4357/abc8e9}

\bibitem[{{Jarrett} {et~al.}(2000){Jarrett}, {Chester}, {Cutri}, {Schneider},
  {Skrutskie}, \& {Huchra}}]{2000AJ....119.2498J}
{Jarrett}, T.~H., {Chester}, T., {Cutri}, R., {et~al.} 2000, \aj, 119, 2498,
  \dodoi{10.1086/301330}

\bibitem[{{Jarrett} {et~al.}(2003){Jarrett}, {Chester}, {Cutri}, {Schneider},
  \& {Huchra}}]{2003AJ....125..525J}
{Jarrett}, T.~H., {Chester}, T., {Cutri}, R., {Schneider}, S.~E., \& {Huchra},
  J.~P. 2003, \aj, 125, 525, \dodoi{10.1086/345794}

\bibitem[{{Jones} {et~al.}(2009){Jones}, {Read}, {Saunders}, {Colless},
  {Jarrett}, {Parker}, {Fairall}, {Mauch}, {Sadler}, {Watson}, {Burton},
  {Campbell}, {Cass}, {Croom}, {Dawe}, {Fiegert}, {Frankcombe}, {Hartley},
  {Huchra}, {James}, {Kirby}, {Lahav}, {Lucey}, {Mamon}, {Moore}, {Peterson},
  {Prior}, {Proust}, {Russell}, {Safouris}, {Wakamatsu}, {Westra}, \&
  {Williams}}]{2009MNRAS.399..683J}
{Jones}, D.~H., {Read}, M.~A., {Saunders}, W., {et~al.} 2009, \mnras, 399, 683,
  \dodoi{10.1111/j.1365-2966.2009.15338.x}

\bibitem[{{Kaisina} {et~al.}(2012){Kaisina}, {Makarov}, {Karachentsev}, \&
  {Kaisin}}]{2012AstBu..67..115K}
{Kaisina}, E.~I., {Makarov}, D.~I., {Karachentsev}, I.~D., \& {Kaisin}, S.~S.
  2012, Astrophysical Bulletin, 67, 115, \dodoi{10.1134/S1990341312010105}

\bibitem[{{Kalirai} {et~al.}(2010){Kalirai}, {Beaton}, {Geha}, {Gilbert},
  {Guhathakurta}, {Kirby}, {Majewski}, {Ostheimer}, {Patterson}, \&
  {Wolf}}]{2010ApJ...711..671K}
{Kalirai}, J.~S., {Beaton}, R.~L., {Geha}, M.~C., {et~al.} 2010, \apj, 711,
  671, \dodoi{10.1088/0004-637X/711/2/671}

\bibitem[{{Karachentsev} \& {Kashibadze}(2021)}]{2021AN....342..999K}
{Karachentsev}, I., \& {Kashibadze}, O. 2021, Astronomische Nachrichten, 342,
  999, \dodoi{10.1002/asna.20210018}

\bibitem[{{Karachentsev} {et~al.}(2025){Karachentsev}, {Chazov}, \&
  {Karachentseva}}]{2025arXiv250821494K}
{Karachentsev}, I.~D., {Chazov}, M.~I., \& {Karachentseva}, V.~E. 2025, arXiv
  e-prints, arXiv:2508.21494, \dodoi{10.48550/arXiv.2508.21494}

\bibitem[{{Karachentsev} {et~al.}(2024){Karachentsev}, {Karachentseva},
  {Kaisin}, \& {Zhang}}]{2024A&A...684L..24K}
{Karachentsev}, I.~D., {Karachentseva}, V.~E., {Kaisin}, S.~S., \& {Zhang},
  C.-P. 2024, \aap, 684, L24, \dodoi{10.1051/0004-6361/202449672}

\bibitem[{{Karachentsev} {et~al.}(2000{\natexlab{a}}){Karachentsev},
  {Karachentseva}, {Suchkov}, \& {Grebel}}]{2000A&AS..145..415K}
{Karachentsev}, I.~D., {Karachentseva}, V.~E., {Suchkov}, A.~A., \& {Grebel},
  E.~K. 2000{\natexlab{a}}, \aaps, 145, 415, \dodoi{10.1051/aas:2000249}

\bibitem[{{Karachentsev} {et~al.}(2009){Karachentsev}, {Kashibadze}, {Makarov},
  \& {Tully}}]{2009MNRAS.393.1265K}
{Karachentsev}, I.~D., {Kashibadze}, O.~G., {Makarov}, D.~I., \& {Tully}, R.~B.
  2009, \mnras, 393, 1265, \dodoi{10.1111/j.1365-2966.2008.14300.x}

\bibitem[{{Karachentsev} \& {Kutkin}(2005)}]{2005AstL...31..299K}
{Karachentsev}, I.~D., \& {Kutkin}, A.~M. 2005, Astronomy Letters, 31, 299,
  \dodoi{10.1134/1.1922528}

\bibitem[{{Karachentsev} {et~al.}(2013){Karachentsev}, {Makarov}, \&
  {Kaisina}}]{2013AJ....145..101K}
{Karachentsev}, I.~D., {Makarov}, D.~I., \& {Kaisina}, E.~I. 2013, \aj, 145,
  101, \dodoi{10.1088/0004-6256/145/4/101}

\bibitem[{{Karachentsev} {et~al.}(2023){Karachentsev}, {Makarova},
  {Koribalski}, {Anand}, {Tully}, \& {Kniazev}}]{2023MNRAS.518.5893K}
{Karachentsev}, I.~D., {Makarova}, L.~N., {Koribalski}, B.~S., {et~al.} 2023,
  \mnras, 518, 5893, \dodoi{10.1093/mnras/stac3284}

\bibitem[{{Karachentsev} {et~al.}(2015){Karachentsev}, {Tully}, {Makarova},
  {Makarov}, \& {Rizzi}}]{2015ApJ...805..144K}
{Karachentsev}, I.~D., {Tully}, R.~B., {Makarova}, L.~N., {Makarov}, D.~I., \&
  {Rizzi}, L. 2015, \apj, 805, 144, \dodoi{10.1088/0004-637X/805/2/144}

\bibitem[{{Karachentsev} {et~al.}(2000{\natexlab{b}}){Karachentsev},
  {Karachentseva}, {Dolphin}, {Geisler}, {Grebel}, {Guhathakurta}, {Hodge},
  {Sarajedini}, {Seitzer}, \& {Sharina}}]{2000A&A...363..117K}
{Karachentsev}, I.~D., {Karachentseva}, V.~E., {Dolphin}, A.~E., {et~al.}
  2000{\natexlab{b}}, \aap, 363, 117.
\newblock \doarXiv{astro-ph/0010146}

\bibitem[{{Karachentseva}(1968)}]{1968CoBAO..39...61K}
{Karachentseva}, V.~E. 1968, Communications of the Byurakan Astrophysical
  Observatory, 39, 61

\bibitem[{{Karachentseva} \& {Karachentsev}(2000)}]{2000A&AS..146..359K}
{Karachentseva}, V.~E., \& {Karachentsev}, I.~D. 2000, \aaps, 146, 359,
  \dodoi{10.1051/aas:2000275}

\bibitem[{{Karachentseva} {et~al.}(1999){Karachentseva}, {Karachentsev}, \&
  {Richter}}]{1999A&AS..135..221K}
{Karachentseva}, V.~E., {Karachentsev}, I.~D., \& {Richter}, G.~M. 1999, \aaps,
  135, 221, \dodoi{10.1051/aas:1999173}

\bibitem[{{Kennicutt} {et~al.}(2008){Kennicutt}, {Lee}, {Funes}, {J.}, {Sakai},
  \& {Akiyama}}]{2008ApJS..178..247K}
{Kennicutt}, Jr., R.~C., {Lee}, J.~C., {Funes}, J.~G., {et~al.} 2008, \apjs,
  178, 247, \dodoi{10.1086/590058}

\bibitem[{{Kim} {et~al.}(2016){Kim}, {Rey}, {Bureau}, {Yoon}, {Chung},
  {Jerjen}, {Lisker}, {Jeong}, {Sung}, {Lee}, {Lee}, \&
  {Chung}}]{2016ApJ...833..207K}
{Kim}, S., {Rey}, S.-C., {Bureau}, M., {et~al.} 2016, \apj, 833, 207,
  \dodoi{10.3847/1538-4357/833/2/207}

\bibitem[{{Kirby} {et~al.}(2011){Kirby}, {Cohen}, {Smith}, {Majewski}, {Sohn},
  \& {Guhathakurta}}]{2011ApJ...727...79K}
{Kirby}, E.~N., {Cohen}, J.~G., {Smith}, G.~H., {et~al.} 2011, \apj, 727, 79,
  \dodoi{10.1088/0004-637X/727/2/79}

\bibitem[{{Kourkchi} \& {Tully}(2017)}]{2017ApJ...843...16K}
{Kourkchi}, E., \& {Tully}, R.~B. 2017, \apj, 843, 16,
  \dodoi{10.3847/1538-4357/aa76db}

\bibitem[{{Kreckel} {et~al.}(2011){Kreckel}, {Peebles}, {van Gorkom}, {van de
  Weygaert}, \& {van der Hulst}}]{2011AJ....141..204K}
{Kreckel}, K., {Peebles}, P.~J.~E., {van Gorkom}, J.~H., {van de Weygaert}, R.,
  \& {van der Hulst}, J.~M. 2011, \aj, 141, 204,
  \dodoi{10.1088/0004-6256/141/6/204}

\bibitem[{{Lee} {et~al.}(2006){Lee}, {Skillman}, {Cannon}, {Jackson}, {Gehrz},
  {Polomski}, \& {Woodward}}]{2006ApJ...647..970L}
{Lee}, H., {Skillman}, E.~D., {Cannon}, J.~M., {et~al.} 2006, \apj, 647, 970,
  \dodoi{10.1086/505573}

\bibitem[{{Lee} {et~al.}(2011){Lee}, {Gil de Paz}, {Kennicutt}, {Bothwell},
  {Dalcanton}, {Jos{\'e} G. Funes S.}, {Johnson}, {Sakai}, {Skillman},
  {Tremonti}, \& {van Zee}}]{2011ApJS..192....6L}
{Lee}, J.~C., {Gil de Paz}, A., {Kennicutt}, Jr., R.~C., {et~al.} 2011, \apjs,
  192, 6, \dodoi{10.1088/0067-0049/192/1/6}

\bibitem[{{Lee} {et~al.}(1993){Lee}, {Freedman}, \&
  {Madore}}]{1993AJ....106..964L}
{Lee}, M.~G., {Freedman}, W.~L., \& {Madore}, B.~F. 1993, \aj, 106, 964,
  \dodoi{10.1086/116697}

\bibitem[{{Li} {et~al.}(2022){Li}, {Ji}, {Pace}, {Erkal}, {Koposov}, {Shipp},
  {Da Costa}, {Cullinane}, {Kuehn}, {Lewis}, {Mackey}, {Simpson}, {Zucker},
  {Ferguson}, {Martell}, {Bland-Hawthorn}, {Balbinot}, {Tavangar},
  {Drlica-Wagner}, {De Silva}, \& {Simon}}]{2022ApJ...928...30L}
{Li}, T.~S., {Ji}, A.~P., {Pace}, A.~B., {et~al.} 2022, \apj, 928, 30,
  \dodoi{10.3847/1538-4357/ac46d3}

\bibitem[{{Lim} {et~al.}(2017){Lim}, {Mo}, {Lu}, {Wang}, \&
  {Yang}}]{2017MNRAS.470.2982L}
{Lim}, S.~H., {Mo}, H.~J., {Lu}, Y., {Wang}, H., \& {Yang}, X. 2017, \mnras,
  470, 2982, \dodoi{10.1093/mnras/stx1462}

\bibitem[{{Madore} \& {Freedman}(1995)}]{1995AJ....109.1645M}
{Madore}, B.~F., \& {Freedman}, W.~L. 1995, \aj, 109, 1645,
  \dodoi{10.1086/117391}

\bibitem[{{Mager} {et~al.}(2008){Mager}, {Madore}, \&
  {Freedman}}]{2008ApJ...689..721M}
{Mager}, V.~A., {Madore}, B.~F., \& {Freedman}, W.~L. 2008, \apj, 689, 721,
  \dodoi{10.1086/592563}

\bibitem[{{Mainzer} {et~al.}(2014){Mainzer}, {Bauer}, {Cutri}, {Grav},
  {Masiero}, {Beck}, {Clarkson}, {Conrow}, {Dailey}, {Eisenhardt}, {Fabinsky},
  {Fajardo-Acosta}, {Fowler}, {Gelino}, {Grillmair}, {Heinrichsen}, {Kendall},
  {Kirkpatrick}, {Liu}, {Masci}, {McCallon}, {Nugent}, {Papin}, {Rice},
  {Royer}, {Ryan}, {Sevilla}, {Sonnett}, {Stevenson}, {Thompson}, {Wheelock},
  {Wiemer}, {Wittman}, {Wright}, \& {Yan}}]{2014ApJ...792...30M}
{Mainzer}, A., {Bauer}, J., {Cutri}, R.~M., {et~al.} 2014, \apj, 792, 30,
  \dodoi{10.1088/0004-637X/792/1/30}

\bibitem[{{Makarov} \& {Karachentsev}(2011)}]{2011MNRAS.412.2498M}
{Makarov}, D., \& {Karachentsev}, I. 2011, \mnras, 412, 2498,
  \dodoi{10.1111/j.1365-2966.2010.18071.x}

\bibitem[{{Makarov} {et~al.}(2011){Makarov}, {Karachentsev}, \&
  {Uklein}}]{2011EAS....48..289M}
{Makarov}, D., {Karachentsev}, I., \& {Uklein}, R. 2011, in EAS Publications
  Series, Vol.~48, EAS Publications Series, ed. M.~{Koleva}, P.~{Prugniel}, \&
  I.~{Vauglin}, 289--294, \dodoi{10.1051/eas/1148066}

\bibitem[{{Makarov} {et~al.}(2025){Makarov}, {Makarov}, {Makarova}, \&
  {Libeskind}}]{2025A&A...698A.178M}
{Makarov}, D., {Makarov}, D., {Makarova}, L., \& {Libeskind}, N. 2025, \aap,
  698, A178, \dodoi{10.1051/0004-6361/202554778}

\bibitem[{{Makarov} {et~al.}(2014){Makarov}, {Prugniel}, {Terekhova},
  {Courtois}, \& {Vauglin}}]{2014A&A...570A..13M}
{Makarov}, D., {Prugniel}, P., {Terekhova}, N., {Courtois}, H., \& {Vauglin},
  I. 2014, \aap, 570, A13, \dodoi{10.1051/0004-6361/201423496}

\bibitem[{{Makarov} \& {Uklein}(2012)}]{2012AstBu..67..135M}
{Makarov}, D.~I., \& {Uklein}, R.~I. 2012, Astrophysical Bulletin, 67, 135,
  \dodoi{10.1134/S1990341312020022}

\bibitem[{{Makarova} {et~al.}(2026){Makarova}, {Purytin}, {Tully}, \&
  {Anand}}]{2026arXiv260418183M}
{Makarova}, L.~N., {Purytin}, D.~G., {Tully}, R.~B., \& {Anand}, G.~S. 2026,
  arXiv e-prints, arXiv:2604.18183.
\newblock \doarXiv{2604.18183}

\bibitem[{{Makarova} {et~al.}(2023){Makarova}, {Tully}, {Anand}, {Lambert},
  {Sharina}, {Koribalski}, \& {Kraan-Korteweg}}]{2023ApJ...943..139M}
{Makarova}, L.~N., {Tully}, R.~B., {Anand}, G.~S., {et~al.} 2023, \apj, 943,
  139, \dodoi{10.3847/1538-4357/acb048}

\bibitem[{{McCall}(2014)}]{2014MNRAS.440..405M}
{McCall}, M.~L. 2014, \mnras, 440, 405, \dodoi{10.1093/mnras/stu199}

\bibitem[{{McIntyre} {et~al.}(2011){McIntyre}, {Minchin}, {Momjian}, {Henning},
  {Kaur}, \& {Parton}}]{2011ApJ...739L..26M}
{McIntyre}, T.~P., {Minchin}, R.~F., {Momjian}, E., {et~al.} 2011, \apjl, 739,
  L26, \dodoi{10.1088/2041-8205/739/1/L26}

\bibitem[{{McQuinn} {et~al.}(2015){McQuinn}, {Mitchell}, \&
  {Skillman}}]{2015ApJS..218...29M}
{McQuinn}, K. B.~W., {Mitchell}, N.~P., \& {Skillman}, E.~D. 2015, \apjs, 218,
  29, \dodoi{10.1088/0067-0049/218/2/29}

\bibitem[{{McQuinn} {et~al.}(2020){McQuinn}, {Berg}, {Skillman}, {Adams},
  {Cannon}, {Dolphin}, {Salzer}, {Giovanelli}, {Haynes}, {Hirschauer},
  {Janoweicki}, {Klapkowski}, \& {Rhode}}]{2020ApJ...891..181M}
{McQuinn}, K. B.~W., {Berg}, D.~A., {Skillman}, E.~D., {et~al.} 2020, \apj,
  891, 181, \dodoi{10.3847/1538-4357/ab7447}

\bibitem[{{Moore} {et~al.}(1998){Moore}, {Lake}, \&
  {Katz}}]{1998ApJ...495..139M}
{Moore}, B., {Lake}, G., \& {Katz}, N. 1998, \apj, 495, 139,
  \dodoi{10.1086/305264}

\bibitem[{{Nazarova} {et~al.}(2025){Nazarova}, {Cannon}, {Karachentsev},
  {Makarov}, {Chazov}, {Schisgal}, \& {St. John}}]{2025AJ....170...23N}
{Nazarova}, A.~E., {Cannon}, J.~M., {Karachentsev}, I.~D., {et~al.} 2025, \aj,
  170, 23, \dodoi{10.3847/1538-3881/add888}

\bibitem[{{Nazarova} {et~al.}(2026){Nazarova}, {Makarov}, {Karachentsev},
  {Zhang}, {Chazov}, \& {Zhu}}]{2026arXiv260409082N}
{Nazarova}, A.~E., {Makarov}, D.~I., {Karachentsev}, I.~D., {et~al.} 2026,
  arXiv e-prints, arXiv:2604.09082.
\newblock \doarXiv{2604.09082}

\bibitem[{{Neuzil} {et~al.}(2020){Neuzil}, {Mansfield}, \&
  {Kravtsov}}]{2020MNRAS.494.2600N}
{Neuzil}, M.~K., {Mansfield}, P., \& {Kravtsov}, A.~V. 2020, \mnras, 494, 2600,
  \dodoi{10.1093/mnras/staa898}

\bibitem[{{Nolthenius}(1993)}]{1993ApJS...85....1N}
{Nolthenius}, R. 1993, \apjs, 85, 1, \dodoi{10.1086/191753}

\bibitem[{{Oemler}(1974)}]{1974ApJ...194....1O}
{Oemler}, Jr., A. 1974, \apj, 194, 1, \dodoi{10.1086/153216}

\bibitem[{{Olex} {et~al.}(2024){Olex}, {Knebe}, {Libeskind}, {Makarov}, \&
  {Gottl{\"o}ber}}]{2024PASA...41...18O}
{Olex}, E., {Knebe}, A., {Libeskind}, N.~I., {Makarov}, D.~I., \&
  {Gottl{\"o}ber}, S. 2024, \pasa, 41, e018, \dodoi{10.1017/pasa.2024.21}

\bibitem[{{Pace}(2025)}]{2025OJAp....8E.142P}
{Pace}, A.~B. 2025, The Open Journal of Astrophysics, 8, 142,
  \dodoi{10.33232/001c.144859}

\bibitem[{{Paturel} {et~al.}(1996){Paturel}, {Garnier}, {Petit}, \&
  {Marthinet}}]{1996A&A...311...12P}
{Paturel}, G., {Garnier}, R., {Petit}, C., \& {Marthinet}, M.~C. 1996, \aap,
  311, 12

\bibitem[{{Peng} {et~al.}(2015){Peng}, {Maiolino}, \&
  {Cochrane}}]{2015Natur.521..192P}
{Peng}, Y., {Maiolino}, R., \& {Cochrane}, R. 2015, \nat, 521, 192,
  \dodoi{10.1038/nature14439}

\bibitem[{{Pomar{\`e}de} {et~al.}(2017){Pomar{\`e}de}, {Hoffman}, {Courtois},
  \& {Tully}}]{2017ApJ...845...55P}
{Pomar{\`e}de}, D., {Hoffman}, Y., {Courtois}, H.~M., \& {Tully}, R.~B. 2017,
  \apj, 845, 55, \dodoi{10.3847/1538-4357/aa7f78}

\bibitem[{{Reid} {et~al.}(2019){Reid}, {Pesce}, \&
  {Riess}}]{2019ApJ...886L..27R}
{Reid}, M.~J., {Pesce}, D.~W., \& {Riess}, A.~G. 2019, \apjl, 886, L27,
  \dodoi{10.3847/2041-8213/ab552d}

\bibitem[{{Rizzi} {et~al.}(2007){Rizzi}, {Tully}, {Makarov}, {Makarova},
  {Dolphin}, {Sakai}, \& {Shaya}}]{2007ApJ...661..815R}
{Rizzi}, L., {Tully}, R.~B., {Makarov}, D., {et~al.} 2007, \apj, 661, 815,
  \dodoi{10.1086/516566}

\bibitem[{{Rizzi} {et~al.}(2017){Rizzi}, {Tully}, {Shaya}, {Kourkchi}, \&
  {Karachentsev}}]{2017ApJ...835...78R}
{Rizzi}, L., {Tully}, R.~B., {Shaya}, E.~J., {Kourkchi}, E., \& {Karachentsev},
  I.~D. 2017, \apj, 835, 78, \dodoi{10.3847/1538-4357/835/1/78}

\bibitem[{{Sansom} {et~al.}(2015){Sansom}, {Thirlwall}, {Deakin},
  {V{\"a}is{\"a}nen}, {Kniazev}, \& {van Loon}}]{2015MNRAS.450.1338S}
{Sansom}, A.~E., {Thirlwall}, J.~J., {Deakin}, M.~A., {et~al.} 2015, \mnras,
  450, 1338, \dodoi{10.1093/mnras/stv690}

\bibitem[{{Saviane} {et~al.}(2008){Saviane}, {Momany}, {Da Costa}, {Rich}, \&
  {Hibbard}}]{2008ApJ...678..179S}
{Saviane}, I., {Momany}, Y., {Da Costa}, G.~S., {Rich}, R.~M., \& {Hibbard},
  J.~E. 2008, \apj, 678, 179, \dodoi{10.1086/533408}

\bibitem[{{Schechter}(1976)}]{1976ApJ...203..297S}
{Schechter}, P. 1976, \apj, 203, 297

\bibitem[{{Schweizer} {et~al.}(2008){Schweizer}, {Burns}, {Madore}, {Mager},
  {Phillips}, {Freedman}, {Boldt}, {Contreras}, {Folatelli}, {Gonz{\'a}lez},
  {Hamuy}, {Krzeminski}, {Morrell}, {Persson}, {Roth}, \&
  {Stritzinger}}]{2008AJ....136.1482S}
{Schweizer}, F., {Burns}, C.~R., {Madore}, B.~F., {et~al.} 2008, \aj, 136,
  1482, \dodoi{10.1088/0004-6256/136/4/1482}

\bibitem[{{Searle} \& {Sargent}(1972)}]{1972ApJ...173...25S}
{Searle}, L., \& {Sargent}, W. L.~W. 1972, \apj, 173, 25,
  \dodoi{10.1086/151398}

\bibitem[{{Sharina} {et~al.}(2008){Sharina}, {Karachentsev}, {Dolphin},
  {Karachentseva}, {Tully}, {Karataeva}, {Makarov}, {Makarova}, {Sakai},
  {Shaya}, {Nikolaev}, \& {Kuznetsov}}]{2008MNRAS.384.1544S}
{Sharina}, M.~E., {Karachentsev}, I.~D., {Dolphin}, A.~E., {et~al.} 2008,
  \mnras, 384, 1544, \dodoi{10.1111/j.1365-2966.2007.12814.x}

\bibitem[{{Shaya} \& {Tully}(1984)}]{1984ApJ...281...56S}
{Shaya}, E.~J., \& {Tully}, R.~B. 1984, \apj, 281, 56, \dodoi{10.1086/162074}

\bibitem[{{Shaya} {et~al.}(2017){Shaya}, {Tully}, {Hoffman}, \&
  {Pomar{\`e}de}}]{2017ApJ...850..207S}
{Shaya}, E.~J., {Tully}, R.~B., {Hoffman}, Y., \& {Pomar{\`e}de}, D. 2017,
  \apj, 850, 207, \dodoi{10.3847/1538-4357/aa9525}

\bibitem[{{Shaya} {et~al.}(2022){Shaya}, {Tully}, {Pomar{\`e}de}, \&
  {Peel}}]{2022ApJ...927..168S}
{Shaya}, E.~J., {Tully}, R.~B., {Pomar{\`e}de}, D., \& {Peel}, A. 2022, \apj,
  927, 168, \dodoi{10.3847/1538-4357/ac4f66}

\bibitem[{{Skillman} {et~al.}(2017){Skillman}, {Monelli}, {Weisz}, {Hidalgo},
  {Aparicio}, {Bernard}, {Boylan-Kolchin}, {Cassisi}, {Cole}, {Dolphin},
  {Ferguson}, {Gallart}, {Irwin}, {Martin}, {Mart{\'\i}nez-V{\'a}zquez},
  {Mayer}, {McConnachie}, {McQuinn}, {Navarro}, \&
  {Stetson}}]{2017ApJ...837..102S}
{Skillman}, E.~D., {Monelli}, M., {Weisz}, D.~R., {et~al.} 2017, \apj, 837,
  102, \dodoi{10.3847/1538-4357/aa60c5}

\bibitem[{{Smoot} \& {Lubin}(1979)}]{1979ApJ...234L..83S}
{Smoot}, G.~F., \& {Lubin}, P.~M. 1979, \apjl, 234, L83, \dodoi{10.1086/183114}

\bibitem[{{Somerville}(2002)}]{2002ApJ...572L..23S}
{Somerville}, R.~S. 2002, \apjl, 572, L23, \dodoi{10.1086/341444}

\bibitem[{{Thilker} {et~al.}(2010){Thilker}, {Bianchi}, {Schiminovich}, {Gil de
  Paz}, {Seibert}, {Madore}, {Wyder}, {Rich}, {Yi}, {Barlow}, {Conrow},
  {Forster}, {Friedman}, {Martin}, {Morrissey}, {Neff}, \&
  {Small}}]{2010ApJ...714L.171T}
{Thilker}, D.~A., {Bianchi}, L., {Schiminovich}, D., {et~al.} 2010, \apjl, 714,
  L171, \dodoi{10.1088/2041-8205/714/1/L171}

\bibitem[{{Tremonti} {et~al.}(2004){Tremonti}, {Heckman}, {Kauffmann},
  {Brinchmann}, {Charlot}, {White}, {Seibert}, {Peng}, {Schlegel}, {Uomoto},
  {Fukugita}, \& {Brinkmann}}]{2004ApJ...613..898T}
{Tremonti}, C.~A., {Heckman}, T.~M., {Kauffmann}, G., {et~al.} 2004, \apj, 613,
  898, \dodoi{10.1086/423264}

\bibitem[{{Trentham} \& {Tully}(2009)}]{2009MNRAS.398..722T}
{Trentham}, N., \& {Tully}, R.~B. 2009, \mnras, 398, 722,
  \dodoi{10.1111/j.1365-2966.2009.15189.x}

\bibitem[{{Tully}(1986)}]{1986ApJ...303...25T}
{Tully}, R.~B. 1986, \apj, 303, 25, \dodoi{10.1086/164049}

\bibitem[{{Tully}(1987)}]{1987ApJ...321..280T}
---. 1987, \apj, 321, 280, \dodoi{10.1086/165629}

\bibitem[{{Tully}(1988)}]{1988ngc..book.....T}
---. 1988, {Nearby galaxies catalog} (Cambridge and New York, Cambridge
  University Press, 1988, 221 p.)

\bibitem[{{Tully}(2015{\natexlab{a}})}]{2015AJ....149...54T}
---. 2015{\natexlab{a}}, \aj, 149, 54, \dodoi{10.1088/0004-6256/149/2/54}

\bibitem[{{Tully}(2015{\natexlab{b}})}]{2015AJ....149..171T}
---. 2015{\natexlab{b}}, \aj, 149, 171, \dodoi{10.1088/0004-6256/149/5/171}

\bibitem[{{Tully} {et~al.}(2025){Tully}, {Anand}, {Karachentsev}, \&
  {Shaya}}]{2025hst..prop18070T}
{Tully}, R.~B., {Anand}, G.~S., {Karachentsev}, I.~D., \& {Shaya}, E.~J. 2025,
  {SNAP Observations of All Unobserved Galaxies within 10 Mpc}, HST Proposal.
  Cycle 33, ID. \#18070

\bibitem[{{Tully} {et~al.}(2014){Tully}, {Courtois}, {Hoffman}, \&
  {Pomar{\`e}de}}]{2014Natur.513...71T}
{Tully}, R.~B., {Courtois}, H., {Hoffman}, Y., \& {Pomar{\`e}de}, D. 2014,
  \nat, 513, 71, \dodoi{10.1038/nature13674}

\bibitem[{{Tully} \& {Fisher}(1987)}]{1987nga..book.....T}
{Tully}, R.~B., \& {Fisher}, J.~R. 1987, {Nearby galaxies Atlas} (Cambridge:
  University Press, 1987)

\bibitem[{{Tully} {et~al.}(2026{\natexlab{a}}){Tully}, {Kourkchi}, \&
  {Neill}}]{2026arXiv260605780T}
{Tully}, R.~B., {Kourkchi}, E., \& {Neill}, J.~D. 2026{\natexlab{a}}, arXiv
  e-prints, arXiv:2606.05780, \dodoi{10.48550/arXiv.2606.05780}

\bibitem[{{Tully} {et~al.}(2019){Tully}, {Pomar{\`e}de}, {Graziani},
  {Courtois}, {Hoffman}, \& {Shaya}}]{2019ApJ...880...24T}
{Tully}, R.~B., {Pomar{\`e}de}, D., {Graziani}, R., {et~al.} 2019, \apj, 880,
  24, \dodoi{10.3847/1538-4357/ab2597}

\bibitem[{{Tully} {et~al.}(2008){Tully}, {Shaya}, {Karachentsev}, {Courtois},
  {Kocevski}, {Rizzi}, \& {Peel}}]{2008ApJ...676..184T}
{Tully}, R.~B., {Shaya}, E.~J., {Karachentsev}, I.~D., {et~al.} 2008, \apj,
  676, 184, \dodoi{10.1086/527428}

\bibitem[{{Tully} {et~al.}(2006){Tully}, {Rizzi}, {Dolphin}, {Karachentsev},
  {Karachentseva}, {Makarov}, {Makarova}, {Sakai}, \&
  {Shaya}}]{2006AJ....132..729T}
{Tully}, R.~B., {Rizzi}, L., {Dolphin}, A.~E., {et~al.} 2006, \aj, 132, 729,
  \dodoi{10.1086/505466}

\bibitem[{{Tully} {et~al.}(2026{\natexlab{b}}){Tully}, {Karachentseva},
  {Karachentsev}, {Anand}, {Chazov}, {Kaisan}, {Kourkchi}, {Makarova},
  {Pustilnik}, {Shaya}, \& {Tepliakova}}]{2026arXiv260904437T}
{Tully}, R.~B., {Karachentseva}, V.~E., {Karachentsev}, I.~D., {et~al.}
  2026{\natexlab{b}}, arXiv e-prints, arXiv:2609.04437.
\newblock \doarXiv{2609.04437}

\bibitem[{{van den Bergh}(1959)}]{1959PDDO....2..147V}
{van den Bergh}, S. 1959, Publications of the David Dunlap Observatory, 2, 147

\bibitem[{{van der Marel} {et~al.}(2012{\natexlab{a}}){van der Marel}, {Besla},
  {Cox}, {Sohn}, \& {Anderson}}]{2012ApJ...753....9V}
{van der Marel}, R.~P., {Besla}, G., {Cox}, T.~J., {Sohn}, S.~T., \&
  {Anderson}, J. 2012{\natexlab{a}}, \apj, 753, 9,
  \dodoi{10.1088/0004-637X/753/1/9}

\bibitem[{{van der Marel} {et~al.}(2012{\natexlab{b}}){van der Marel},
  {Fardal}, {Besla}, {Beaton}, {Sohn}, {Anderson}, {Brown}, \&
  {Guhathakurta}}]{2012ApJ...753....8V}
{van der Marel}, R.~P., {Fardal}, M., {Besla}, G., {et~al.} 2012{\natexlab{b}},
  \apj, 753, 8, \dodoi{10.1088/0004-637X/753/1/8}

\bibitem[{{Weisz} {et~al.}(2014){Weisz}, {Dolphin}, {Skillman}, {Holtzman},
  {Gilbert}, {Dalcanton}, \& {Williams}}]{2014ApJ...789..148W}
{Weisz}, D.~R., {Dolphin}, A.~E., {Skillman}, E.~D., {et~al.} 2014, \apj, 789,
  148, \dodoi{10.1088/0004-637X/789/2/148}

\bibitem[{{Weisz} {et~al.}(2011){Weisz}, {Dalcanton}, {Williams}, {Gilbert},
  {Skillman}, {Seth}, {Dolphin}, {McQuinn}, {Gogarten}, {Holtzman}, {Rosema},
  {Cole}, {Karachentsev}, \& {Zaritsky}}]{2011ApJ...739....5W}
{Weisz}, D.~R., {Dalcanton}, J.~J., {Williams}, B.~F., {et~al.} 2011, \apj,
  739, 5, \dodoi{10.1088/0004-637X/739/1/5}

\bibitem[{{Werner} {et~al.}(2004){Werner}, {Roellig}, {Low}, {Rieke}, {Rieke},
  {Hoffmann}, {Young}, {Houck}, {Brandl}, {Fazio}, {Hora}, {Gehrz}, {Helou},
  {Soifer}, {Stauffer}, {Keene}, {Eisenhardt}, {Gallagher}, {Gautier}, {Irace},
  {Lawrence}, {Simmons}, {Van Cleve}, {Jura}, {Wright}, \&
  {Cruikshank}}]{2004ApJS..154....1W}
{Werner}, M.~W., {Roellig}, T.~L., {Low}, F.~J., {et~al.} 2004, \apjs, 154, 1,
  \dodoi{10.1086/422992}

\bibitem[{{Worthey}(1994)}]{1994ApJS...95..107W}
{Worthey}, G. 1994, \apjs, 95, 107, \dodoi{10.1086/192096}

\bibitem[{{Wright} {et~al.}(2010){Wright}, {Eisenhardt}, {Mainzer}, {Ressler},
  {Cutri}, {Jarrett}, {Kirkpatrick}, {Padgett}, {McMillan}, {Skrutskie},
  {Stanford}, {Cohen}, {Walker}, {Mather}, {Leisawitz}, {Gautier}, {McLean},
  {Benford}, {Lonsdale}, {Blain}, {Mendez}, {Irace}, {Duval}, {Liu}, {Royer},
  {Heinrichsen}, {Howard}, {Shannon}, {Kendall}, {Walsh}, {Larsen}, {Cardon},
  {Schick}, {Schwalm}, {Abid}, {Fabinsky}, {Naes}, \&
  {Tsai}}]{2010AJ....140.1868W}
{Wright}, E.~L., {Eisenhardt}, P. R.~M., {Mainzer}, A.~K., {et~al.} 2010, \aj,
  140, 1868, \dodoi{10.1088/0004-6256/140/6/1868}

\bibitem[{{Wu} {et~al.}(2014){Wu}, {Tully}, {Rizzi}, {Dolphin}, {Jacobs}, \&
  {Karachentsev}}]{2014AJ....148....7W}
{Wu}, P.-F., {Tully}, R.~B., {Rizzi}, L., {et~al.} 2014, \aj, 148, 7,
  \dodoi{10.1088/0004-6256/148/1/7}

\bibitem[{{Zhang} {et~al.}(2024){Zhang}, {Zhu}, {Jiang}, {Cheng}, {Wang},
  {Wang}, {Xu}, {Liu}, {Yu}, {Qian}, {Yu}, {Ai}, {Jing}, {Xu}, {Liu}, {Guan},
  {Sun}, {Yang}, {Huang}, {Hao}, \& {FAST Collaboration}}]{2024SCPMA..6719511Z}
{Zhang}, C.-P., {Zhu}, M., {Jiang}, P., {et~al.} 2024, Science China Physics,
  Mechanics, and Astronomy, 67, 219511, \dodoi{10.1007/s11433-023-2219-7}

\end{thebibliography}
\bibliographystyle{aasjournal}

\end{document}